\documentclass[backref,twocolumn,breaklinks,colorlinks]{aastex7}   
\hypersetup{linkcolor=blue,citecolor=blue,filecolor=cyan,urlcolor=magenta}

\usepackage{enumitem}
\usepackage{mathtools}
\usepackage{graphicx}
\usepackage{amssymb}
\usepackage{amsmath}
\usepackage{ulem}
\usepackage{epstopdf}
\usepackage{color}
\usepackage{aas_macros}
\usepackage{enumitem}  
\usepackage{hyperref}
\usepackage{xcolor}
\newcommand{\AngledUpArrow}[1]{%
  \ensuremath{\rotatebox[origin=c]{#1}{$\uparrow$}}%
}

\newcommand{\AngledDownArrow}[1]{%
  \ensuremath{\rotatebox[origin=c]{#1}{$\downarrow$}}%
}

\begin{document}
\tighten

\author[0009-0001-8783-1004]{Dimitrios Skiathas}
\affiliation{Southeastern Universities Research Association, Washington, DC 20005, USA}
\affiliation{Astrophysics Science Division, NASA Goddard Space Flight Center, Greenbelt, MD 20771, USA.}
\affiliation{Center for Research and Exploration in Space Science and Technology, NASA/GSFC, Greenbelt, MD 20771, USA}
\affiliation{Department of Physics, University of Patras, Rio, 26504, Greece}
\email[show]{dimitrios.skiathas@nasa.gov}

\author[0000-0003-1080-5286]{Constantinos Kalapotharakos}
\affiliation{Astrophysics Science Division, NASA Goddard Space Flight Center, Greenbelt, MD 20771, USA.}
\email{Constantinos.kalapotharakos@nasa.gov}

\author[0000-0002-9249-0515]{Zorawar Wadiasingh}
\affiliation{Department of Astronomy, University of Maryland, College Park, MD 20742, USA}
\affiliation{Astrophysics Science Division, NASA Goddard Space Flight Center, Greenbelt, MD 20771, USA.}
\affiliation{Center for Research and Exploration in Space Science and Technology, NASA/GSFC, Greenbelt, Maryland 20771, USA}
\email{zorowar.wadiasingh@nasa.gov}
\author[0000-0002-7435-7809]{Demosthenes Kazanas}
\affiliation{Astrophysics Science Division, NASA Goddard Space Flight Center, Greenbelt, MD 20771, USA.}
\email{demos.kazanas@nasa.gov}
\author[0000-0001-6119-859X]{Alice K. Harding }
\affiliation{Theoretical Division, Los Alamos National Laboratory, Los Alamos, NM 87545, USA}
\email{ahardingx@yahoo.com}
\author[0009-0009-0746-7790]{Paul T. Kolbeck}
\affiliation{Center for Experimental Nuclear Physics and Astrophysics and Department of Physics, University of Washington, Seattle, WA 98195, USA}
\email{kolbeckpaul@gmail.com}

\title{Magnetosphere Evolution and Precursor-Driven Electromagnetic Signals in Merging Binary Neutron Stars}

\begin{abstract}
We detail new force-free simulations to investigate magnetosphere evolution and precursor electromagnetic (EM) signals from binary neutron stars. Our simulations fully follow a representative inspiral motion, capturing the intricate magnetospheric dynamics and their impact on EM outflows. We explore a range of stellar magnetic moment orientations and relative strengths, finding that the magnetospheres and Poynting flux evolution are strongly configuration dependent. The Poynting flux exhibits pulsations at twice the orbital frequency, $2\Omega$, and is highly anisotropic, following a power-law dependence on orbital frequency. The index ranges from 1 to 6, shaped by the intricate dynamics of the magnetospheres. Furthermore, we present the first computation of (1) the EM forces acting on the star surfaces, revealing the presence of torques that, for highly magnetized stars, could influence the orbital dynamics or break the crust; (2) the high-energy emission signals from these systems by adopting the established isolated pulsar theory. Assuming curvature radiation in the radiation-reaction limit, we find that photons could reach TeV--PeV energies in the last $\sim\rm{ms}$ for magnetic field strengths $10^{10}-10^{15}$~G. However, our analysis of single photon magnetic pair production suggests that these photons are unlikely to escape, with the MeV band emerging as a promising observational window for precursor high-energy emission. In this framework, we construct high-energy emission skymaps and light curves, exploring observational implications. Finally, we propose potential precursor radio emission and delayed afterglow echoes from magnetized outflows, which may contribute to late-time rebrightening in short gamma-ray bursts or to orphan afterglows.
\end{abstract}


\section{Introduction} 
\label{sec:intro}

Binary neutron star (BNS) mergers are some of the most extraordinary events of our Universe \citep[e.g.,][]{2020LRR....23....4B}. During the final moments of their inspiral, significant amounts of energy ( $\sim 0.01\, M_\odot c^2$) are liberated in the form of gravitational waves (GWs). The joint detection GW170817/GRB170817 \citep{2017PhRvL.119p1101A,2017ApJ...848L..14G} established the long-predicted connection between the BNS mergers and some gamma-ray bursts \citep[GRBs;][]{1989Natur.340..126E}, with the subsequent `kilonova' proving them as r-process sites of heavy elements.

Although the vast majority of the electromagnetic (EM) emission is produced after the merger and, consequently, a large number of studies are conducted on that stage, involving both observational discussions and numerical simulations, less attention has been given to precursor EM emission at earlier stages just prior to the merger \citep[see][for a review]{Suvorov2024}. 

It is well known that pulsars are broadband emitters from radio up to GeV and TeV $\gamma$-rays \citep{2023NatAs...7.1341H,3PC}. In rotation-powered pulsars, today's consensus holds that high-energy emission originates predominantly in the equatorial current sheet extending beyond the light cylinder in force-free (FF) configurations \citep{2010ApJ...715.1282B,2010MNRAS.404..767C,Kalapotharakos2014,2016MNRAS.457.2401C,2017ApJ...842...80K,Kalapotharakos2018,2018ApJ...855...94P,2019ApJ...874..166C,Kalapotharakos2023,2024arXiv241202307C}. This region, where strong electric fields accelerate particles, plays a central role in the production of high-energy photons. Additionally, it appears that the efficiency of electron-positron pair production within the magnetosphere regulates the overall efficiency of high-energy emission, acting as a critical factor in determining the observed luminosity and spectral characteristics. Similar physical processes ought to operate in the BNS magnetospheres, which, just prior to coalescence, attain orbital periods of milliseconds. Considered a system in the far-field, the conditions due to orbital motion are not too dissimilar from an isolated pulsar with a significant multipolar field structure.

However, prior to the coalescence, the two NS magnetospheres strongly interact, and the concept of a well-defined light cylinder, applicable only in the case of synchronized stars, no longer holds. There is rapid evolution of the EM fields and global magnetosphere structure, with intrinsic time dependence. Field lines bend and reconnect asymmetrically between the two stars, potentially powering precursors and influencing the inspiral orbital dynamics, particularly if either NS possesses a strong, i.e., magnetar-like magnetic field.

These precursor signals contain additional information about the nature of the merging objects and complement those derived from GWs and the postmerger signals. These include improved localization, binary parameters, viewing geometry and orientation, and NS equation of state.

There are several GRBs, likely originating from mergers, that exhibit precursors \citep{2010ApJ...723.1711T,2019ApJ...884...25Z,2020PhRvD.102j3014C,2020ApJ...902L..42W}. However, in the absence of joint GW detections, it remains challenging to assess whether they are truly precursors to the GRB, components of the GRB prompt emission, or associated with phenomena that occur after the merger but prior to the main GRB. Understanding the beaming patterns of precursors and their evolution up to the merger can bring valuable insight into these events.

In this direction, there are a few studies discussing the precursor signals originating from magnetospheric interactions of BNSs. These studies propose estimates for EM luminosities, or Poynting fluxes using simplified considerations, invoking the unipolar inductor model \cite{Goldreich&Lynden-Bell1969}, assuming the simpler case of binaries with one low and one high field companion \citep{Hansen&Lyutikov2001,Lai2012,Piro2012,Gourgouliatos&Lynden-Bell2019}, or in more general binaries of similar fields \citep{1996A&A...312..937L,Medvedev&Loeb,Lyutikov2019,Cooper2023}. Nonetheless, the rich complexity of the nontrivial magnetospheric interactions and the plethora of the involved parameters (even for circular inspirals) require numerical simulations. 

\cite{Palenzueala2013b,Palenzueala2013a} used resistive magnetohydrodynamics (MHD) simulations to study the magnetospheres of inspiral BNSs and calculated the Poynting flux evolution for three cases of magnetic field configurations with aligned, antialigned and unequal magnetic moments, followed by cases with perpendicular magnetic moments \citep{2014PhRvD..90d4007P}, highlighting the magnetosphere diversity and rich phenomenology of their EM emissions. \citet{Most&Philippov2020,Most&Philippov2022} using special-relativistic FF electrodynamics simulations, considered cases with fixed separation between the neutron stars (NSs) and explored magnetic field topologies including the antialigned case, introducing also inclined dipoles and quadrupole fields on the stars. Their study focused on energy dissipation within BNS magnetospheres and the generation of precursor flares during the late inspiral. More recently, \cite{ Mahlmann&Beloborodov}, with similar simulations, investigated the rise of Kelvin-Helmholtz instability at the interface between two NSs with aligned magnetic moments as well as cases where the second star has an inclined magnetic moment,  claiming that dissipation occurs even without flux tubes connecting the two stars. \cite{Ortiz2022} adopted an enclosing surface approximation (a perfectly conducting sphere around the area where the two NSs orbit) to simulate the magnetosphere of synchronized NSs in a binary system at constant separation. Although not a physically realistic scenario, this approach enabled them to numerically calculate accurate fields far away from the system center allowing them to calculate the high-energy skymaps and light curves assuming emissions originate around the separatrix layer and the current sheet. In general, such simulations \citep[see also][]{2019A&A...622A.161C,2020PhRvD.101f3017C} are essential for capturing qualitative details that cannot be addressed by analytic estimates offering quantitative insights that extend beyond the limitations of simple magnetostatic and vacuum approximations \citep[e.g.,][]{1990MNRAS.244..731K,2000ApJ...537..327I,2003PhRvD..68l4006V,2022GReGr..54..146L,2023A&A...675A..32A,2024RAA....24k5002T}.

In this work, we comprehensively survey the parameter space of allowed magnetic field configurations of BNS during their inspirals. Using numerical FF MHD simulations, we model the magnetosphere structure, we calculate the outward-propagated EM flux and its sky distribution for a variety of cases involving different star magnetic field ratios and alignments. In addition to the directly observed EM outflows, the nonuniform Poynting flux distribution indicates the potential for ``echoes," where the outward EM flux interacts with material in the local environment and may be observed postmerger.

We also present the first calculation of the backreaction EM torques experienced by each NS component.  We analyze our results to determine the evolution of the Poynting flux and its dependence on the orbital frequency up to the merger, finding nontrivial dependence varying for different set of parameters. Furthermore, invoking observationally validated theoretical approaches from isolated pulsar studies, we investigate the production and emission of high-energy photons and examine the emission patterns and spectral properties. 

This paper is structured as follows: In \S\ref{sec:methods}, we describe the numerical methods used in this work, along with parameters and simulation setup. In \S\ref{sec:results}, we present our results, beginning with a detailed description of the magnetosphere structure and Poynting flux dependence. We then compute the forces on the NSs and the resulting torques, discussing their potential impact on orbital dynamics. Additionally, we examine high-energy emissions, deriving their beaming patterns and light curves while also addressing gamma-ray attenuation due to magnetic pair production. Finally, in \S\ref{sec:summary}, we conclude with a discussion of the observational implications of our results, while \S\ref{sec:Summary and Future Directions} provides a concise summary and outlines pathways for advancing future modeling efforts.

\begin{figure*}
\centering
    \includegraphics[width=0.7\textwidth]{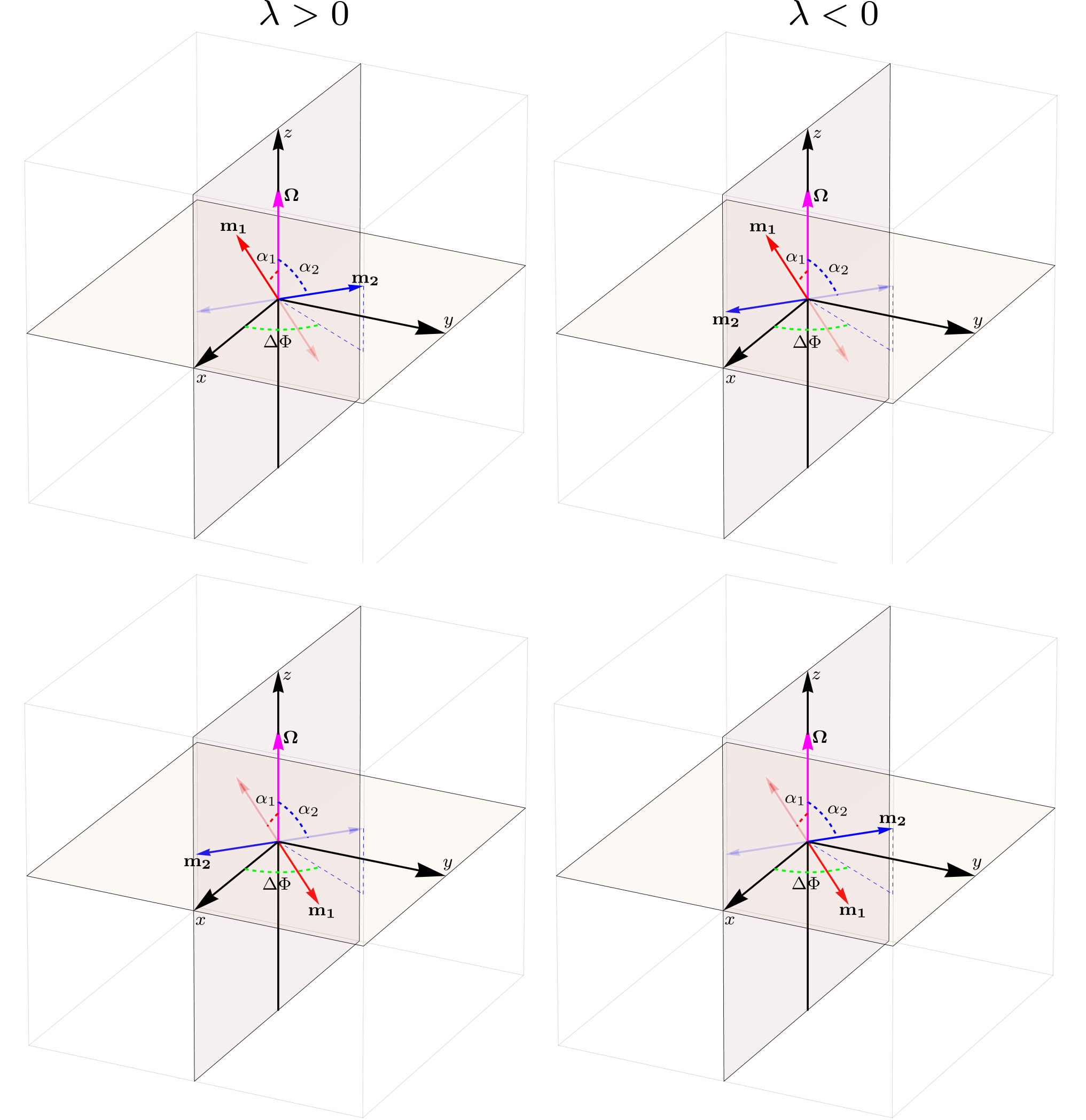}\hfill
    \caption{A schematic diagram illustrating the parameters defining the configuration realizations. Each panel depicts the two magnetic moments (in red and blue) along with their opposite vectors (shown with low opacity). The orbital angular frequency, $ \boldsymbol{\Omega}$, is represented by the magenta vector aligned with the $z$-axis. The primary, i.e., stronger magnetic moment, $m_1$, is always confined to the $x$-$z$ plane, making the azimuthal angle difference, $\Delta\Phi$, between the two magnetic moments equivalent to the azimuthal angle of the second magnetic moment, $m_2$. Each set of parameters corresponds to two possible configurations, which are, however, physically equivalent. The left (right) panels represent the two configurations for the $\lambda>0$ ($\lambda<0$) cases. In our simulations, only one of these equivalent configurations is implemented.}
    \label{FigureA}
\end{figure*}

\section{Methodology and Parameters} 
\label{sec:methods}

\subsection{Force-free MHD Approximation}
\label{sec:FFapprox}

In studies of the magnetospheres of isolated NSs \citep{1999ApJ...511..351C,2006MNRAS.368.1717B,2006MNRAS.367...19K,Spitkovsky06,2006MNRAS.368.1055T,Kalapotharakos&Contopoulos2009,2012MNRAS.424..605P,2016MNRAS.455.4267C,2016MNRAS.462.1894A,2023MNRAS.518.6390S,2024MNRAS.527.6691N}, the FF relativistic ideal MHD framework is typically adopted. This approach is well justified for the global field structure of rotation-powered pulsars because it assumes locally plasma-filled conditions where the energy density of the EM fields dominates over the particle inertia and thermal pressure and that copious pair production occurs. These conditions are generally satisfied by pulsars, which are not very close to the curvature pair-production death line. Notably, the FF formalism for isolated pulsars has revealed the existence of a characteristic equatorial current sheet extending beyond the light cylinder, which has been identified as the likely site of gamma-ray emission observed by Fermi Large Area Telescope \citep[LAT;][]{3PC}. This geometric consistency between the FF solution and gamma-ray observations provides compelling evidence for the validity of the FF framework in describing isolated pulsar magnetospheres.

This framework not only is applicable to isolated pulsars but also extends naturally to the magnetospheres of BNSs. During the inspiral phase, the magnetospheres of the two NSs interact dynamically, yet the fundamental conditions underpinning the FF approximation largely persist. 

Moreover, the FF framework effectively captures the behavior of the twisted and dynamic magnetic field structures resulting from the relative orbital motion resulting from the intense gravitationally driven orbital inspiral. These features are critical for understanding key phenomena such as magnetospheric reconnection, which can power EM precursors and contribute to observable signals such as short GRBs, fast radio transients, and echoes.

Near the merger's final stages, the tidal distortion of the NSs and the ejection of dense matter from potential crust failure will produce localized regions where particle inertia, radiation, and thermal pressure become significant, necessitating full radiation-MHD treatment to capture the dynamics accurately. Such conditions are likely to arise in the immediate vicinity of the distorted NSs. Nevertheless, in this study, we focus exclusively on the FF regime, as it is expected to prevail on larger scales in lower-density regions of the magnetosphere. By emphasizing the EM field structures under the FF approximation, we aim to isolate the dominant large-scale magnetospheric dynamics that are expected to govern much of the inspiral phase and early merger interactions, providing insights into their role in shaping observable EM signals. For this reason, we also neglect general relativistic distortions to fields or modifications to the four-current or any dynamical spacetime effects that primarily influence near-surface regions at the final stages of the merger. These approximations result in code that is more efficient in the parameter exploration and characterization we study in this work.

\subsection{Key Configuration Parameters}
\label{sec:parameters}

The direction and magnitude of the magnetic moments of each NS component are crucial parameters that govern the magnetospheric dynamics. In contrast, parameters like the masses (which govern the orbital dynamics and timescales) and radii (which is most relevant as a length scale) of the stars have a smaller influence on the EM phenomenology and are, therefore, assigned fixed values, i.e., $M_{\star}=1.4 \,M_{\odot}$, $r_{\star}=12\rm \, km$.

The simplest assumption that can be made regarding the EM field of each star is to consider that each of them hosts a dipole magnetic moment, $m$, at its center,
\begin{equation}
\boldsymbol{m}_i=m_i(\cos{\Phi_i}\sin{\alpha_i}\hat{\textbf{x}}+\sin{\Phi_i}\sin{\alpha_i}\hat{\textbf{y}}+\cos{\alpha_i}\hat{\textbf{z}})
\end{equation}
with $i = 1,2$, where $m_1$ and $m_2$ represent the magnetic moment strengths of each star, 
$\alpha_1$ and $\alpha_2$ denote the inclination angles, i.e., the angles between each star’s magnetic moment and the orbital $z$-axis parallel to the orbital angular\footnote{Unless otherwise noted, all frequencies are angular ones.} frequency $\boldsymbol{\Omega}$,
and $\Phi_1$ and $\Phi_2$ correspond to the angles between the projections of the respective magnetic moments onto the $x$-$y$ plane and the $x$-axis.

The internal magnetic field along the direction $\boldsymbol{\hat{r}}$ at a radial distance $r$ from its center is specified as a dipole field in flat spacetime\footnote{See \citep{1974PhRvD..10.3166P,1983ApJ...265.1036W,1986SvA....30..567M,1994ApJ...425..767G,2013MNRAS.433..986P,2016MNRAS.456.4455P} for fields in curved spacetime. More generally, a fluid conducting star would also require a toroidal component for stability \citep{2009MNRAS.395.2162L,2013MNRAS.433.2445A,2013MNRAS.435L..43C}.}:
\begin{equation}
       \textbf{B}_i=\frac{3(\hat{\textbf{r}}_i\cdot\textbf{m}_i)\hat{\textbf{r}}_i-\textbf{m}_i}{r_i^3}.
\end{equation}
The two stars, in the general case, will have magnetic moments of unequal strength, $m_1 \neq m_2$. 

Moreover, the magnetic fields originating from each star are distorted by the presence of the other star's fields, leading to significant changes in the overall magnetospheric structure. Adjusting the direction of the magnetic moment of each star affects the interactions between the two stars' magnetic fields and currents, qualitatively modifying the way the field lines bend, join, and reconnect. For instance, magnetic field lines that would be closed in the case of an isolated star may, due to the influence of the other star, instead of closing back to the first star, connect to the second star or be open. This behavior, driven by magnetospheric interactions, can have a substantial impact on the magnitude and direction of the Poynting flux produced by the system and, thus, on the potentially observable EM signatures. In this context, it is reasonable to consider the two inclination angles, $\alpha_1$ and $\alpha_2$, as key parameters for exploration.

The inclination angles ${\alpha_1, \alpha_2}$ can be parameterized in two ways. One approach allows each inclination angle to range from $0$ to $\pi$.
Alternatively, the inclination angles can be restricted to the range $0$ to $\pi/2$, with $\lambda=\boldsymbol{\Omega}\cdot \boldsymbol{m}_1/\boldsymbol{\Omega}\cdot \boldsymbol{m}_2$, the ratio between the two magnetic moments, taking either positive or negative values. The latter parameterization arises because any configuration with an inclination angle greater than $\pi/2$ is equivalent to one with an inclination angle in the $0$ to $\pi/2$ range but with $\lambda$ flipped in sign and a $180^\circ$ rotation of the star in the $\Phi$ direction.

For this study, we chose the second parameterization as it provides a more intuitive classification of cases where the magnetic moment components along the orbital axis are either aligned or antialigned. By using the signed magnetic moment ratio, positive when the projections of the moments on the $z$-axis point in the same direction and negative otherwise, we can clearly distinguish between these two classes of configurations. This choice also eliminates redundant cases while preserving all unique magnetospheric structures, allowing for a more efficient exploration of the parameter space.

An isolated pulsar located at $\mathbf{r_s}$, with a magnetic field $\textbf{B}$ and spinning at an angular frequency $\boldsymbol{\omega}$ induces an electric field at a point $\mathbf{r}$ given by $\textbf{E}_s=-[\boldsymbol{\omega}\times(\textbf{r}-\textbf{r}_s)]\times\textbf{B}/c$. In the circular binary case, the two stars orbit the system's barycenter with the orbital frequency $\boldsymbol{\Omega}$. As a result, an additional electric field will be induced due to this motion, 
\begin{equation}\label{eq:internalEfield}
\textbf{E}=-(\boldsymbol{\Omega}\times\textbf{r})\times\textbf{B}/c. 
\end{equation}
During inspiral, $\Omega$ increases with time, and the distance $d$ between the stars decreases as $d \propto \Omega^{-2/3}$; thus, the electric field also rises with time. 

Two limiting cases can be considered regarding the spin frequencies of the two stars. In the first case, during the final few rotations of their inspiral, the two stars may become tidally synchronized, resulting in spin frequencies such that $\omega_1 = \omega_2 = \Omega$. In the second case, $\Omega$ is significantly higher than the spin frequencies of the individual stars\footnote{This will not be true if one or both of the NSs is a recycled millisecond pulsar (MSP). In this case, the MSP spin is preferentially parallel to the orbital axes. However, formation channels where one or both members of the BNS is an MSP require dynamical interactions in clusters, rather than isolated binary evolution \citep{2017ApJ...846..170T,2022LRR....25....1M} where two low-kick core-collapse supernovae occur. For instance, in the double pulsar PSR J0737-3039, the fastest member has a spin period of $22$ ms \citep{2004Sci...303.1153L}. }, allowing the stars to be approximated as effectively irrotational, with $\omega_1 = \omega_2 = 0$. This second case is likely for NSs \citep{1992ApJ...400..175B} while the first is for double white dwarfs. Therefore, the simulations presented in this study adopt the irrotational assumption.

In general, when the individual NSs have nonzero spin, the angles $\Phi_1$ and $\Phi_2$ evolve over time. However, in the irrotational case, $\Phi_1$ and $\Phi_2$ remain fixed throughout the evolution, and we must explicitly specify their relative azimuthal orientation, defined as $\Delta\Phi = \Phi_2 - \Phi_1$, to fully determine the system's configuration.

For our simulations, we set the magnetic moment $\boldsymbol{m_1}$ to lie in the $x$-$z$ plane at the beginning of the simulation, ensuring that $\Phi_1 = 0$ without loss of generality. Consequently, $\Delta\Phi$ directly corresponds to the angle between the projection of $\boldsymbol{m_2}$ onto the $x$-$y$ plane and the $x$-axis. Figure~\ref{FigureA} provides a schematic representation illustrating all relevant parameters.

\subsection{Specification of Orbital Inspiral}
\label{sec:orb_motion}

In this work, we consider the gravitational radiation-driven inspiral dynamics of the BNS system. With regard to these dynamics, we neglect any backreaction from the evolving magnetosphere, and we treat the NSs as point masses, thereby ignoring tidal distortions, dynamical tides \citep{2016PhRvL.116r1101H}, or energy transfers related to NS oscillation modes \citep{1994MNRAS.270..611L,1994ApJ...426..688R,1995MNRAS.275..301K}, which may be excited during inspirals.

In particular, for the low-eccentricity inspiral orbits of the two NSs, with masses $M_1$ and $M_2$, we follow the description in \cite{1963PhRv..131..435P},\cite{1964PhRv..136.1224P}, and \cite{1975ctf..book.....L} for the simplest case of point masses and the GW mass quadrupole formula. The length and frequency scales are normalized using a characteristic gravitational radius\footnote{This corresponds to the horizon of the maximally spinning Kerr black hole with the total mass of the system.}, $R_g=G(M_1+M_2)/c^2$ where $G$ is Newton's gravitational constant, and $c$ is the speed of light, and the Keplerian frequency at $R_g$ is $\Omega_g=[G(M_1+M_2)/R_g^3]^{1/2}$, respectively. The separation $d$ between the two NSs is expressed in terms of the orbital angular frequency $\Omega$ as
\begin{equation} 
\label{eq:motion}
    \frac{d}{R_g}=\left(\frac{\Omega}{\Omega_g}\right)^{-2/3}, 
\end{equation}
indicating that the stars follow a circular orbit whose radius decreases over time as $\Omega$ increases.

Solving the corresponding energy loss equation for $\Omega$, a finite time singularity solution can be obtained \citep{1963PhRv..131..435P,1964PhRv..136.1224P,1975ctf..book.....L}:
\begin{equation}
\label{eq:omega}
    \Omega(t)=\Omega_i\left(1-\frac{t}{t_0}\right)^{-3/8}, 
\end{equation}
where $\Omega_i$ is the initial orbital frequency, and 
\begin{equation}
    t_0=\frac{5}{256}\frac{(M_1+M_2)^2}{M_1M_2}\left(\frac{\Omega_i}{\Omega_g}\right)^{-8/3}\Omega_g^{-1},
\end{equation}
is the time until the merger.

This formula provides only a leading-order approximation of the rate of the energy loss from the system. For a more precise modeling, particularly during the later stages of the inspiral, higher-order corrections including tidal effects from the post-Newtonian contributions of Einstein's equations should be employed \citep{PhysRevD.56.3416,2014LRR....17....2B,PhysRevResearch.2.043039,2025arXiv250406918Y}. However, since the main focus of this study is on the magnetosphere structure of the BNSs and associated EM phenomena, we adopt Eq.~(\ref{eq:omega}) to simplify the treatment of the orbital evolution.

In our simulations, the two NSs are treated as solid conducting bodies occupying finite and fixed volumes throughout the entire evolution. Consequently, the inspiral is considered to end when the surfaces of the two stars touch, which occurs at an earlier time, $t_{\rm Final}$, than the formal merger time predicted for point masses from the GW mass quadrupole formula. This corresponds to a separation of $d_{\rm Final} = 2r_\star$, where $r_\star$ is the radius of each NS.
The time $t_{\rm Final}$ reads \citep{Medvedev&Loeb} 
\begin{equation} \label{eq:finaltime}
    t_{\rm Final}=t_0-\frac{5}{256}\left(\frac{d_{\rm Final}}{R_g}\right)^4\frac{(M_1+M_2)^2}{M_1M_2}\Omega_g^{-1},
\end{equation}
where, for the adopted values $r_\star=12 \rm \, km$ and $M_1=M_2=1.4 M_{\odot}$, this corresponds to a duration of $t_0~-~t_{\rm Final}~\sim~1.2 \rm \, ms$. Hereafter, we define $\tau=t-t_0$ as the simulation time indicator, which is always negative, representing the time remaining until the merger. Although $\tau$ increases algebraically as the system evolves toward merger, its absolute value decreases over time.

\subsection{Magnetospheric Evolution and Numerical Implementation}
\label{sec:FDTD}

For our simulations, we use a modified version of the 3D Cartesian Finite Difference Time Domain (FDTD) code developed by \citep{Kalapotharakos&Contopoulos2009,Kalapotharakos2012,Kalapotharakos2014,2017ApJ...842...80K}, implemented in {\tt MPI Fortran 90}. Specifically, the magnetospheric evolution of the electric and magnetic fields, $\mathbf{E}$ and $\mathbf{B}$, is computed by integrating the time-dependent Maxwell equations in flat spacetime

\begin{equation} \label{eq:Maxwell1}
    \frac{\partial\textbf{E}}{\partial t}=c\nabla\times\textbf{B}-4\pi\textbf{J},
\end{equation}
\begin{equation}\label{eq:Maxwell2}
    \frac{\partial\textbf{B}}{\partial t}=-c \nabla\times\textbf{E},
\end{equation}

where $\mathbf{J}$ is the current density.

The FF current density prescription, which ensures that $\mathbf{E} \cdot \mathbf{B} = 0$ reads \citep{Gruzinov1999}
\begin{equation}
\label{eq:ffcurrent}
\textbf{J}=c\rho\frac{\textbf{E}\times\textbf{B}}{B^2}+\frac{c}{4\pi}\frac{\textbf{B}\cdot\nabla\times\textbf{B}-\textbf{E}\cdot\nabla\times\textbf{E}}{B^2}\textbf{B}~.
\end{equation}
where $\rho = \nabla \cdot \mathbf{E} /( 4 \pi)$ is the charge density.

Nonetheless, in this study, we employ the dissipative current density prescription introduced by \cite{Kalapotharakos2012,Kalapotharakos2014}, which reads
\begin{equation} \label{eq:current}
    \textbf{J}=c\rho\frac{\textbf{E}\times\textbf{B}}{B^2+E_0^2}+\sigma\textbf{E}_{||},
\end{equation}
where $\sigma$ denotes the plasma conductivity (assumed spatially uniform and constant with time in this study), and $E_0$ and $B_0$ denote the electric and magnetic field, respectively, in the frame where $\mathbf{E}$ and $\mathbf{B}$ are parallel. They are defined using the Lorentz invariants as $E_0 B_0 = \mathbf{E} \cdot \mathbf{B}$ and $E_0^2 - B_0^2 = E^2 - B^2$. The parallel electric field, $\mathbf{E}_{||}$, is given by
$\mathbf{E}_{||} = \frac{\mathbf{E} \cdot \mathbf{B}}{B^2} \mathbf{B}$.

As the conductivity $\sigma$ varies from $0$ to $\infty$, the magnetosphere structure spans the entire spectrum of solutions, from a vacuum to FF. For practical purposes, we adopt a relatively high value of conductivity, $\sigma = 6\Omega_{\rm i}$, which brings the system close to the FF regime. At the end of each time step, any remaining parallel electric field component is subtracted to further enforce the ideal FF condition.

During the evolution, when using the current prescription of Eq.~\ref{eq:ffcurrent}, there is a possibility for regions to develop where $E > B$ \citep{Spitkovsky06,Kalapotharakos&Contopoulos2009}. Such regions are inconsistent with the FF regime due to the presence of accelerating electric field components. In those regions, the FF approximation breaks down, resulting in a superluminal plasma drift velocity, and the underlying equations become mathematically ill posed. To mitigate this issue, following \cite{Gruzinov2007,Gruzinov2008} and \cite{Li2012}, we modify the Ohm's law by including the $E_0^2$ in the denominator of the $\mathbf{E}\times\mathbf{B}$ term, as shown in Eq.~\ref{eq:current}. With this prescription, the denominator is not allowed to attain small values, facilitating the treatment of the current sheet region \citep[see, e.g., ][]{Kalapotharakos2012,Kalapotharakos2014}, where the magnetic field may vanish.  In real physical conditions, the plasma would be redistributed in regions where $E$ exceeds $B$ to neutralize the corresponding accelerating electric field component. Additionally, in order to prevent the development of $E>B$ regions, we enforce the condition $E \leq B$ throughout the simulation domain at the end of each time step. We do so by reducing $|\textbf{E}|$ in these regions to $98\%$ of $|\textbf{B}|$.

Our FDTD code integrates the Maxwell equations, Equations \ref{eq:Maxwell1}--\ref{eq:Maxwell2}, tracking the inspiral of the two NSs by enforcing field values inside them that correspond to their respective magnetic moments, with the electric fields determined by Eq.~\eqref{eq:internalEfield}. Compared to simulations of an isolated spinning NS, the merging BNS system proved more susceptible to the growth of oscillations (spurious ripples) near regions with steep gradients or discontinuities, such as current sheets. To address this, we implemented a third-order total variation diminishing (TVD) Runge-Kutta scheme \citep{1988JCoPh..77..439S}, replacing the standard third-order Runge-Kutta method used previously. This TVD scheme is well known for its enhanced numerical stability, particularly in handling sharp transitions and avoiding oscillatory artifacts.

In the case of perfect conductors, as the stars are modeled here, the boundary conditions require the continuity of the tangential components of the electric field and the normal components of the magnetic field. Handling spherical boundaries within a Cartesian grid poses significant computational challenges. This issue arises even in the simpler case of an isolated spinning NS. However, in the BNS system, the motion of the stellar boundaries aggravates the problem, often spouting unphysical nonzero values of $\nabla \cdot \mathbf{B}$ near the stellar boundary as the system evolves.

\begin{figure*}
\centering
    \includegraphics[width=1.0\textwidth]{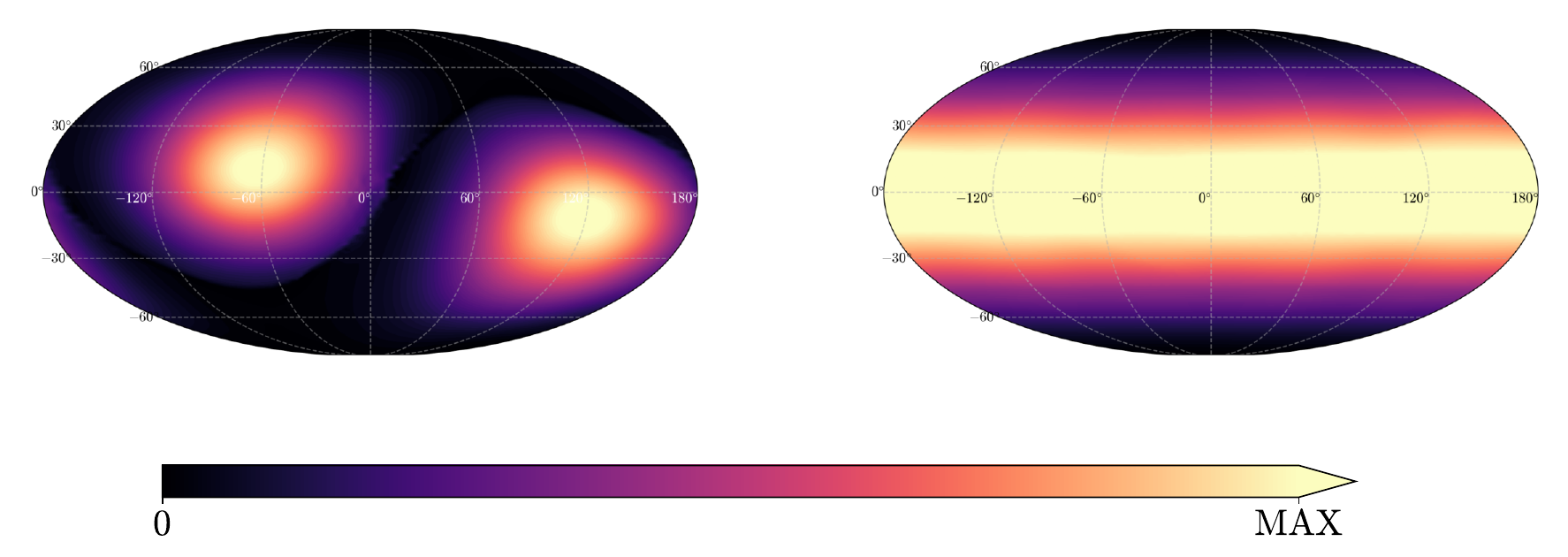}\hfill
    \caption{Mollweide projection of the Poynting flux distribution of an isolated pulsar with $45^{\circ}$ inclination angle, calculated at a sphere of radius 1.3 of the light cylinder. The color indicates the norm of the radial Poynting vector. (\textbf{Left}) The corotating anisotropic pattern is shown at a specific time. (\textbf{Right}) The total Poynting flux integrated over five pulsar rotations.}
    \label{Figure1}
\end{figure*}

\begin{figure}
\centering
\includegraphics[width=\columnwidth]{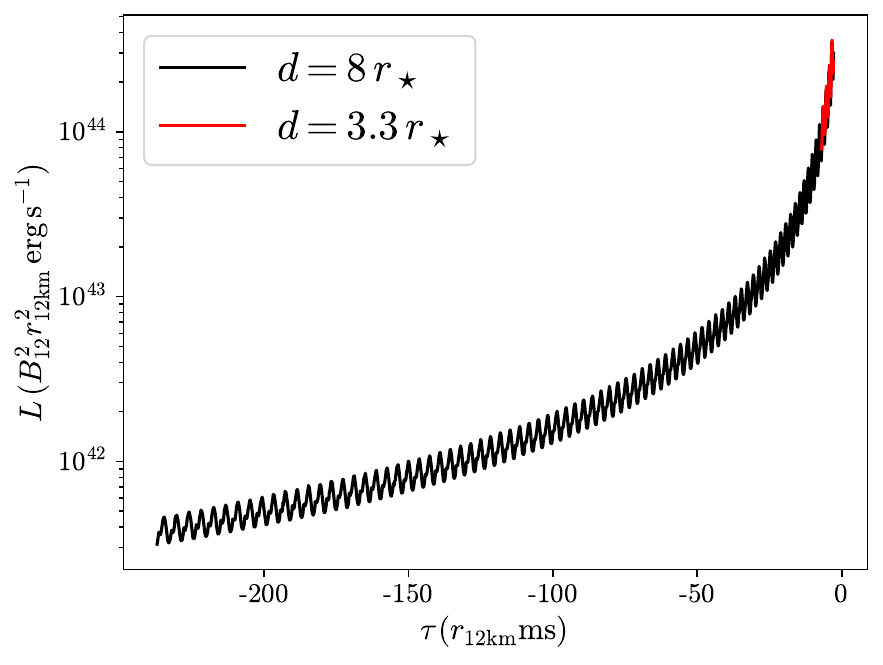}\hfill
    \caption{The total Poynting flux that crosses a surface of radius $13 \,r_\star$ during system evolution starting from the initial separation of $3.3 \,r_\star$  (red line) and  $8 \,r_\star$ (black line). Both simulations show similar flux evolution in the case of $\alpha_1=0^\circ, \,\alpha_2=45^\circ$,  and $\lambda=-1$ (\AngledUpArrow{0}\AngledDownArrow{-45}).}
    \label{Figure3}
\end{figure}
Various techniques have been developed to address numerical errors associated with the magnetic field divergence in MHD simulations. These include the projection method \citep{BRACKBILL1980426}, diffusive method \citep{MARDER198748,VANDERHOLST2007925}, constrained transport method \citep{1988ApJ...332..659E}, eight-wave formulation approach \citep{1993atcf.work..303P}, and the hyperbolic divergence cleaning approach \citep{2002JCoPh.175..645D}; see also \citet{divcleaning} for a comparative review.

In this work, we adopt the diffusive method scheme implemented in \cite{divcleaning}, which adds an additional term to Eq.~\ref{eq:Maxwell2} to eliminate any residual divergence of the magnetic field. In practice, at the end of each time step, we apply the following scheme
\begin{equation}
    \textbf{B}_n=\textbf{B}_n+\eta\Delta t\nabla(\nabla\cdot\textbf{B}_n)
\end{equation}
The parameter $\eta$ has to satisfy \citep[see][and references therein]{divcleaning} 
\begin{equation}\label{ineq}
    \eta\leq \eta_{\rm max}=1.3\frac{(\Delta x)^2}{3\Delta t}
\end{equation}
with $\Delta x$ representing the uniform grid spacing in all three dimensions. In our simulations, we set $\eta = 0.25 \eta_{\text{max}}$, a value that effectively eliminates residual magnetic field divergence around the stars while minimally impacting the field structure in regions far from them. To validate that our results are not impacted by the implemented scheme, we performed a calibration test using an isolated pulsar simulation. Running this test both with and without the divergence cleaning scheme, we found that the global magnetospheric structure remained practically unchanged. Furthermore, the spatial distribution of the normalized divergence measure, $\nabla \cdot \mathbf{B}\; \Delta x / B$, was found to be consistent across all cases, with a median value of approximately $10^{-4}$. This confirms that our divergence control scheme does not introduce artificial effects and that the Poynting flux values and patterns (see Section \S\ref{sec:results}) remain physically reliable.

Finally, to minimize inward reflections of outward-propagating EM waves at the outer boundaries of the simulation domain, we applied the perfectly matched layer (PML) method \citep[][see also \citealt{Kalapotharakos&Contopoulos2009}]{Barenger1996}. 

The evolution of the binary system is tracked from an initial separation, $d=3.3 r_\star$, until the stars contact, which from Equation~\ref{eq:finaltime} corresponds to $t_{\rm Final}= 7.7 \rm \, ms$ with a time step of $dt=0.04 \, r_\star/c$. In our simulations, we have selected a mass-radius ratio, and that defines a specific orbital evolution following the prescription described earlier \footnote{Note that the initial separation is given in units of $r_\star$.}. Although the results presented below can be scaled with the stellar radius $r_{\star}$, in order for our simulations to still be able to describe the derived EM fluxes, for a different radius, the mass should be changed accordingly, resulting to the same mass/radius ratio.  Below, if not stated otherwise, $r_\star=12{\rm km}$ is used, from $\Omega_{\rm i}\sim2451\rm\,\,rad/s \,\,(390 \, \rm Hz)$ to $\Omega_{\rm Final}\sim5195\rm\,\,rad/s \,\,(827 \, \rm Hz)$. The grid consists of $512$ grid points in each direction, covering the entire computational domain with an edge length of $25.6 r_\star$ centered on the binary system. However, the active computational domain, excluding the PML region, has an edge length of $22.3 r_\star$. This region, in practice, is where the simulation's magnetospheric evolution is calculated.

Resolution studies were also conducted (see \autoref{AppendixA}). For a selected subset of configurations, we performed simulations at higher spatial resolution and smaller time steps, utilizing 1024 grid points per dimension, effectively doubling the resolution, or with computational domains twice the size of the standard setup. The results presented in \S\ref{sec:results} were found to be robust, showing no significant variation with enhanced resolution and box size.

\section{Results} 
\label{sec:results}

A series of simulations were run to survey different unique initial configurations for the magnetic fields of the two stars. We explored  $\lambda=\{\pm1, \pm10\}$, $\alpha_1,\alpha_2=\{0^\circ,45^\circ,90^\circ\}$ and $\Delta\Phi=\{0^\circ,45^\circ,90^\circ\}$. Additionally, as changes in the $\Delta\Phi$ parameter only slightly affect the outflow patterns, for $\Delta\Phi=0$, further cases were explored with $\lambda=\{\pm\sqrt{10},\pm100\}$ and the same range for the inclination angles. Lastly, to validate the specific results presented below (see section ~\ref{sec:fit}), we also performed additional simulations for cases with $
\lambda=\{\pm\sqrt{10^3},\pm\sqrt{10^5},\pm10^3\}$, $\alpha_1=\{0^\circ,45^\circ\}$, $\alpha_2=0$, and $\Delta\Phi=0$. This exploration resulted in $132= (3^3+3^2+3)2^2-24$  unique\footnote{When $\alpha_2=0$, the configuration is the same for every $\Delta\Phi$; thus, those cases have been excluded.} simulations.

\begin{figure*}
    \centering
    \includegraphics[width=1.\textwidth]{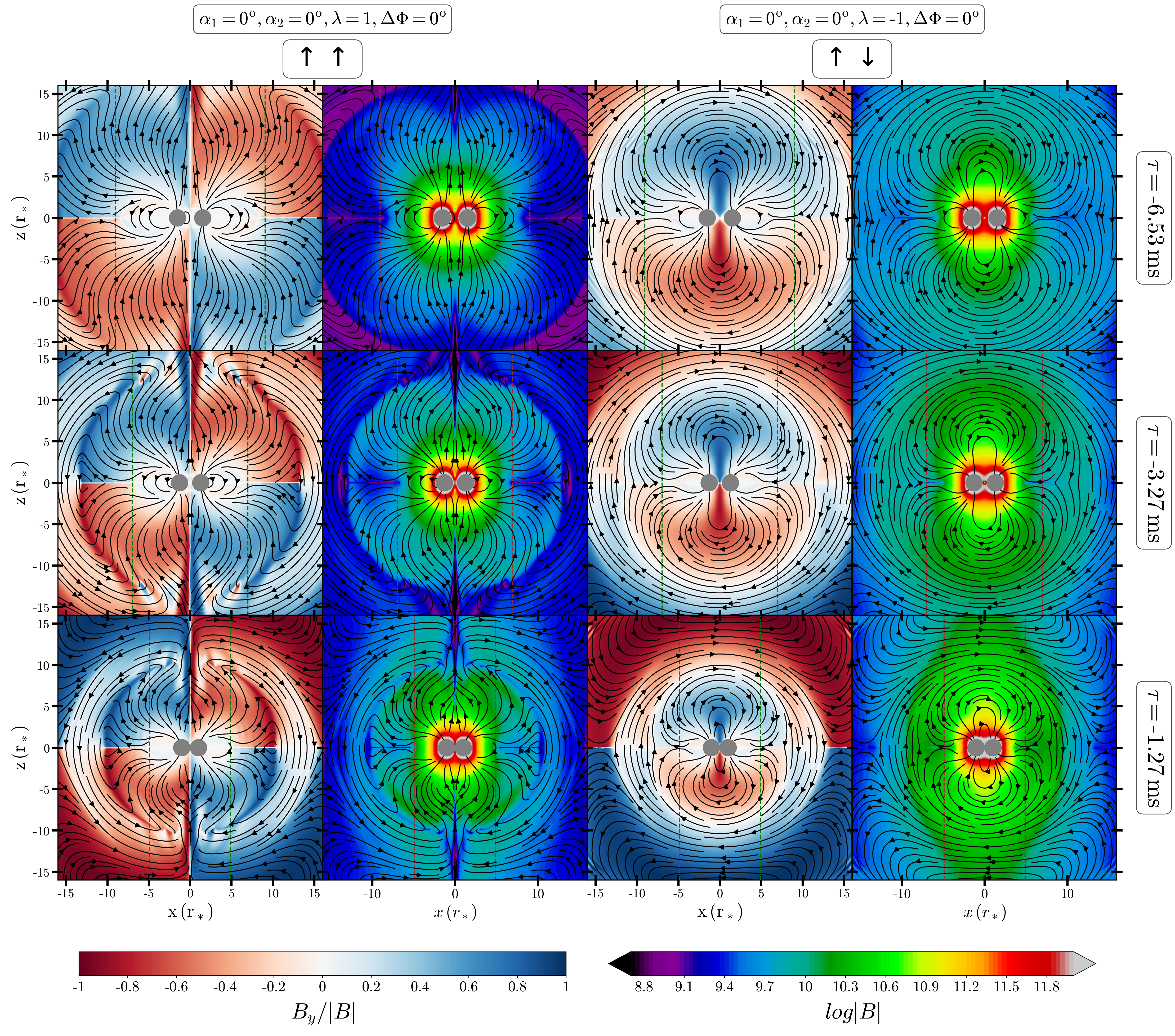}
    \caption{Snapshots of the magnetic field evolution are shown for aligned (\AngledUpArrow{0}\AngledUpArrow{0}, left two columns) and antialigned (\AngledUpArrow{0}\AngledDownArrow{0}, right two columns) magnetic moments on the $z-x$ plane in the corotating frame. Black streamlines indicate the magnetic field direction on the poloidal plane, while the color denotes the normalized magnitude of the magnetic field component perpendicular to it (first and third column) and the magnitude of the magnetic field (second and fourth column). Dashed vertical lines indicate the characteristic distance $c/\Omega$.}
    \label{Figure4}
\end{figure*}

Below, the numerical results are expressed for the magnetic field in units of $B_{12}=B/10^{12}G$ and for stellar radii in units of $r_{12\, \rm km}=r_\star/(1.2\times10^6 \, \, \rm cm)$.

\subsection{Electromagnetic Fluxes, Field Structures, and Emission Anisotropy} 
\label{sec:Magnetospheric_Dynamics}

For a single pulsar with a plasma-filled magnetosphere, as it rotates around its spin axis with an angular frequency $\omega$, energy is lost to low-frequency EM radiation that is mostly sourced from the open magnetic field lines \citep{1999ApJ...511..351C,2006MNRAS.368.1717B,Spitkovsky06,Kalapotharakos&Contopoulos2009,2016MNRAS.457.3384T}. Calculating the magnetic flux along the open magnetic field lines $\Phi_{\rm open}$, one may compute an approximate luminosity as $L\approx\omega^2\Phi_{\rm open}^2/(6\pi^2c)$ for direct comparison with the integrated Poynting flux crossing a surface at a large distance. 

\begin{figure*}
\centering
\includegraphics[width=0.95\textwidth]{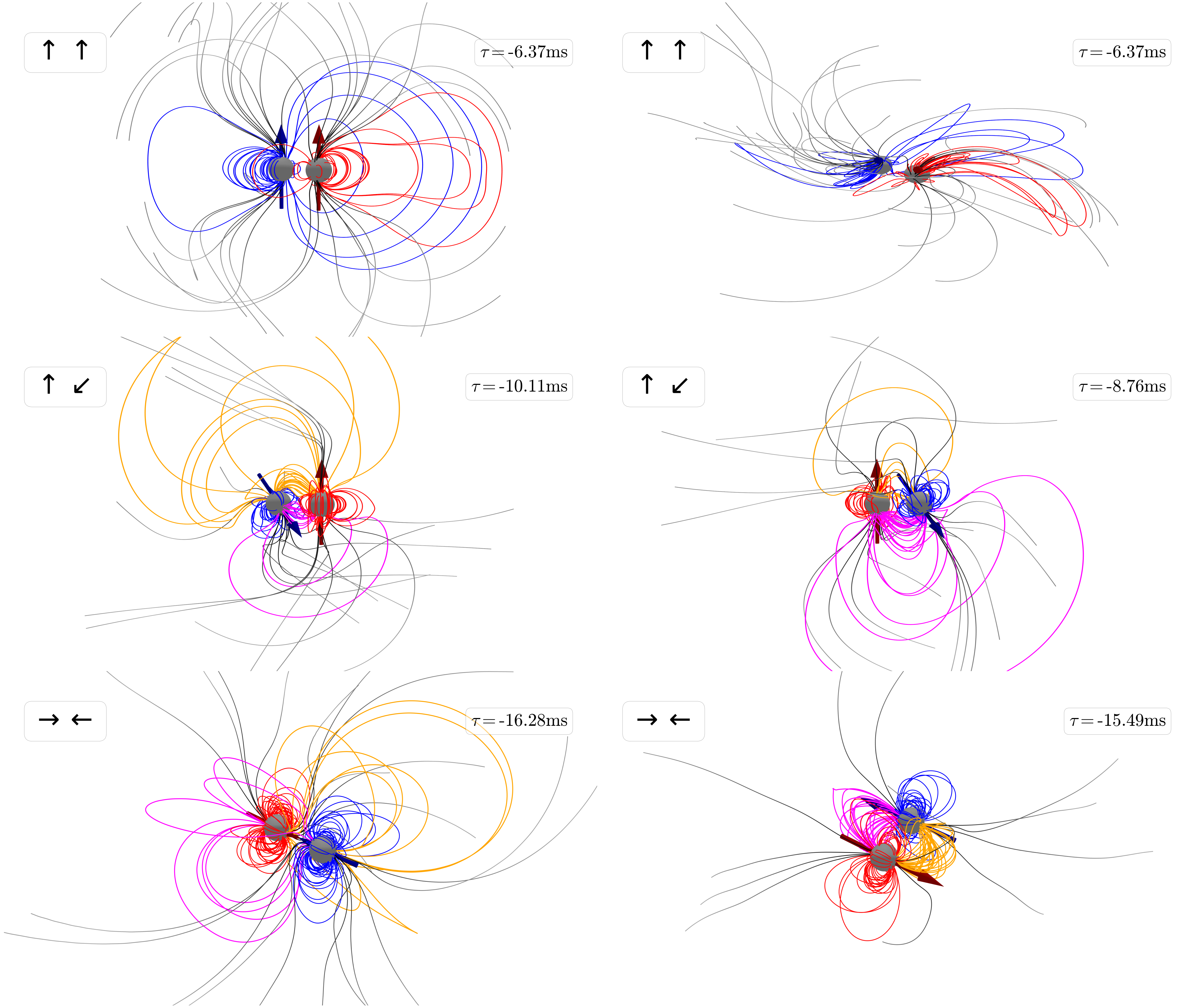}\hfill
    \caption{3D representation of the magnetic field lines for different magnetic moment configurations. \textbf{Top row:} both stars have aligned magnetic moments ($\uparrow \uparrow$), shown from two perspectives,  equatorial plane (left) and top-down along the orbital axis (right). \textbf{Middle row:} a configuration with $\alpha_1=0^{\circ}, \alpha_2=45^{\circ},\Delta\Phi=0^\circ $ and $\lambda=-1$  (\AngledUpArrow{0}\AngledDownArrow{-45}), presented at two snapshots separated by half orbital rotation. \textbf{Bottom row:} A configuration $\alpha_1=90^{\circ}, \alpha_2=90^{\circ}, \Delta\Phi=0^\circ$, and $\lambda=-1$  (\AngledUpArrow{-90}\,\AngledDownArrow{-90}), shown at two snapshots separated by a quarter orbital rotation. The red and blue lines correspond to fields that start and close on the same star. In contrast, the purple lines represent field lines that originate from the north pole of the first star and connect to the south pole of the second star, while the orange lines represent field lines that originate from the north pole of the second star and connect to the south pole of the first star. The black lines correspond to the open magnetic fields. We note that the time coordinate $\tau< t_{\rm Final}=-7.7\rm ms$ indicates that these cases correspond to simulations with an initial separation of $d=4.64 r_\star$, larger than the separation used in the rest of the parameter exploration simulations.}
    \label{Figure5}
\end{figure*}

In particular, the outgoing EM power can be calculated by integrating the radial component of the Poynting vector, $\mathbf{S} = c \mathbf{E} \times \mathbf{B} / (4\pi)$, over a spherical surface of radius $r_F$. Numerical FF simulations \citep{Spitkovsky06}, later validated by MHD and particle-in-cell (PIC) simulations within the FF regime, have shown that the luminosity of a single isolated pulsar with a magnetic inclination angle $\alpha$ follows an approximate relation
\begin{equation}
    L\approx\frac{m^2\omega^4}{c^3}(1+\sin^2\alpha).
\end{equation}
The Poynting flux distribution crossing a spherical surface at a given time (Fig.~\ref{Figure1}, left panel) rotates in sync with the pulsar’s rotation. By integrating this luminosity over a full rotation period, the average luminosity distribution can be determined (Fig.~\ref{Figure1}, right). An observer at infinity may be pointed at a specific set of the azimuthal angle $\phi$ and polar angle $\zeta$, i.e., the so-called observer angle relative to the spin axis. Notably, the average luminosity observed over a complete rotation depends exclusively on $\zeta$.

This averaging over many rotations may be used to derive both the total luminosity and its distribution on any sphere in the case of binary systems.

\begin{figure*}
\centering
    \includegraphics[width=0.9\textwidth]{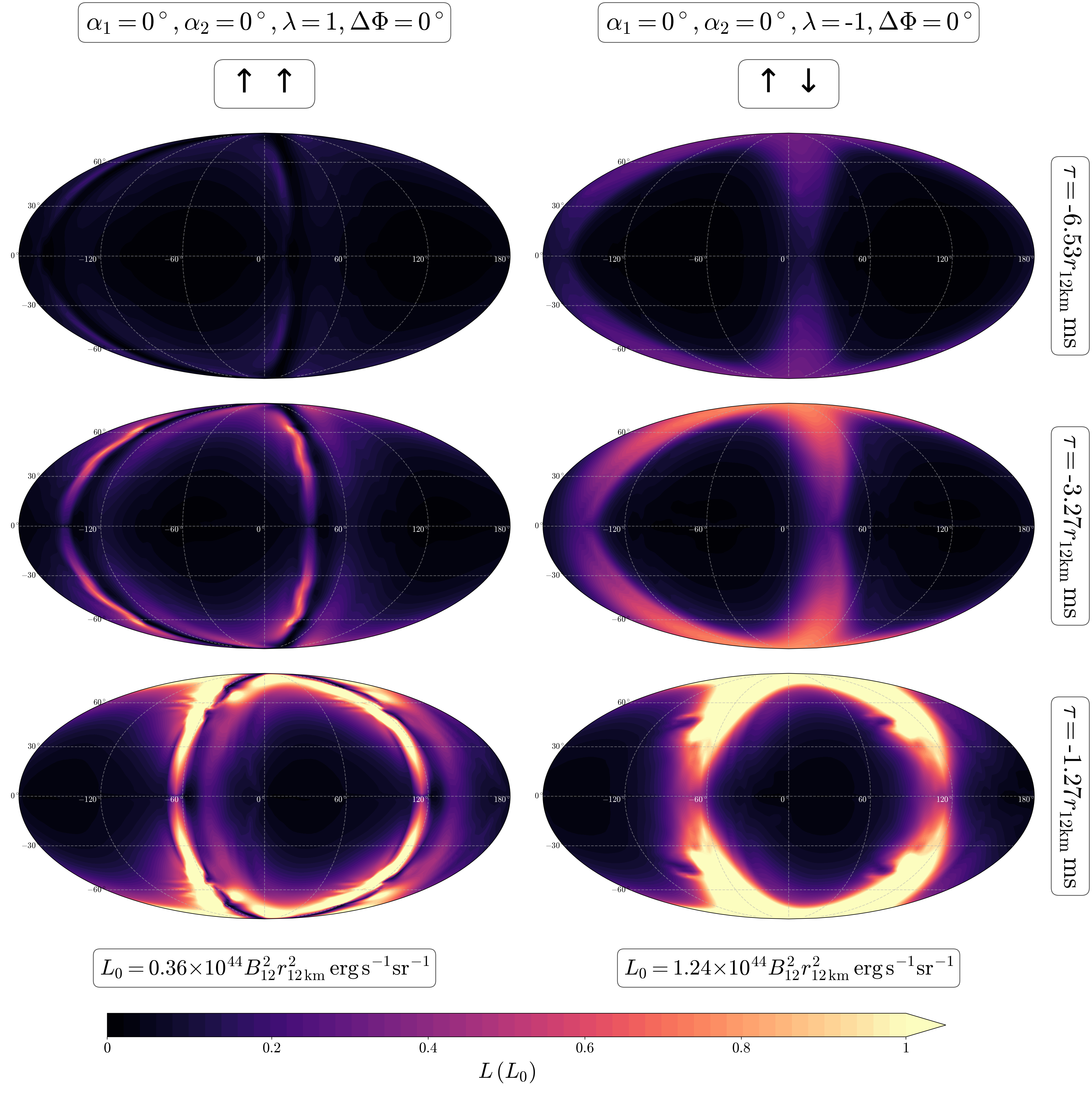}\hfill
    \caption{The evolution of the Poynting flux, normalized to $L_0$ (as indicated), crossing a spherical surface of radius $13r_\star$  around the system for equal aligned  ($\uparrow\uparrow$, left column) and antialigned ($\uparrow\downarrow$, right column) magnetic moments. For the aligned case, $L_0^{\uparrow\uparrow}=0.36\times 10^{44}B_{12}^2r_{12 \,\rm km}^2\,\rm erg\,s^{-1}\, sr^{-1}$, while, for the antialigned case, $L_0^{\uparrow\downarrow}=1.24\times10^{44}B_{12}^2r_{12 \,\rm km}^2\,\rm erg\,s^{-1}\,sr^{-1}$. Here, $B_{12}$ and $r_{12 \,\rm km}$ are the surface magnetic field (in units of $10^{12} \rm \, G$) and the stellar radius (in units of $1.2 \times 10^6\rm \,  cm$).}
    \label{Figure6}
\end{figure*}

The binary system, starting from a large initial separation, evolves in a low orbital frequency regime as dictated by Eq.~\eqref{eq:motion}, resulting in a gradual evolution toward merger. For a synchronized system (where the stars have the same spin and orbital frequencies) in this regime, the light cylinder is mathematically well defined at a distance $R_{\rm LC} = c / \Omega$. In this configuration, the magnetosphere structure corotates with the system (neglecting the radial inspiral motion), and the Poynting flux distribution forms a fixed, rotating pattern. However, in this study, the stars are effectively irrotational, and no strictly defined single light cylinder exists. Instead, the length scale $c/\Omega$ can be interpreted as an instantaneous light cylinder associated with the evolving field configuration and orbital frequency $\Omega$. Yet, $c/\Omega$ is a crucial length scale for the magnetosphere's evolution, and the active computational domain size is chosen so that this length scale is always well within it.

For the initial orbital frequency $\Omega_i$, we have $c/\Omega_i = 10.2 r_\star$, while, at the final time $t = t_{\text{final}}$, it decreases to $c/\Omega_f = 4.8 r_\star$ owing to the inspiral.

Additionally, a larger initial separation necessitates a larger computational domain to evaluate the system's evolution accurately.  
In Figure \ref{Figure3}, we compare the evolution of the Poynting flux for an initial separation of $d = 8 r_\star$ with that for the standard separation $d = 3.3 r_\star$ adopted for the majority of the runs in this study. Thus, at earlier times, the system's evolution can be inferred by extrapolating backward from the results presented below. 

\begin{figure}
    \includegraphics[width=\columnwidth]{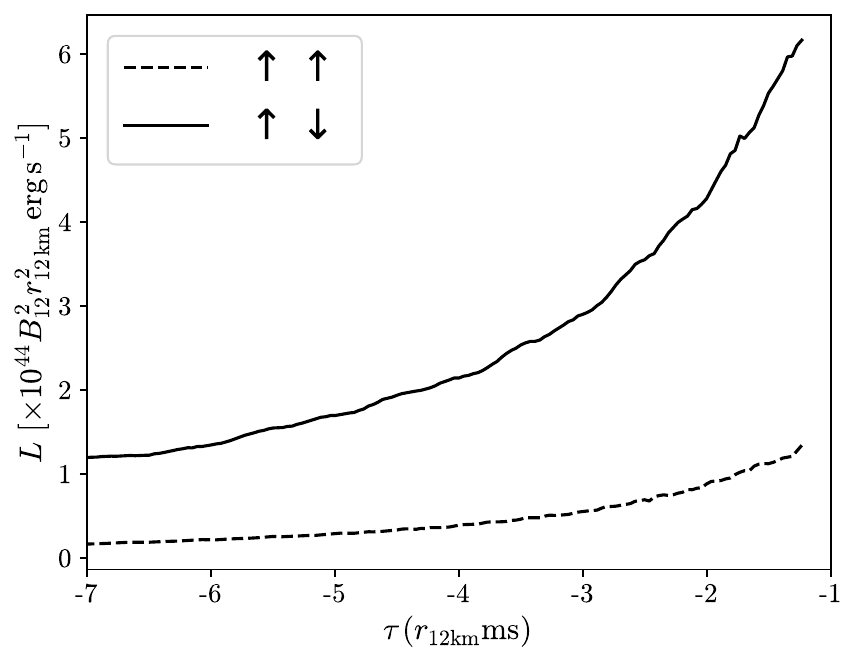}\hfill
    \caption{The total Poynting flux crossing a spherical surface of radius $13r_\star$ throughout the system evolution for equal antialigned ($\uparrow\downarrow$, solid line) and  aligned ($\uparrow\uparrow$, dashed line) magnetic moments.}
    \label{Figure7}
\end{figure}
\subsubsection{Aligned and Antialigned Magnetic Moments}
\label{sec:aligned}

In Figure~\ref{Figure4}, we present slices of the magnetosphere's magnetic field structure at the indicated snapshots. This comparison involves cases where the magnetic moments of both stars are either aligned ($\uparrow \uparrow$, left two columns) or antialigned ($\uparrow \downarrow$, right two columns). 
In Fig.~\ref{Figure4}, the slices are taken in the corotating poloidal plane, which follows the binary’s motion and contains the centers of both stars. Arrowed lines represent the corresponding poloidal magnetic field lines, and the color scale indicates the magnitude of the magnetic field component perpendicular to the poloidal plane in the first and third columns, while, in the second and fourth columns, it indicates the total magnetic field strength on a logarithmic scale.

The primary distinction between these two cases lies in the behavior of the magnetic field lines in the region around the two stars. If each star were isolated, these magnetic field lines would belong to either the closed or open magnetic field regions of their respective magnetospheres. In the presence of both stars, these lines do not maintain the same behavior. In the antialigned case, some magnetic field lines connect the two stars directly. In contrast, in the aligned case, these field lines are bent out of the poloidal plane in the azimuthal direction, closing within the opposite hemisphere of the star they originate from. This behavior is more evident in a 3D representation of the magnetic field lines, as shown in the top row of Figure~\ref{Figure5}. The left column shows the magnetic field lines from the equatorial plane, while the right column provides a top-down view. Due to the symmetry, the field lines of each star would have the same directions, and thus, no line is allowed to connect the two stars \citep{Palenzueala2013b,Mahlmann&Beloborodov}.
Furthermore, as each star's magnetic field lines are repelled from the region dominated by the other star's magnetosphere, they become compressed in the leading direction of motion. In contrast, comparatively fewer field lines populate the trailing side toward the companion. Due to retardation effects, the field lines are consistently swept back, resulting in an ``S''-shaped morphology when viewed from above, i.e., along the $z$-axis (see left panel of the top row in Fig.~\ref{Figure5}).
The motion of the stars also results in a constant opening and closing of field lines, producing a stable flow of EM flux along the open field lines that increases following the increase of the fields as the separation decreases. 

Isolated FF NS magnetospheres reach a steady state in the corotating frame; magnetic field lines that cross the light cylinder remain open, and by definition, no closed lines extend beyond it. However, binary systems generally do not permit such steady configurations.  Field lines that are open, i.e., crossing the cylindrical radius $c/\Omega$ at one instant, may become closed, and vice versa. 

Thus, the distinction between open and closed field lines is dynamic rather than fixed. 

Richness also emerges as certain lines that momentarily extend beyond the cylindrical radius $c/\Omega$ bend backward and close on one of the two stars. This behavior is illustrated in Figure~\ref{Figure4}, which also shows the instantaneous length scale $c/\Omega$, marked by vertical dashed lines. As the system evolves, these lines progressively shift inward, reflecting the changing dynamics of the magnetosphere.

Figure~\ref{Figure6} presents the distributions of the Poynting flux on a spherical surface with a radius of $13r_\star$ for the aligned $\uparrow\uparrow$ (left-hand column) and antialigned $\uparrow\downarrow$ (right-hand column) cases at the indicated snapshots. While the sky distributions are qualitatively similar, the magnitude of the total flux at a given time \citep[in agreement with][]{Palenzueala2013b} is higher in the antialigned case compared to the aligned case. In particular, we find it to be approximately 4 times higher than the aligned case. The time evolution of the total flux for both cases is shown in Figure \ref{Figure7}.

This difference arises from the magnetospheric solution realized by the system, as discussed earlier. When the magnetic moments of the two stars are aligned $\uparrow \uparrow$, the magnetic field lines originating closer to the poles of the stars diverge from the orbital axis toward the equator before curving back to it with some of them connecting with the opposite magnetic pole of the same star. A detailed evolution of the magnetic field lines during the evolution, for a similar configuration, is described in \cite{Mahlmann&Beloborodov}.  This structure diminishes the time-averaged open flux $\Phi_{\rm open}$, thereby reducing the emitted EM power. Additionally, it produces a flux distribution with the dark zones, similar to that shown in Figure~\ref{Figure6} \citep[and similar to Figure 13 presenting the Poynting flux distribution in][]{Palenzueala2013b}, corresponding to the interface separating the two interacting magnetospheres.

\begin{figure*}
\centering
    \includegraphics[width=0.9\textwidth]{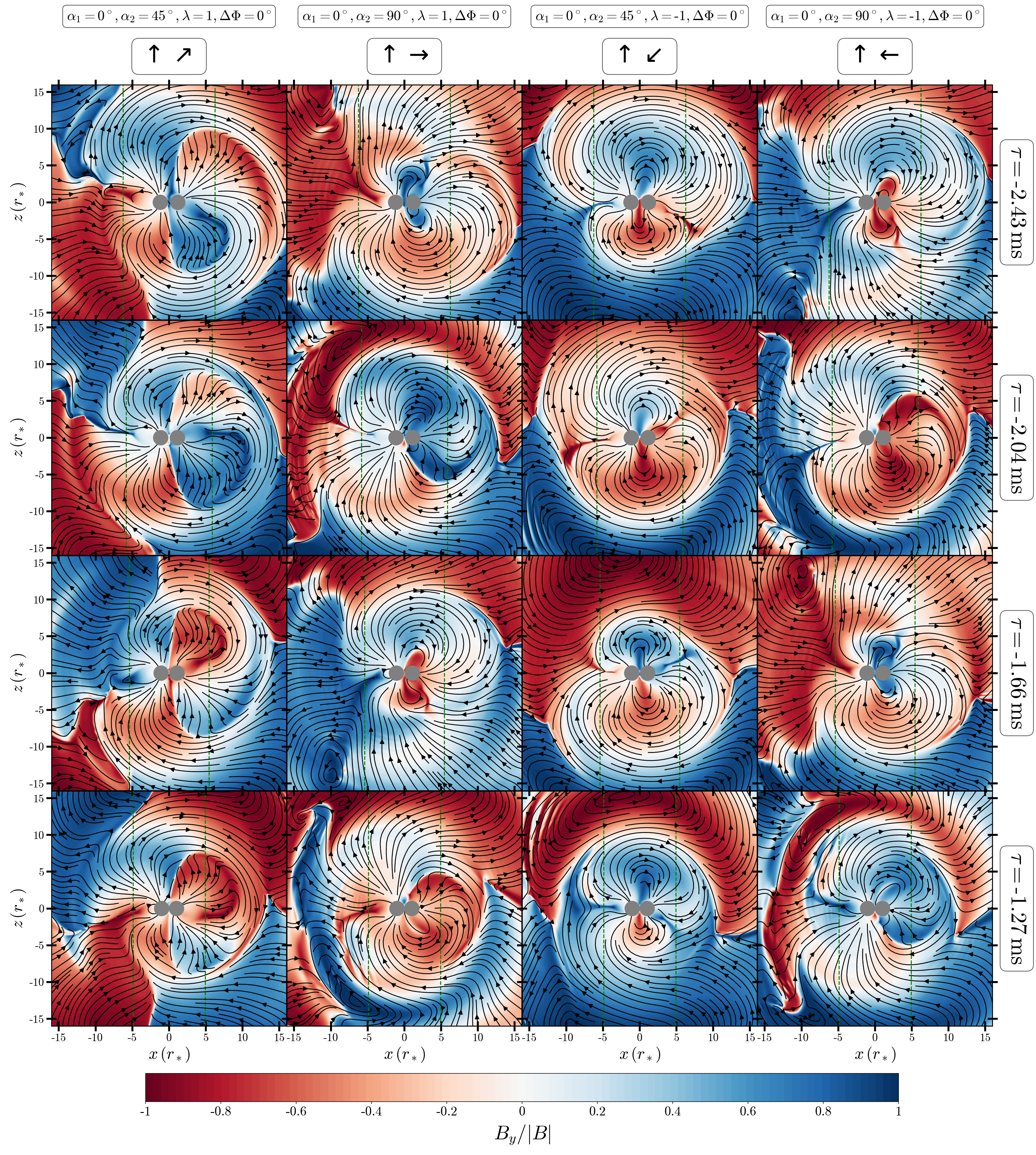}
    \caption{Similar to Figure~\ref{Figure4}. The first and second columns correspond to $\lambda=1$ with the right star having inclination angles of $\alpha_2=45^\circ$ (\AngledUpArrow{0}\AngledUpArrow{-45}) and $\alpha_2=90^\circ$ (\AngledUpArrow{0}\AngledUpArrow{-90}), respectively. Similarly, the third and fourth columns represent $\lambda=-1$, with $\alpha_2=45^\circ$ (\AngledUpArrow{0}\AngledDownArrow{-45}) and $\alpha_2=90^\circ$ (\AngledUpArrow{0}\AngledDownArrow{-90}), respectively.}
    \label{Figure8}
\end{figure*}

On the other hand, antialigned magnetic moments $\uparrow\downarrow$ lead to a significantly different global solution with higher Poynting flux. This is somewhat surprising, especially considering that, in the vacuum case, antialigned moments result in partial cancellation of dipolar components in favor of higher multipoles \citep[also see][]{Palenzueala2013b}. Instead, in this scenario, magnetic field lines from the inner hemispheres of each star (the sides facing one another) connect directly between the stars. Some field lines originating from the outer hemispheres curve toward the magnetic poles of the other star and connect to the outer hemisphere of that star. However, field lines that originate closer to the equator tend to be longer, and those originating at sufficiently high polar angles may not curve inward enough to connect to the second star before crossing the equator. Once they cross the equator, they no longer connect with the other star, as they encounter the pole that has the same polarity. Consequently, these lines remain open and escape to infinity. This magnetosphere structure results in a significantly higher $\Phi_{\rm open}$ than the aligned case and a Poynting flux distribution without the dark zone observed in the aligned case. It is similar to the one reported in \cite{Most&Philippov2022}.

\begin{figure}
    \centering
        \includegraphics[width=\columnwidth]{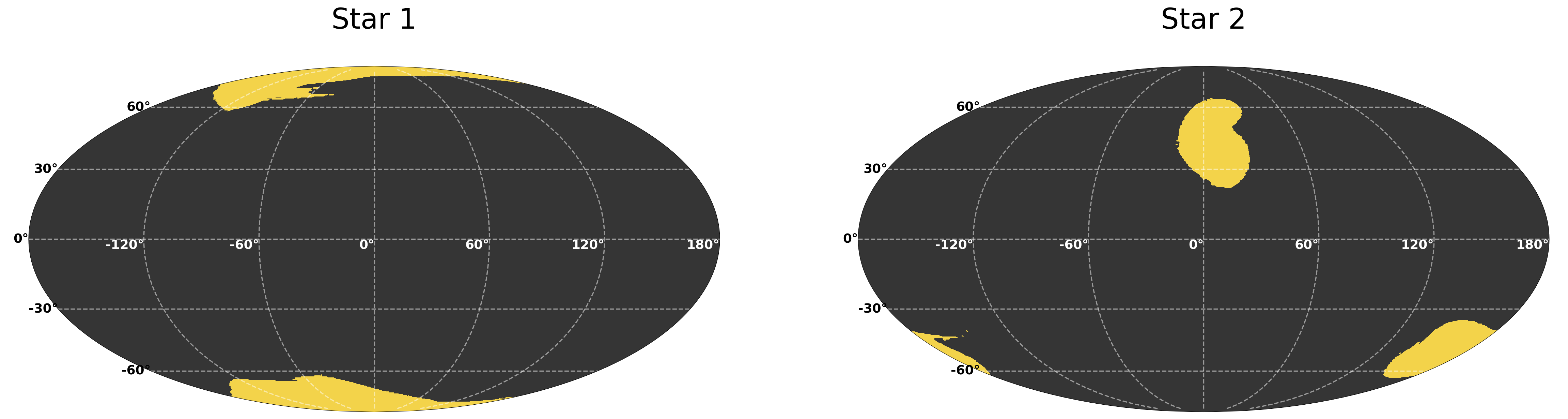}
        \includegraphics[width=\columnwidth]{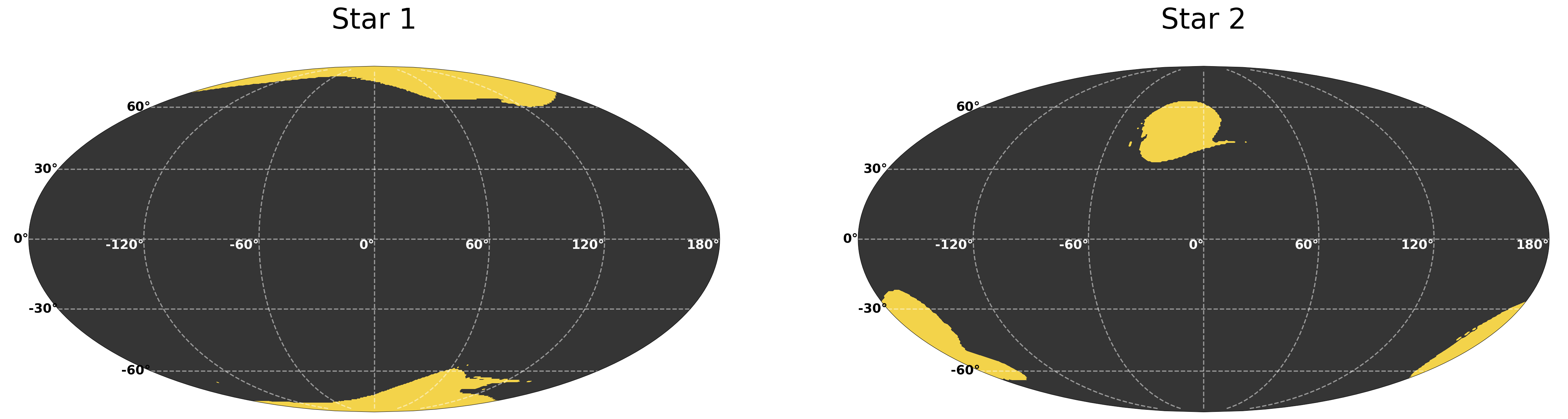}
    \caption{The polar caps, i.e., instantaneous zones of open flux, of the two stars in a Mollweide projection in the case with $\alpha_1=0^\circ,\,\alpha_2=45^\circ,\,\lambda=-1$, and $\Delta\Phi=0^\circ$(\AngledUpArrow{0}\AngledDownArrow{-45}) similar to the middle row of Figure ~\ref{Figure5}. The top panel corresponds to $\tau=-9.91$~ms, while the bottom panel shows $\tau=-8.6$~ms, with the two snapshots separated by half an orbital period.}
    \label{Figure31}
\end{figure}

\subsubsection{Inclined Magnetic Moments}
\label{sec:misaligned}

Inclined magnetic moments introduce significant richness into the solutions, breaking both the latitudinal symmetry and the reflection symmetry of the emergent Poynting flux across the axial plane that passes through the center of the system, perpendicular to the line connecting the two stars. Figure~\ref{Figure8} depicts configurations where the stars have non-zero-inclination angles for both the quasi-aligned ($\lambda > 0$, first and second columns) and quasi-antialigned ($\lambda < 0$, third and fourth columns) solutions. In the first (\AngledUpArrow{0}\AngledUpArrow{-45}) and third columns (\AngledUpArrow{0}\AngledDownArrow{-45}), a star has an inclination angle of $45^\circ$, whereas, in the second (\AngledUpArrow{0}\AngledUpArrow{-90}) and fourth columns (\AngledUpArrow{0}\AngledDownArrow{-90}), it has an inclination angle of $90^\circ$. Notably, the second and fourth columns correspond to configurations that are mirror images of each other under a rotation by $\pi$ around the axis, defined by the line connecting the two stars.
Additionally, in the corotating frame, the two stars appear to rotate because they are irrotational in the observer frame. 

Unlike the fully aligned or antialigned cases (\S\ref{sec:aligned}), the presence of inclined magnetic moments introduces a more pronounced impact of orbital motion on the magnetosphere's structure. The resulting magnetospheres evolve dynamically, exhibiting periodic variations and eruptions of Poynting flux linked to the orbital phases. As the stars orbit each other, magnetic fields originating from one star are carried along with its motion, anchored on its surface. As a result, the system reorganizes during a single rotation (see the middle and bottom rows in Figure ~\ref{Figure5}). Field lines can dynamically evolve throughout the inspiral; some may connect one star to the other, while others that initially link the two stars may break and reconfigure. As a result, field lines anchored to a star's surface can transition between being open, closed, or connecting with the companion star. 

Thus, the polar caps, defined as footpoints of field lines extending to infinity (i.e., open flux), evolve over time, as shown in Figure~\ref{Figure31}. This evolution follows the orbital periodicity, with the variability becoming increasingly pronounced as the separation between the stars decreases during inspiral. The polar cap regions are found by integrating magnetic field lines using a fourth-order Runge-Kutta method, classifying as open those that cross a cylinder of radius $25r_\star$ \citep[if sufficiently large, the value of this cylindrical radius does not seem to impact significantly the size and shape of the polar caps; see also][]{Ortiz2022} and height of $50r_\star$. In Figure~\ref{Figure31}, the polar caps of both stars are shown at two snapshots separated by half a rotation, for the case with $\lambda=-1$, $\alpha_1=0^\circ$, and $\alpha_2=45^\circ$ (\AngledUpArrow{0}\AngledDownArrow{-45}).

The fluctuating size of the polar caps influences the open flux $\Phi_{\rm open}$, causing the energy flux to vary periodically with the orbital frequency and chirps during inspiral. The $\Phi_{\rm open}$ behavior can be quantified by calculating the radial component of Poynting flux through a large spherical surface.

\begin{figure*}
\centering
    \includegraphics[width=0.9\textwidth]{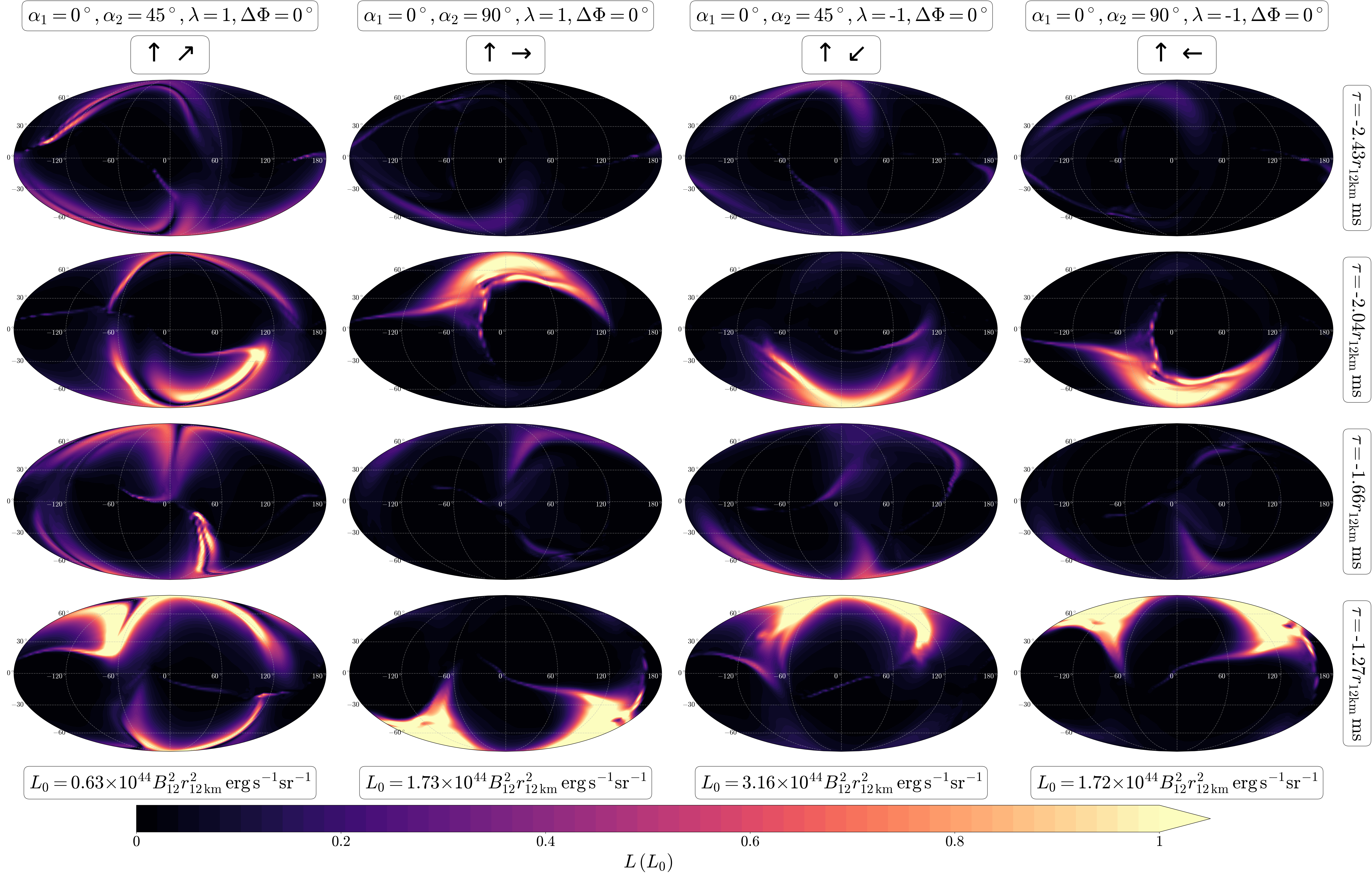}
    \caption{Snapshots of the Poynting flux evolution over a full rotation of the system, directly corresponding to the magnetospheric configurations and snapshots presented in Figure~\ref{Figure6}.} 
    \label{Figure10}
\end{figure*}

Figure~\ref{Figure10} shows the flux distributions on a spherical surface with radius $ r=13 r_\star$ corresponding to the magnetospheric structures and snapshots shown in Figure~\ref{Figure8}. Consistent with the discussion above, unlike the zero-inclination cases discussed previously, the flux distribution in these cases is no longer symmetric between the northern and southern hemispheres. While the flux pattern may appear symmetric over a full orbital period, the instantaneous flux evolves throughout the inspiral, leading to unequal fluxes in the two hemispheres at any given moment \citep{Most&Philippov2022}. This asymmetry in the instantaneous and time-averaged Poynting flux pattern indicates the presence of EM forces and torques acting on the system, influencing the motion of the NSs and potentially affecting the postmerger remnant. This aspect is further explored in Section \ref{sec:torques}. 

The reflection symmetry noted earlier between the second and fourth columns of Figure~\ref{Figure8} is also evident in the Poynting flux distributions in Figure~\ref{Figure10}, where the flux patterns for the corresponding cases are mirrored across the equatorial plane. Additionally, the flux reaching the surface where the calculations are made exhibits a phase shift equivalent to half a rotation.

\begin{figure*}
    \centering
    \includegraphics[width=0.49\textwidth]{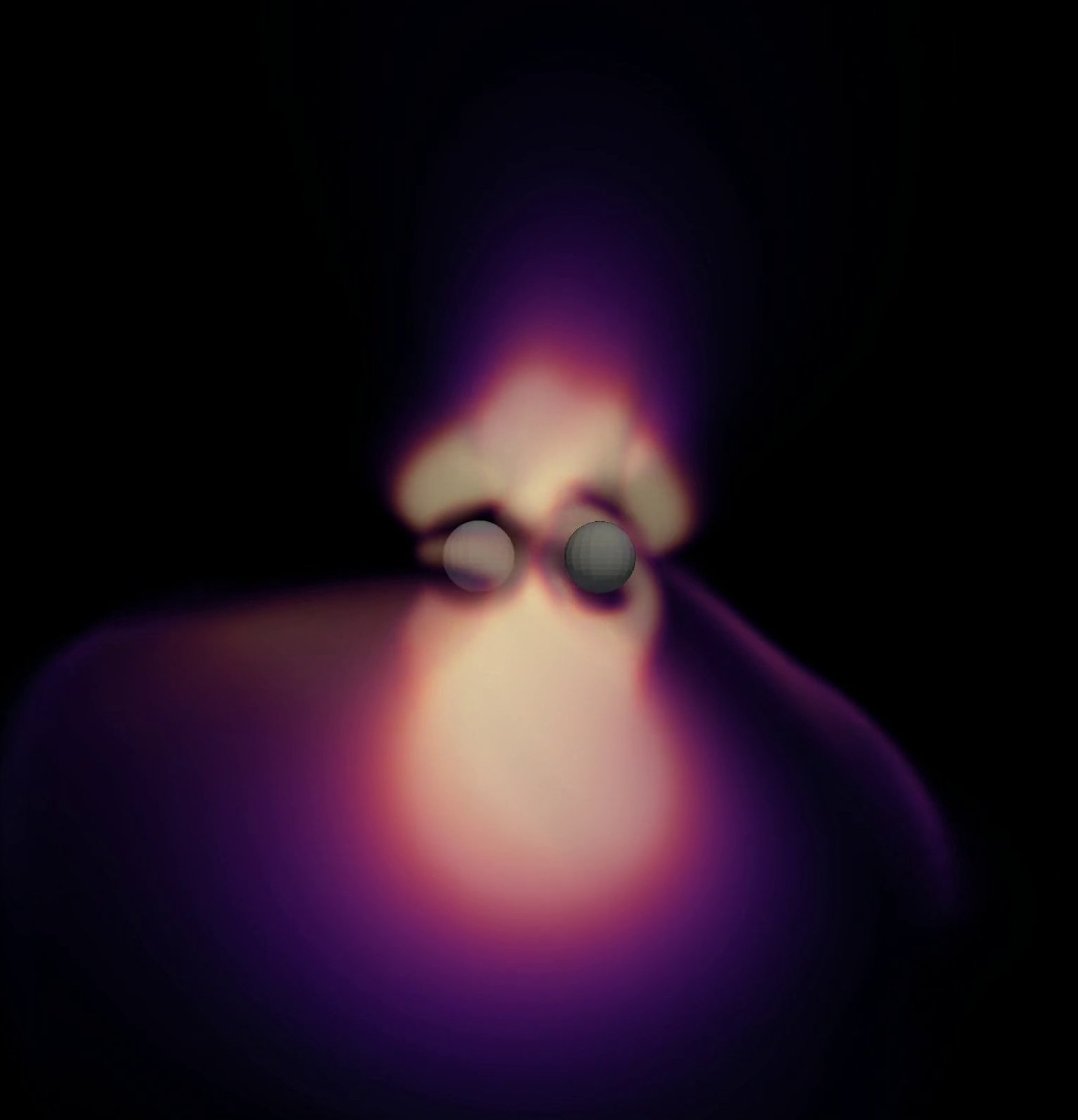}\hfill
    \includegraphics[width=0.49\textwidth]{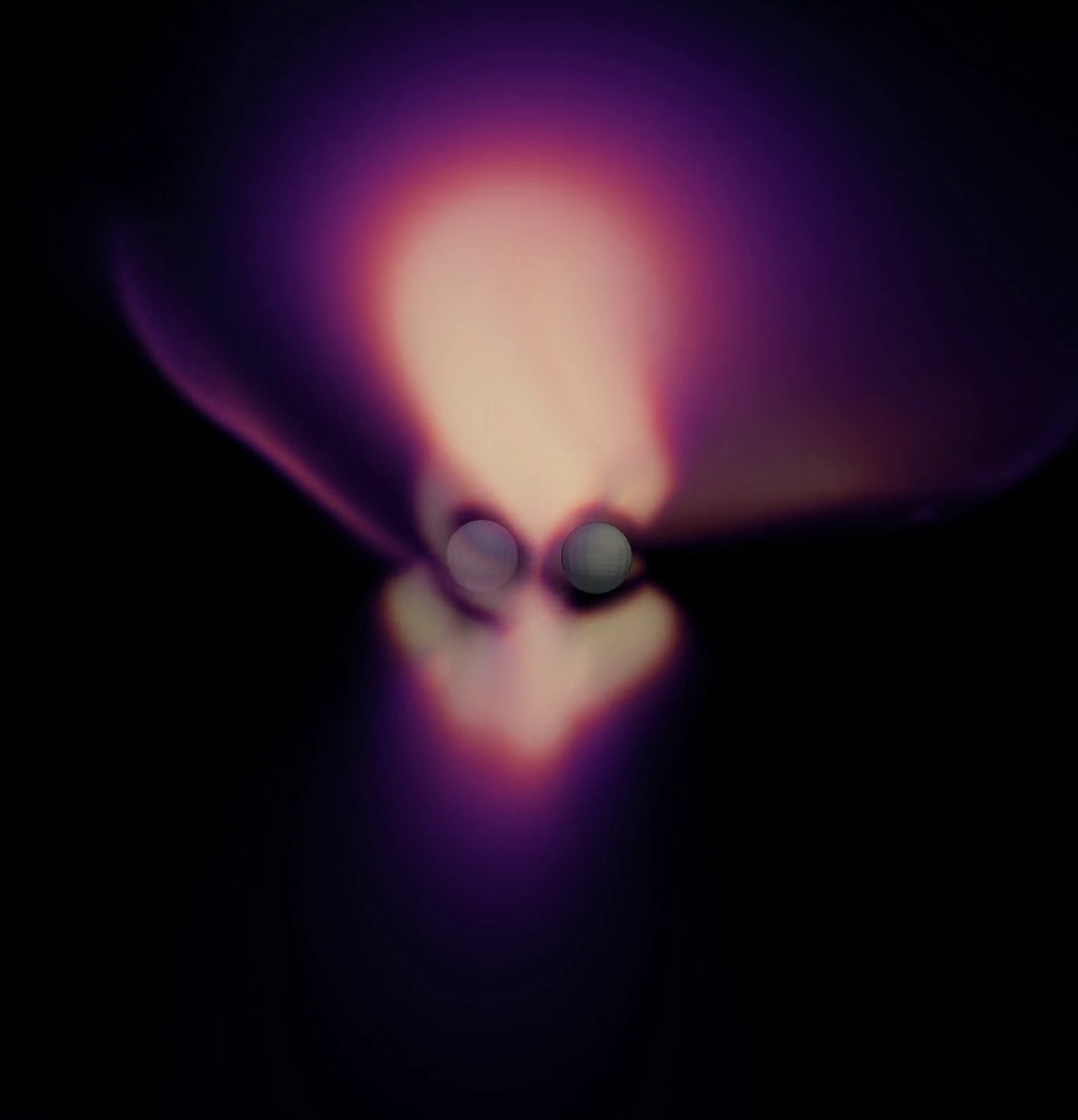}\hfill
    \caption{3D volume rendering of the Poynting flux for the case with $\alpha_1=0^\circ,\,\alpha_2=45^\circ,\,\lambda=-1$ and $\Delta\Phi=0^\circ$ (\AngledUpArrow{0}\AngledDownArrow{-45}) similar to the middle row of Figure ~\ref{Figure5}. (\textbf{Left}) Snapshot at $\tau=-9.91$~ms, and (\textbf{right}) snapshot for $\tau=-8.6$~ms, the two instances are separated by half an orbital period. The color scale represents the local Poynting flux values, while only regions where the Poynting flux multiplied with $r^2$ exceeds a threshold are shown,  effectively accounting for the radial decrease in flux with distance and highlighting areas of intrinsically strong emission.}
    \label{Figure30}
\end{figure*}

To better understand the Poynting flux anisotropy, Figure ~\ref{Figure30} presents a volume-rendered visualization of the differential Poynting flux for the $\lambda=-1$ and $\alpha_1=0^\circ,\alpha_2=45^\circ$ (\AngledUpArrow{0}\AngledDownArrow{-45}). The flux was calculated on multiple concentric spherical surfaces at increasing radii from the system's center at time intervals separated by a half of rotation. This visualization highlights the dynamic evolution of the system's intrinsic emission anisotropy.

\begin{figure*}
\centering    \includegraphics[width=0.5\textwidth]{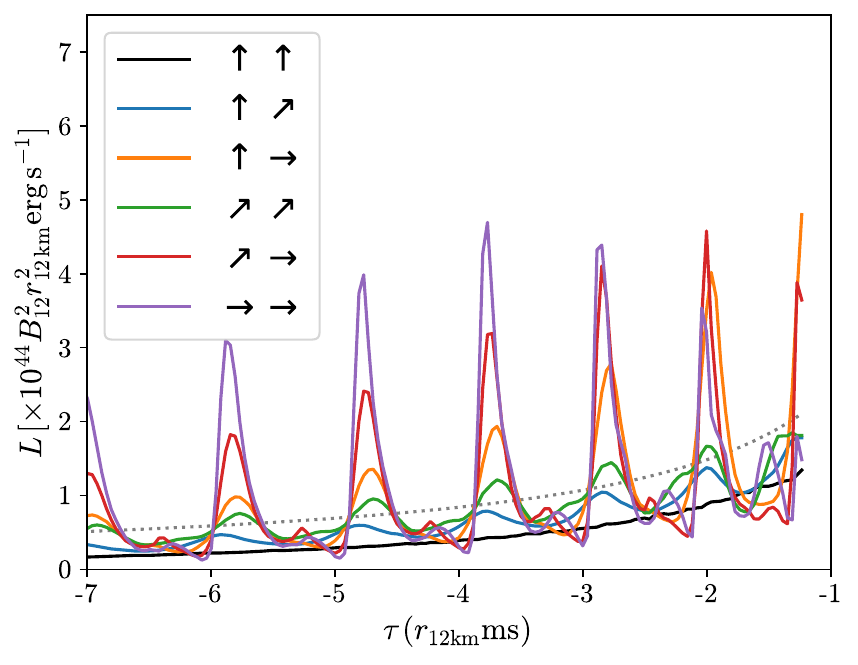}\hfill
\includegraphics[width=0.5\textwidth]{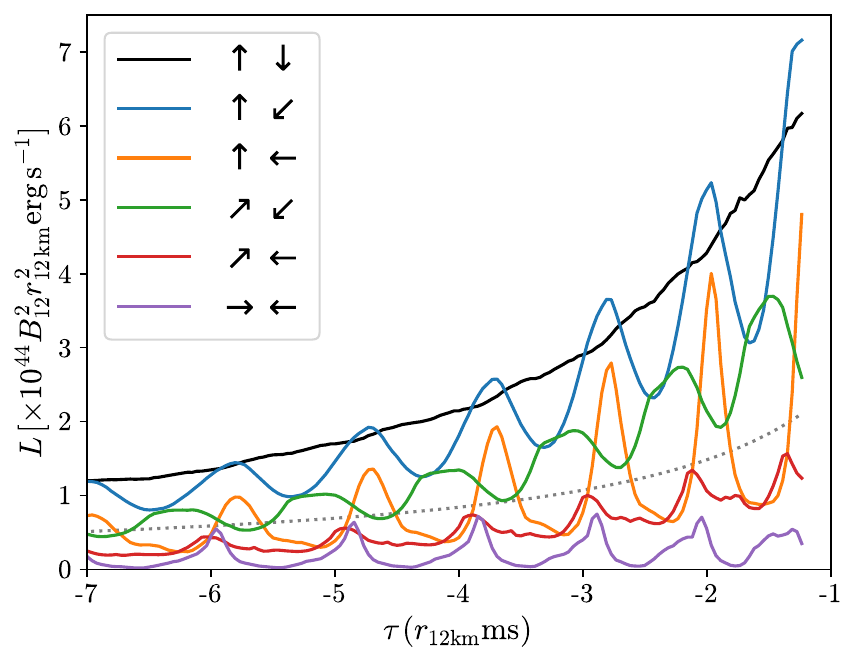}\hfill
    \caption{The total Poynting flux crossing a spherical surface with radius $13r_\star$ for various inclination angle combinations of the two stars, shown for cases with $\lambda=1$ (\textbf{left}) and $\lambda=-1$ (\textbf{right}). The dotted line, in both plots, is the orbital average Poynting flux $\overline{\langle L \rangle}$  derived from the scaling law of Eq.~\ref{eq:Lfit} for $|\lambda|=1$. }
    \label{Figure12}
\end{figure*}

The variability and dependence on magnetic configuration of the EM flux are also evident in Figure \ref{Figure12}, which shows the total flux passing through the spherical surface with $r=13r_{\star}$ as a function of time. For $\lambda > 0$ (left panel), the total flux increases when the stars have non-zero-inclination angles and rises further as the angles increase. In contrast, for $\lambda < 0$ (right panel), the total flux decreases with increasing inclination angles. Notably, when both stars have inclination angles of $90^\circ$ and $\lambda < 0$, the total flux can drop to nearly zero at specific times during the evolution. This is illustrated in the bottom row of Figure ~\ref{Figure5}, where two snapshots are shown, separated by an orbital phase of $\pi/2$. When the magnetic poles of the two stars with the same polarity face each other (left panel), field lines bend away, twist, and predominantly connect each star’s opposite magnetic poles, allowing a modest number of field lines to remain open. On the other hand, after a time evolution corresponding to the rotation of $\pi/2$ (right panel), the magnetic moments are oriented side by side, pointing in opposite directions. This configuration leads to even fewer open field lines and a significantly lower escaping flux compared to the first snapshot.

This behavior reflects the fact that the aligned and antialigned configurations represent two extreme cases of the angle between the magnetic moments, directly influencing the degree of connectivity between the stars and the amount of open flux $\Phi_{\rm open}$.

For completeness, it is worth noting how the above results change in the opposite limit of tidally locked (synchronously rotating) stars, where $\omega_1 = \omega_2 = \Omega$. The differences relative to the irrotational cases depend on the magnetic configuration. In the synchronized and aligned case, the overall magnetospheric structure and resulting Poynting flux would be nearly identical to the aligned irrotational configuration. This is because the relative orientation between the magnetic moments and the orbital motion remains constant, resulting in minimal time variation in the magnetosphere. In the synchronized 
 and antialigned configuration, some magnetic field lines connect the two stars, reducing the number of open field lines available to transport the EM energy. This leads to a lower Poynting flux compared to the corresponding aligned case, in contrast to what was found in the irrotational scenario. However, for cases where at least one of the stars has an inclined magnetic moment, the differences are more pronounced. The synchronization prevents the time-varying reconfiguration of field lines seen in the irrotational cases. As a result, the Poynting flux evolution arises solely from the shrinking orbital separation, which increases the polar cap area and flux magnitude.  

\subsubsection{Unequal Magnetization Configurations}
\label{sec:highlambda}

From isolated binary evolution, the double NSs are not expected to possess equal magnetic moments. We consider such asymmetry now. Interestingly, in the extreme limit of an unmagnetized star merging with a magnetized one, the unmagnetized conductor ``swallows" magnetic flux.

\begin{figure*}
\centering
    \includegraphics[width=0.67\textwidth]{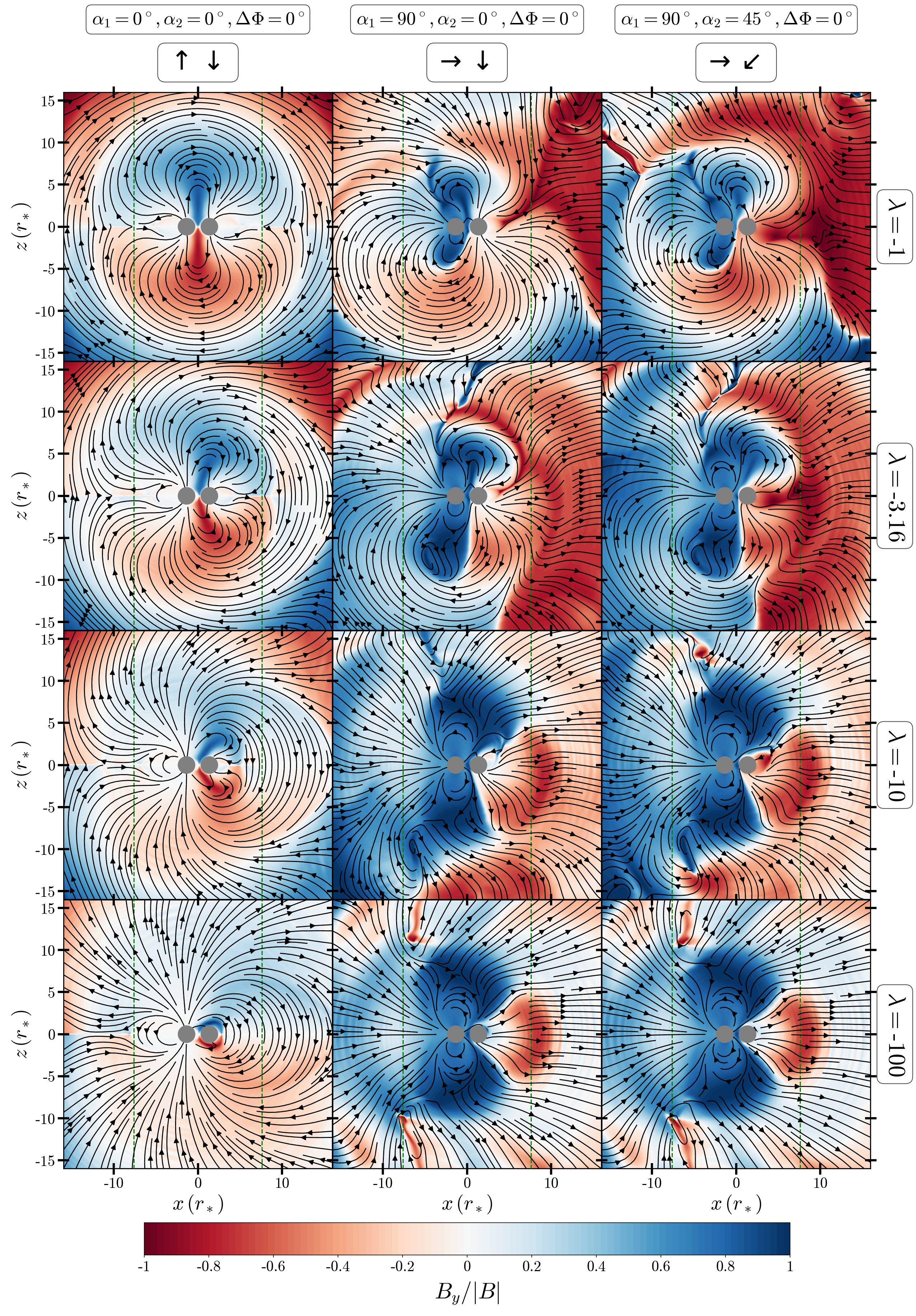}
    \caption{The magnetospheric structure at $\tau = -4.27$ ms, similar to Figure~\ref{Figure8}, is shown with increasing absolute values of the magnetic moment ratio from top to bottom. The left column illustrates the impact of the magnetic moment ratio in the antialigned case ($\alpha_1 = 0^\circ, \alpha_2 = 0^\circ$) (\AngledUpArrow{0}\AngledDownArrow{0}). The middle and right columns follow the same pattern, but for the configurations $\alpha_1 = 90^\circ, \alpha_2 = 0^\circ, \Delta\Phi = 0^\circ$ (\AngledUpArrow{-90}\AngledDownArrow{0}, middle) and $\alpha_1 = 90^\circ, \alpha_2 = 45^\circ, \Delta\Phi = 0^\circ$ (\AngledUpArrow{-90}\AngledDownArrow{-45}, right), respectively.
    }
    \label{Figure13}
\end{figure*}

Figure~\ref{Figure13} shows the evolution of the magnetosphere in cases with $\lambda<0$, specifically focusing on configurations where one star has a smaller magnetic moment than the other, meaning $\lambda < -1$. For comparison, the first row presents the case of $\lambda = -1$.  The left-hand column depicts the (\AngledUpArrow{0}\AngledDownArrow{0}) configuration, the middle column shows (\AngledUpArrow{-90}\AngledDownArrow{0}), and the right-hand column displays (\AngledUpArrow{-90}\AngledDownArrow{-45}). In all cases, the global magnetospheric structure is predominantly influenced by the star with the stronger magnetic moment (the left-hand star in all the panels), while the weaker star primarily affects the field in its immediate vicinity. As $\lambda$ becomes more negative, i.e., the disparity in magnetic moments increases, the influence of the weaker star diminishes further. When $\lambda\ll 1$, the inclination angle of the star with the smaller magnetic moment has minimal impact on the overall magnetospheric structure.

The distribution of orbit-averaged emission in cases with  $|\lambda|>1$ becomes increasingly concentrated around the equator \citep[see also Figures 13 and 4 in][respectively]{Palenzueala2013b,Most&Philippov2022}, as shown in the left column of Figure~\ref{Figure14}. As $|\lambda|$ increases, the emission progressively resembles that of an isolated pulsar with the same inclination angle as the dominant star, similar to Fig.~\ref{Figure1}. 

Figure ~\ref{Figure33} illustrates the total energy flux crossing a spherical surface with radius $r=13r_{\star}$ as a function of time for various configurations with $\lambda<0$, specifically for the case where $\alpha_1 = 0^\circ$, $\alpha_2 = 0^\circ$ (\AngledUpArrow{0}\AngledDownArrow{0}), and $\Delta\Phi = 0^\circ$. The total flux exhibits significant variation, with smaller values of $\lambda$ corresponding to higher flux levels.

It is important to note that the way $\lambda$ is varied here affects the normalization in Figure \ref{Figure33}: we adjust $\lambda$ by reducing the magnetization of one star while keeping the other fixed.

\subsubsection{Anisotropic Emission Patterns}

\label{sec:anisoemission}

\begin{figure*}
\centering
    \includegraphics[width=0.85\textwidth]{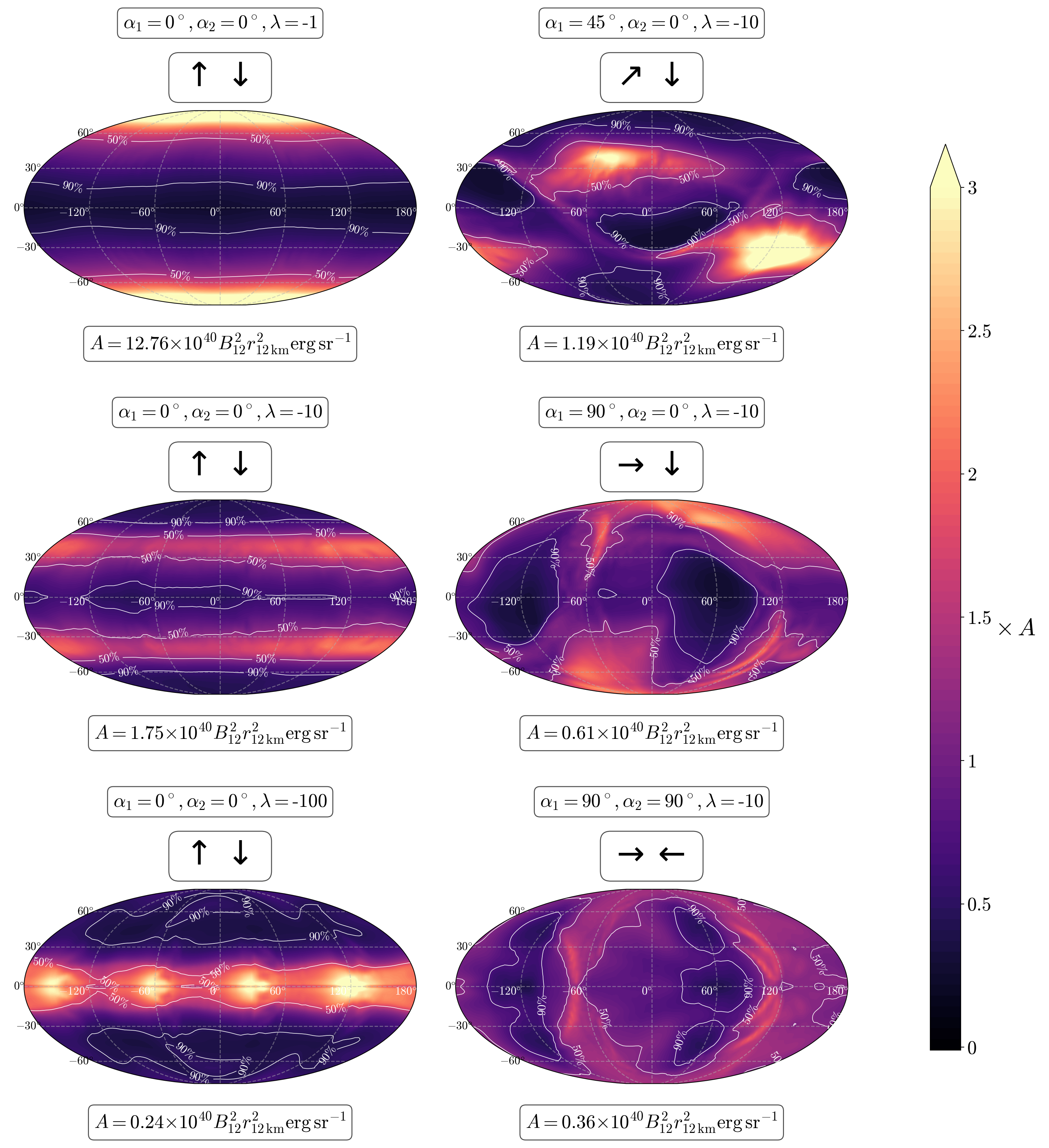}
    \caption{The distribution of the  Poynting flux across a spherical surface of radius $13r_\star$, integrated over the last 7.7 ms of the inspiral. White lines indicate the areas containing $50\%$ and $90\%$ of the total (integrated over the entire surface) flux. \textbf{(Left)} The two stars have antialigned magnetic moments $\uparrow\downarrow$, with magnetic moment ratios $\lambda = -1,-10,-100$ from top to bottom, respectively. \textbf{(Right)} Cases with $\lambda=-10$ and $\alpha_1=45^\circ,\alpha_2=0^\circ$ (\AngledUpArrow{-45}\AngledDownArrow{0}), $\alpha_1=90^\circ,\alpha_2=0^\circ$ (\AngledUpArrow{-90}\AngledDownArrow{0}) and $\alpha_1=90^\circ, \alpha_2=90^\circ, \Delta\Phi=0^\circ$ (\AngledUpArrow{-90}\AngledDownArrow{-90}) from top to bottom, respectively.  
    }
    \label{Figure14}
\end{figure*}

Beyond the cases mentioned in the previous section, we have systematically explored all possible configurations corresponding to the combinations of the adopted parameter values. 
In general, the sky distributions of the total, time-integrated fluence, derived from the corresponding Poynting flux, show significant variability across the parameter space, and are typically highly anisotropic (see the right column in Figure~\ref{Figure14}). Although Figure~\ref{Figure14} presents cases with $\lambda < 1$, we find that cases with $\lambda > 1$ display similar anisotropic energy distributions. While larger $|\lambda|$ values generally lead to more pronounced anisotropies, certain configurations with $|\lambda|=1$ also exhibit substantial directional concentration of the outgoing energy. 

In particular, in general, $50\%$ of the escaping EM energy is in $<50\%$ of the total sky area, while, in some configurations, it is directed toward $<15\%$ of the sky (see Figure~\ref{Figure15}).  

In specific configurations, the emission anisotropy is more uniform, similar to an isolated pulsar. In these cases, the time-averaged energy distribution is independent of the azimuthal angle; thus, potential observers located at different azimuths would receive similar amounts of fluence with the observer's polar angle, $\zeta$, being the only determining parameter. 

\begin{figure}
\centering
    \includegraphics[width=\columnwidth]{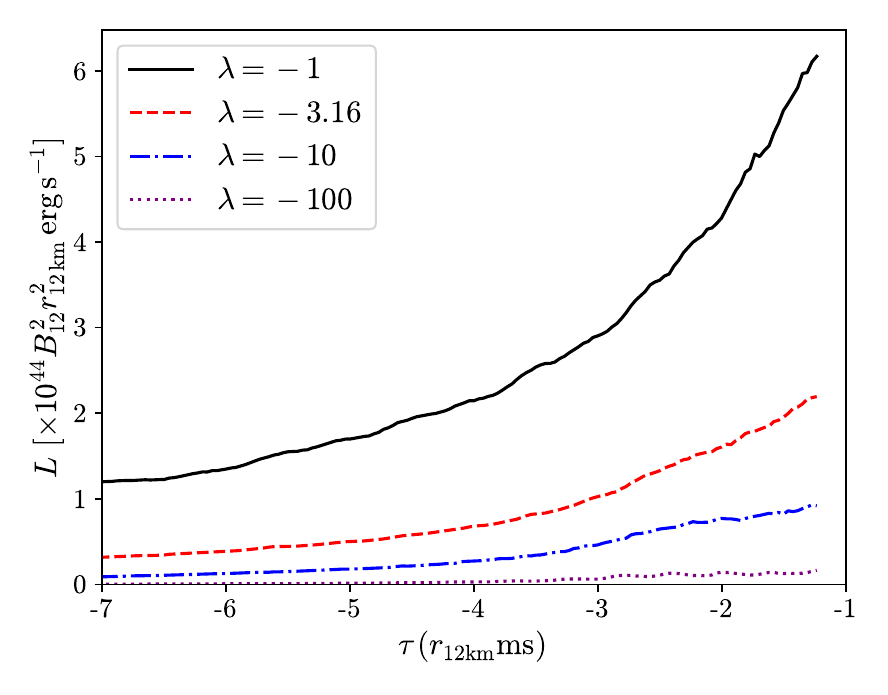}\hfill
    \caption{The total Poynting flux crossing a spherical surface radius $13 r_\star$ throughout the system's evolution for cases with $\alpha_1=\alpha_2=0$ and $\lambda<0$ ($\uparrow\downarrow$) and progressively increasing $|\lambda|$ values.}
    \label{Figure33}
\end{figure}

\begin{figure}
\centering
    \includegraphics[width=\columnwidth]{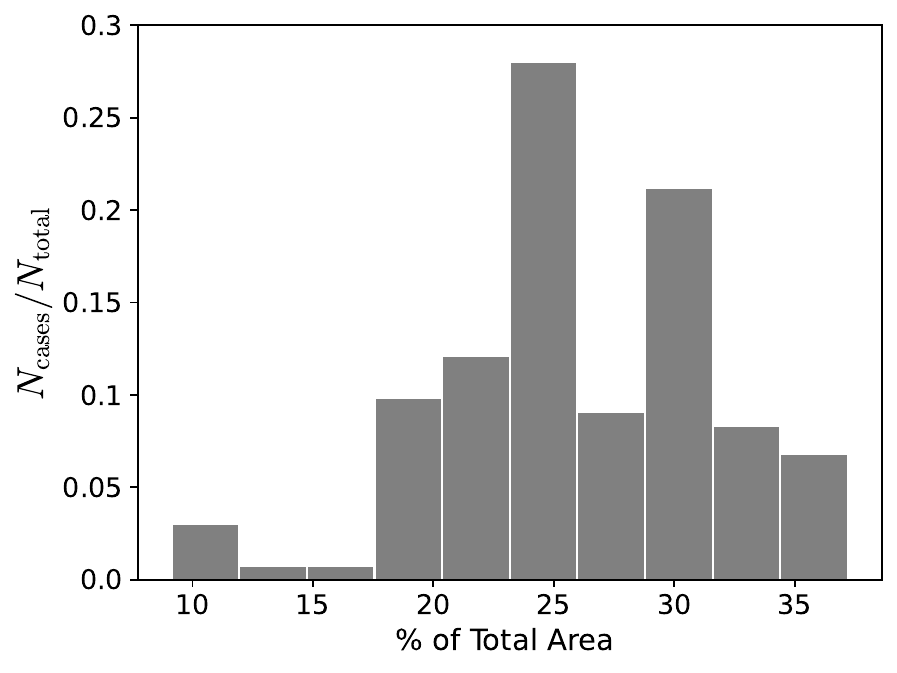}\hfill
    \caption{Histogram of emission anisotropy, showing the distribution of the sky area fraction that contains 50\% of the total flux across the explored parameter combinations.}    \label{Figure15}
\end{figure}

Yet, in the majority of cases explored in our study, the fluence distribution deviates significantly from that of a single isolated pulsar, exhibiting strong azimuthal anisotropy (see the right column in Figure~\ref{Figure14}). This anisotropy is shaped not only by the evolving magnetospheric configurations throughout an orbital period but also by the continuously increasing EM flux, which influences different directions with varying intensities over time. Thus, energy transport is highly anisotropic, with interesting implications (see \S\ref{sec:summary}).

\subsubsection{Scaling Laws for Poynting Luminosity and the Binary Braking Index}
\label{sec:fit}

\begin{figure}
\centering
    \includegraphics[width=\columnwidth]{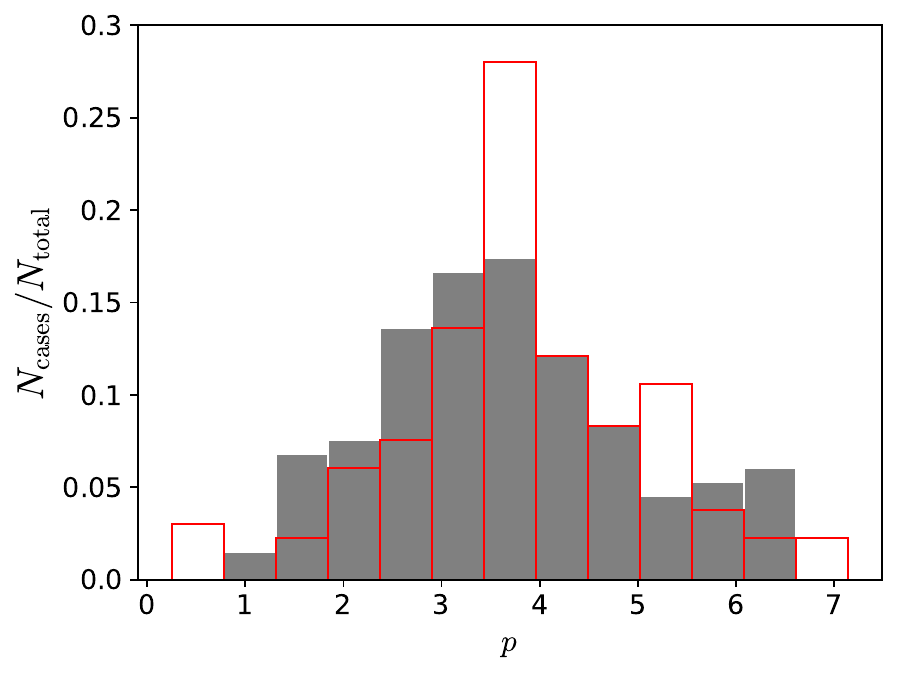}
    \caption{Histogram of the power-law index $p$ obtained from fitting of the relation $\langle L \rangle \propto \Omega^p$ shown in gray for all data points and in red only for the peak values. The distribution of $p$ spans a wide range from $\sim 1$ to $>6$, in contrast to the $p\equiv4$ value for an isolated dipolar pulsar.} 
    \label{Figure16}
\end{figure}

Simulating the evolution of the magnetosphere for a variety of configurations allows us to investigate how the escaping EM flux depends on the orbital frequency, $\Omega$, and consequently on time via Eq.~\ref{eq:omega}. The simplest way to quantify this dependency is to assume a power-law relation between the luminosity and the orbital frequency analogous to an isolated pulsar, such that $\langle L \rangle \propto \Omega^p$, where $p$ is the power-law index. The classical dipolar isolated pulsar magnetosphere in either vacuum or FF regimes follows $p\equiv4$ for the far-field, and it has often been assumed by {\it heuristic arguments} arguments that this law roughly (e.g., $p\sim 3.3-4.7$; see discussion) holds in the double NS case. By fitting this function to the evolving total Poynting flux from each BNS simulation, we obtain the distribution of the values of the exponent $p$, shown in Figure \ref{Figure16}. 

Notably, the parameter $p$ now spans a broad range of values spanning from 1 to more than 6.  While there are many cases around $p=4$, the binary system shows cases where $p$ deviates significantly, either higher or lower, due to the complex interactions between the two stars' magnetospheres. This surprisingly large deviation from the conventional dipole braking index $n \equiv p-1 = 3$ highlights the richness of near-field effects permitted in the double NS case that modify far-fields and escape radiation.
\begin{figure*}
       \centering 
    \includegraphics[width=0.49\textwidth]{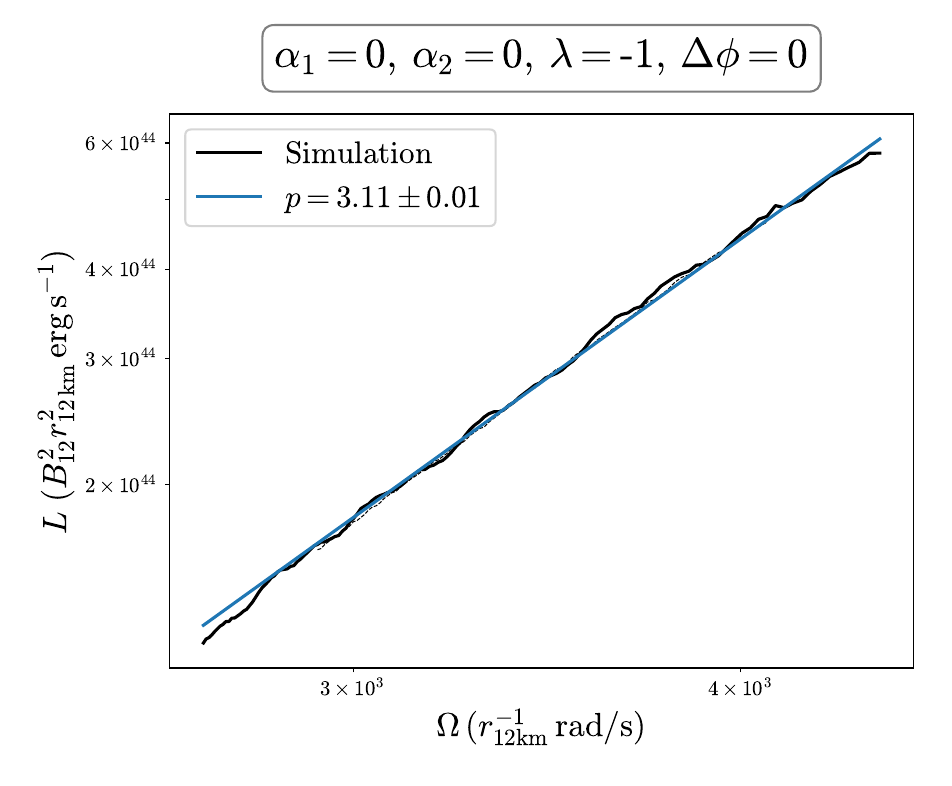}
    \includegraphics[width=0.49\textwidth]{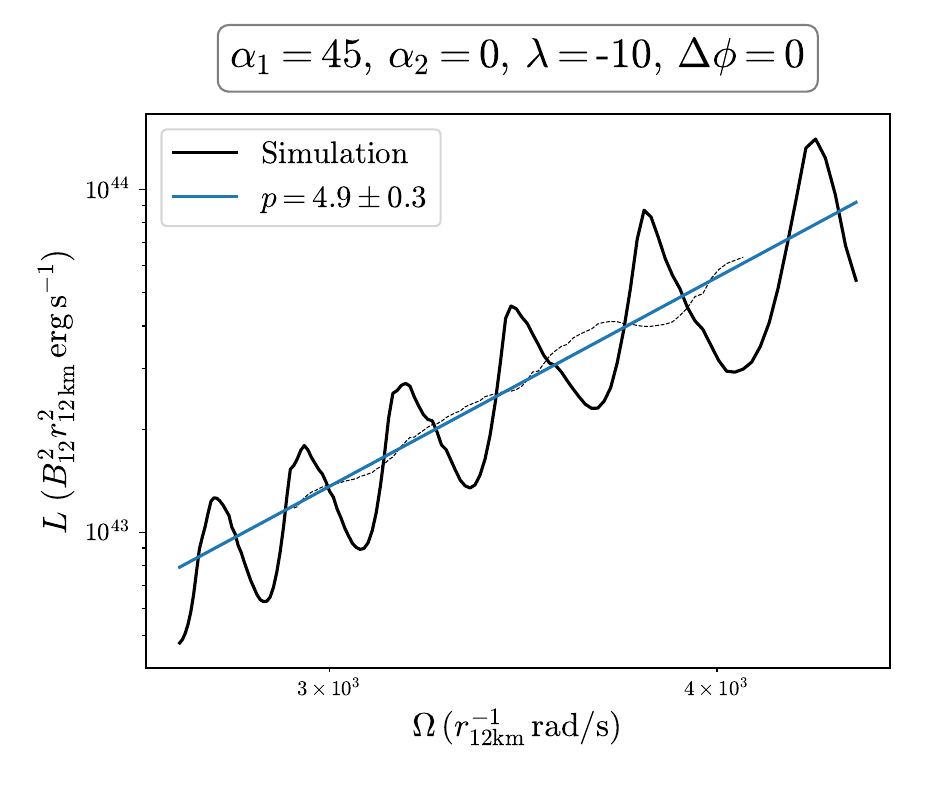}
    \includegraphics[width=0.49\textwidth]{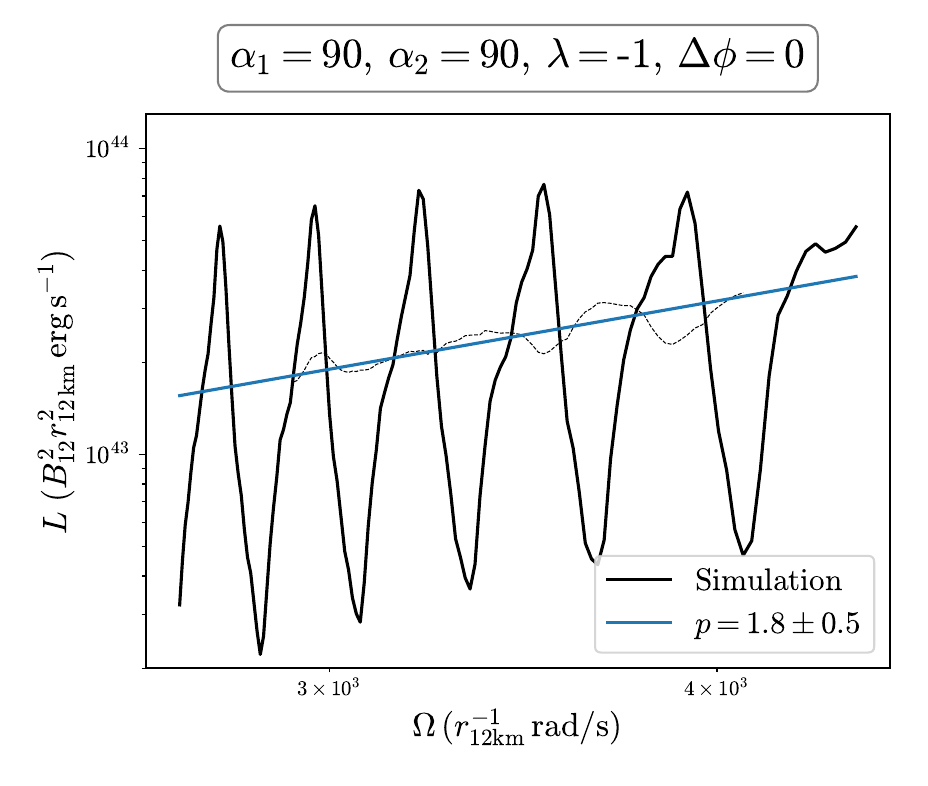}
    \includegraphics[width=0.49\textwidth]{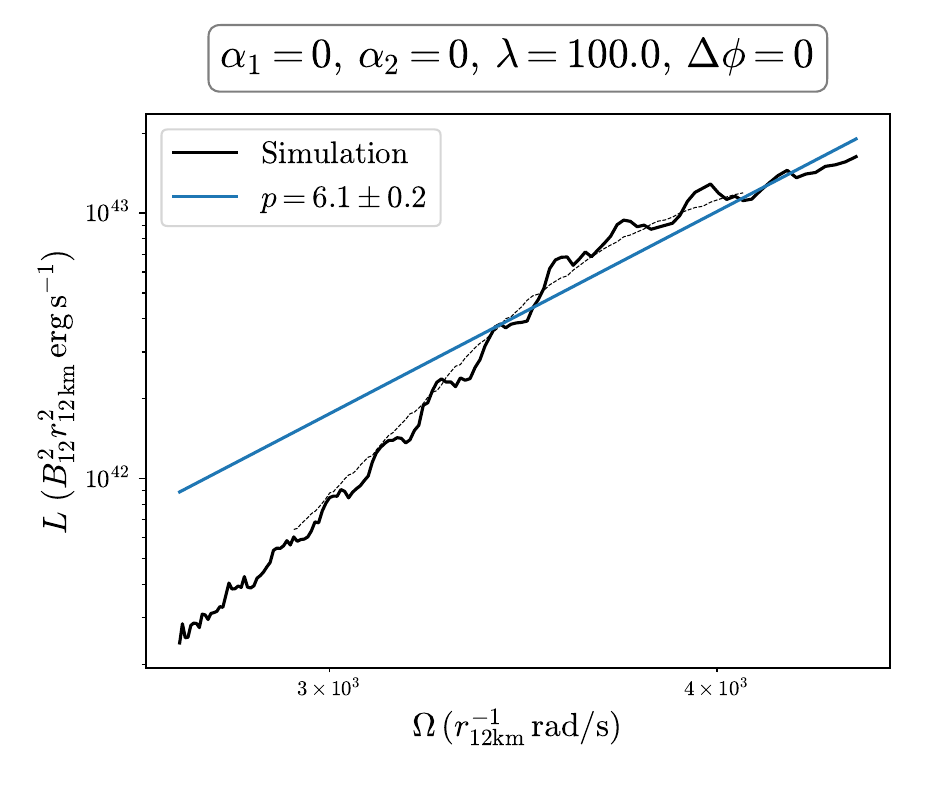}
    \caption {Examples fit of the Poynting flux luminosity with a power law $\langle L\rangle\propto\Omega^p$. \textbf{Top row}: (\textbf{left}) the case with $\alpha_1=0^\circ,\alpha_2=0^\circ$, and $\lambda=-1$.  (\textbf{Right}) The case with $\alpha_1=45^\circ,\alpha_2=0^\circ$, and $\lambda=10$ demonstrates that, although the power law cannot describe the periodic fluctuations, it captures well the trend of the luminosity. \textbf{Bottom row}: (\textbf{left}) the case with $\alpha_1=90^\circ,\alpha_2=90^\circ,\lambda=-1$, and  $\Delta\Phi=0^\circ$ highlights that, although the flux can vary significantly, dropping to very low values and affecting the flux dependence on $\Omega$ the power law still describes quite well the averaged trend. (\textbf{Right}) The case with $\alpha_1=0^\circ,\alpha_2=0^\circ$, and $\lambda=100$, which consists of an example of not that accurate description of the Poynting flux luminosity's evolution with the power law. In all the cases, the dotted line indicates the moving average for comparison.}
    \label{fit_examples}
\end{figure*}

Figure ~\ref{fit_examples} showcases characteristic examples of the Poynting flux luminosity fits.
While the power-law model captures the gross secular orbit-averaged trend of the luminosity, it does not account for the periodic fluctuations around the fitted $L(\Omega)$ values. This is particularly evident for high inclination angles (bottom left panel in Figure~\ref{fit_examples}), where, as described in the previous section, the total flux fluctuates dramatically, occasionally dropping to near-zero levels. This extreme variability weakens the overall dependence of the flux on $\Omega$, causing the value of $p$ to decrease, often approaching values close to 1, as seen in Figure \ref{Figure16}. Importantly, a fitted $p\sim1$ does not necessarily imply an induced split-monopolar or wind configuration \citep[e.g.,][]{1999ApJ...525L.125H,2006MNRAS.368.1717B}. Instead, it reflects the presence of intense, episodic bursts of Poynting flux that produce similar luminosities from one orbit to the next.

In some cases, the evolution of the Poynting flux appears to transition between two distinct regimes (bottom right panel in Figure~\ref{fit_examples}), resulting in a fitted $p$ that represents an average rather than a single consistent trend. For instance, in the configuration shown in the bottom right panel, where both magnetic moments are aligned (to each other and to $\pmb{\Omega}$), and one star's field dominates, no field lines connect the two stars. At first, only the region of the open field lines of the primary star is influenced mostly by the secondary, with some bending around the companion and closing back onto the primary. This phase corresponds to the initial rise behavior in Poynting flux. However, as the inspiral progresses, the companion moves into the closed field region of the primary, leading to a transition where more field lines open. Since no lines can directly connect the two stars in this scenario, these newly opened field lines increase the Poynting flux faster, modifying its overall evolution.

Despite these variations, we find that, in the majority of cases, the average Poynting flux evolution follows a single well-defined power-law trend over the fitting interval in our simulations. Instances where the flux transitions between two regimes, as seen in the bottom right panel of Figure~\ref{fit_examples}, remain relatively rare.

\begin{figure}
    \centering 
    \includegraphics[width=0.49\textwidth]{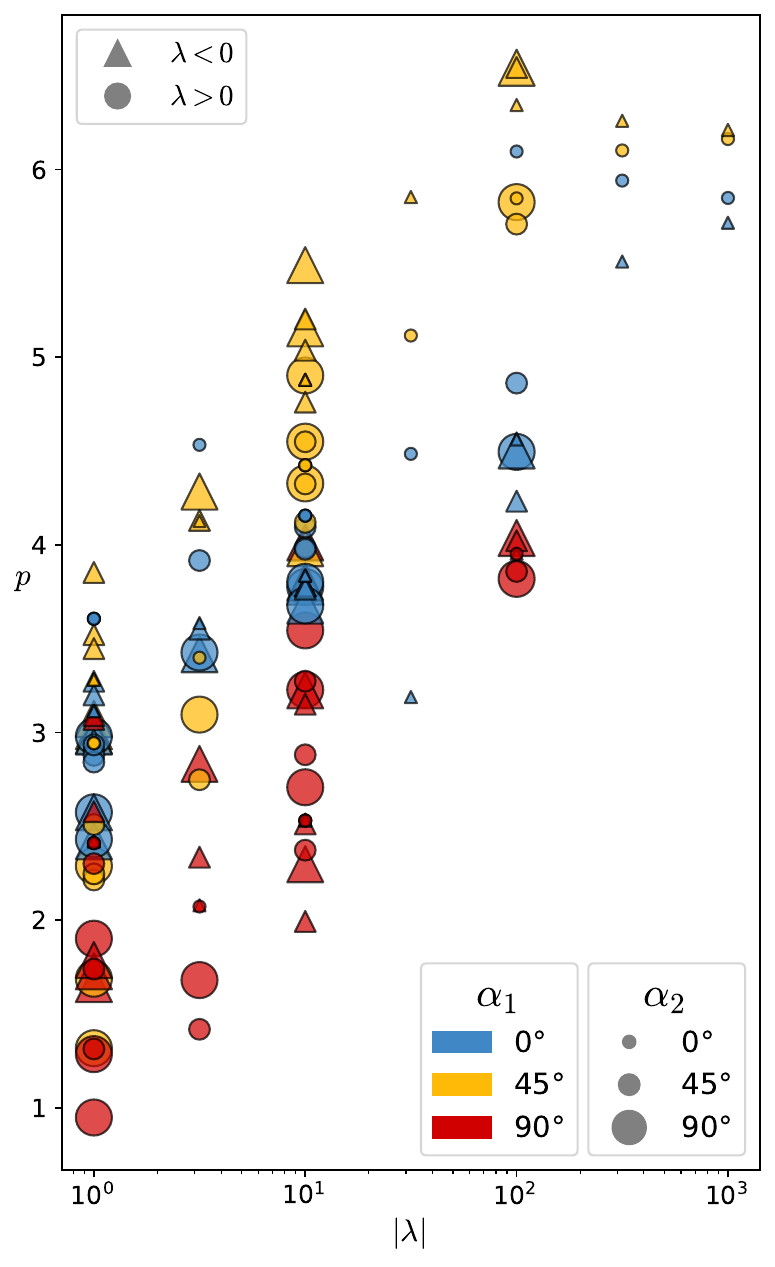}
    \caption{Scatter plot showing the power $p$ derived from the power-law fitting $\langle L\rangle\propto\Omega^p$ for the different simulated cases. On $x$-axis is $\lambda$ in logarithmic scale while the color and the size of each point indicate the two inclination angles $\alpha_1$ and $\alpha_2$ according to the legends. The circles are cases with $\lambda>0$, while the triangles are with $\lambda<0$. The different values of the exponent $p$ for the same $\alpha_1$, $\alpha_2$ and $\lambda$ combinations is due different $\Delta\Phi$ angles.}
    \label{p_vs_parameters}
\end{figure}

In Figure~\ref{p_vs_parameters} the parameter $p$ found is presented for all the simulations; in respect to $|\lambda|$ in $x$-axis (different markers for $\lambda>0$ and $\lambda<0$), the angle $\alpha_1$ (of the dominating field) is indicated by the point color, and the angle $\alpha_2$ is indicated by the point size. The figure demonstrates the significant variability in $p$ and shows a clear correlation between its average value and $|\lambda|$. The average value of $p$ is increased with $\lambda$ up to $|\lambda|=100.$ Note here that the high value of $p$ does not mean that, in this case, the system produces more flux, but it means the flux increases faster. 
 
Additionally, as $|\lambda|$ increases, the effect of the less magnetized star on the magnetosphere structure should eventually be saturated. In this direction, simulations for $|\lambda|=\sqrt{10^3},\sqrt{10^5},10^3$ are conducted and verify that beyond $|\lambda|=100$ the value of $p$ stops increasing. 

On the other hand, the dependence of $p$ on the magnetic moment orientation is not that clear. Although for $|\lambda|\geq10$, there seems to be a tendency for $p$ to be high for cases where the dominant star has an inclination angle $\alpha_1=45^\circ$ and less for cases with $\alpha_1=90^\circ$; this is not always consistent as the specific orientation of the magnetic moments is important, and the power law lacks the ability to describe accurately the periodic variability in the flux. 
Note that in Figure ~\ref{p_vs_parameters} for $|\lambda|=1,10$ we have multiple points with the same combinations of $\alpha_1$ and $\alpha_2$ due to the different $\Delta\Phi$ these cases correspond to. 

\begin{figure}
    \includegraphics[width=0.5\textwidth]{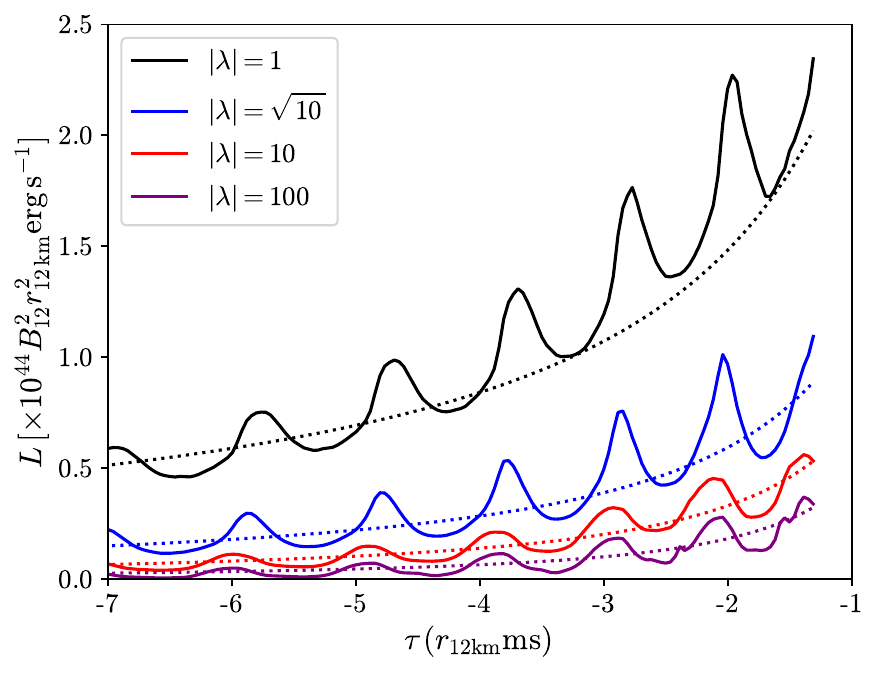}\hfill
    \caption{The time evolution ensemble average of Poynting flux from all simulated cases with varying $|\lambda|$ (solid line of different colors) in comparison with the calculated $\overline{\langle L \rangle} $ from the scaling law of the Eq.~(\ref{eq:Lfit}) for the same value of $\lambda$ (dotted).}
    \label{Figure50}
\end{figure}

Empirically fitting the gross trend from our simulations $\langle L \rangle \approx~\tilde{L} \Omega^p$, we find an expected correlation between $\lambda$ and the overall normalization $\tilde{L}$ of the Poynting luminosity. Namely, $\tilde{L}/(10^{42} \, B_{12}^2 r_{12 \, \rm km}^2 \rm\, erg \, s^{-1})~\sim~0.8\exp{(3.99/|\lambda|^{0.34})}$\footnote{We assume that, for $|\lambda|\gg10^3$, $p$ and $\tilde{L}$ asymptomatically approach the ensemble average values derived for $|\lambda|=10^3$.} with considerable scatter above and below this relation of $\sim4$ depending on inclination angles and phase $\Delta \Phi$. Nevertheless, the trend is robust, with $\tilde{L}$ decreasing, reaching an asymptotic value for $|\lambda|\gg1$ corresponding to an unmagnetized companion. Second, we also notice the correlation in Figure \ref{p_vs_parameters} for the average index, $\overline{p}$ (averaged over all the cases with equal $|\lambda|$), with $|\lambda|$. We find $\overline{p} \sim (6\pm 0.2) - (3.6 \pm 0.1)|\lambda|^{-0.21 \pm 0.02}$. Indeed, $\overline{p}$ is also correlated with $\tilde{L}$; thus, $\overline{p}$ may be eliminated in favor of $\lambda$. 
Putting it all together yields a rough scaling law:
\begin{align}
 \overline{\langle L \rangle} & \sim (8\pm4)\times10^{41} \, B_{12}^2 r_{12 \, \rm km}^2 \,e^{(3.99\pm0.06) |\lambda|^{-0.34\pm0.02}} \notag \\[-2.5pt]
 & \times \left(\frac{\Omega}{\Omega_{
 \rm i}}\right)^{(6\pm0.2) - (3.6\pm0.1)|\lambda|^{-0.21\pm0.02}} \hspace{-6mm} \rm erg \, s^{-1}    
 \label{eq:Lfit}
\end{align}
where the $ \langle \cdot \cdot \cdot\rangle$ denotes an orbital average, implicitly encoding the pulsations in Figure~\ref{Figure12}, while $\overline{\cdot\cdot\cdot}$ denotes the average over various combinations of $\alpha_1$, $\alpha_2$, $\Delta \Phi$, and the sign of $\lambda$ (see the dotted lines in Figures~\ref{Figure12} and \ref{Figure50}). Thus, Eq.~\eqref{eq:Lfit} primarily captures the dependence of $\overline{\langle L \rangle}$ on $\Omega$ as function of $|\lambda|$.

In Figure~\ref{Figure50}, the relation Eq.~\ref{eq:Lfit} is compared (for different values of $|\lambda|$) with the Poynting flux derived from averaging over all the simulated cases with the same $|\lambda|$. 

\begin{figure}
\centering
    \includegraphics[width=\columnwidth]{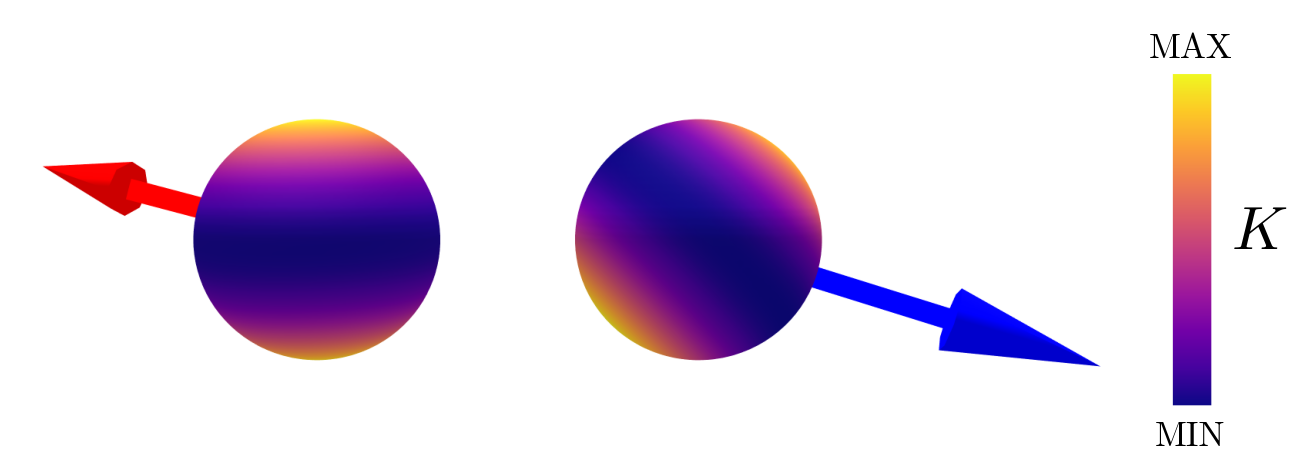}\hfill
    \caption{ A snapshot of the forces that develop at each star for the case with $\alpha_1=0^\circ, \, \alpha_2=45^\circ$, and $\lambda=1$ (\AngledUpArrow{0}\AngledUpArrow{-45}). The color scale indicates the quantity $K$ (see section~\ref{cc}).}
    \label{Figure18}
\end{figure}
Applying the power-law fitting procedure only to the peaks of the total flux reduces the spread in the $p$ distribution, yet it still yields a comparable range of values (see red histograms in Figure~\ref{Figure16}).
\begin{figure}
\centering
    
    \includegraphics[width=\columnwidth]{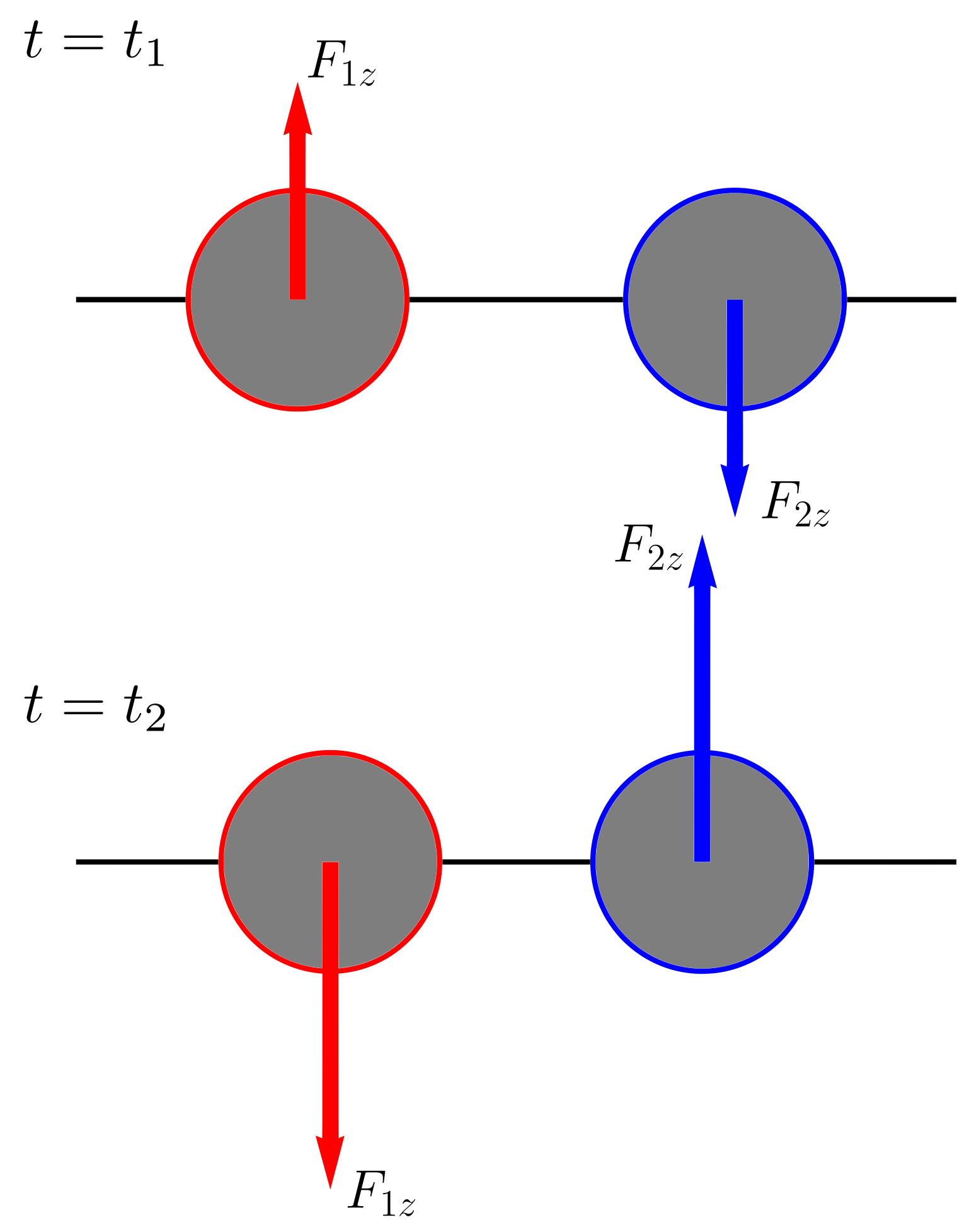}\hfill
    \caption{Two schematic snapshots of the BNS system on the corotating frame during its evolution, illustrating the forces exerted along the $z$-axis on each star. The plotted forces indicate the development of an EM torque. As shown, the magnitude of the forces increases between the two snapshots while their direction reverses, demonstrating that the torque amplitude grows over time and changes sign every half an orbital period. This periodic behavior suggests a cumulative effect that could influence the orbital motion of the system.}
    \label{Figure18pro}
\end{figure}

Additionally, for a few cases, we have verified the scaling laws of Poynting flux on $\Omega$ by calculating the magnetic flux, $\Phi_{\text{open}}$, along open magnetic field lines. This is done by identifying the polar caps on each stellar surface and then calculating the magnetic flux along the open field lines for each star $\Phi_{\text{open},1}$ and $\Phi_{\text{open},2}$. The total open flux is then given by $\Phi_{\text{open}}=\Phi_{\text{open},1} + \Phi_{\text{open},2}$. Our results confirm that the Poynting flux scales as $\Phi_{\text{open}}^2\Omega^2$, and thus, the open magnetic flux scales as $\Phi_{\text{open}} \propto \Omega^{(p-2)/2}$.

\subsection{Forces and Torques} 
\label{sec:torques}

In Section \ref{sec:misaligned}, we noted that, when the magnetic moment of one of the two stars is not parallel to the orbital axis, the EM energy outflow becomes asymmetric between the north and the south orbital hemispheres. 
This behavior implies the existence of backreaction forces on the stars, generating torques that could eventually have an impact on the system's orbital dynamics as well as individual NS spins. 

To investigate this, we calculate the force density of the magnetosphere using Maxwell's stress tensor, $\mathbf{\sigma}^M$, within a thin shell surrounding each star. The tensor is defined as
\begin{equation}
    \sigma^M_{i,j}=\frac{1}{4\pi}\left[E_iE_j+B_iB_j-\frac{1}{2}(E^2+B^2)\delta_{ij}\right]
    \label{eq:Maxwellstresstensor}
\end{equation}
and the force density\footnote{\cite{2021MNRAS.501.4479P} used the Laplace force to calculate the torque for an NS with an off-centered dipole magnetic moment. However, that approach cannot be used here because flux evolves over time \citep{2014PhyU...57..799B}. Therefore, in Eq. \ref{force_density}, the term involving $\partial\textbf{L}_{\rm Poynting}/\partial t$ is required.} reads
\begin{equation}\label{force_density}
    \textbf{f}=\nabla\cdot\pmb{\sigma}^M-\frac{1}{c^2}\frac{\partial\textbf{L}_{\rm Poynting}}{\partial t}.
\end{equation}
By integrating $\mathbf{f}$ over the shell's volume, the total force on each star can be calculated. While this approach might appear to depend on the choice of the shell’s volume, we have explicitly verified that the computed forces remain sufficiently stable regardless of the shell thickness, confirming that our results do not suffer from integration volume artifacts.

To further validate this approach, we applied the method to single, isolated pulsar cases. For a centered magnetic dipole moment, the net force was found to be zero, as expected. Moreover, the opposite forces exerted on the stellar surface produced a torque that led to a spin-down consistent with the corresponding EM spin-down power. For the case of an off-centered dipole moment in vacuum, where the so-called rocket effect \citep{1975ApJ...201..447H,2021MNRAS.501.4479P,2023MNRAS.522.5879A} induces a net force,  our calculations successfully reproduced the expected results, confirming the accuracy of our implementation.

In the BNS system, we find the EM forces can be several orders of magnitude smaller than gravitational forces, consistent with expectations of vacuum and magnetostatic calculations \citep[e.g.,][]{2000ApJ...537..327I,2022GReGr..54..146L}. Even for magnetar-like stars with strong magnetic fields ($\sim10^{15}-10^{16}$ G), the EM forces still remain at least 2 to 3 orders of magnitude smaller than gravitational forces, being 2 orders of magnitude smaller only during the final moments before the merger. Yet, such ${\cal O}(1\%)$ effects are possibly discernible in third-generation GW detectors for loud mergers with waveform signal-to-noise exceeding $10^3$ \citep{2024PhRvD.110h3040B}.
Interestingly, in cases where the Poynting flux is asymmetric between the two hemispheres, the forces develop a component along the $z$-axis, perpendicular to the orbital plane (see Figure~\ref{Figure18}). These forces generate a torque along the $y$-axis in the corotating frame. This torque not only alternates in sign but also increases in magnitude over time, gradually tilting the orbital plane. Figure~\ref{Figure18pro} schematically illustrates the increasing force components at their peak amplitudes, which drive the torque at two snapshots where the torque has an opposite sign. As a result, the stars exhibit oscillatory motion relative to an orbital plane that gradually tilts over time.

\begin{figure*}
    \includegraphics[width=0.5\textwidth]{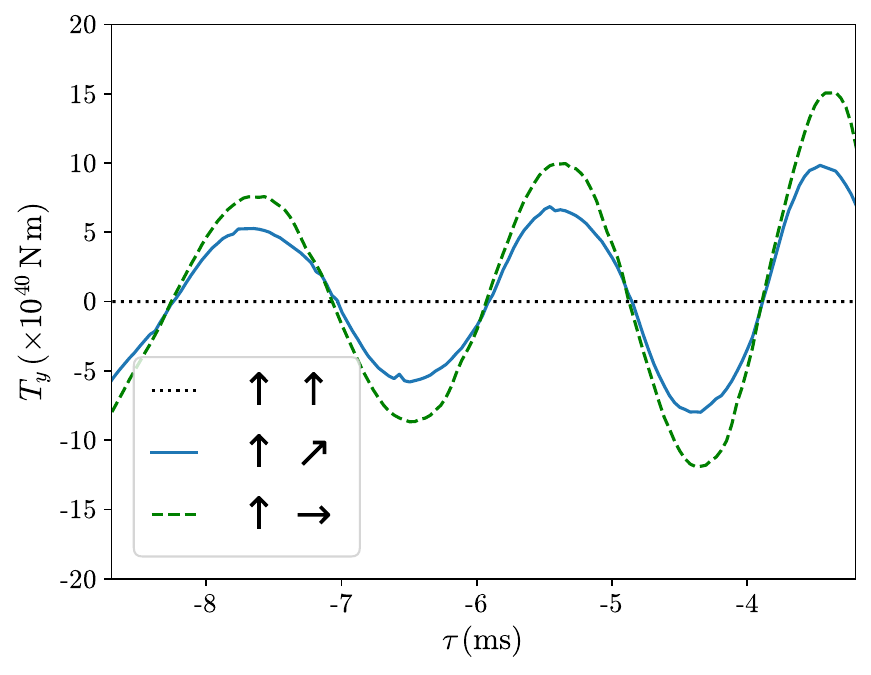}\hfill
    \includegraphics[width=0.5\textwidth]{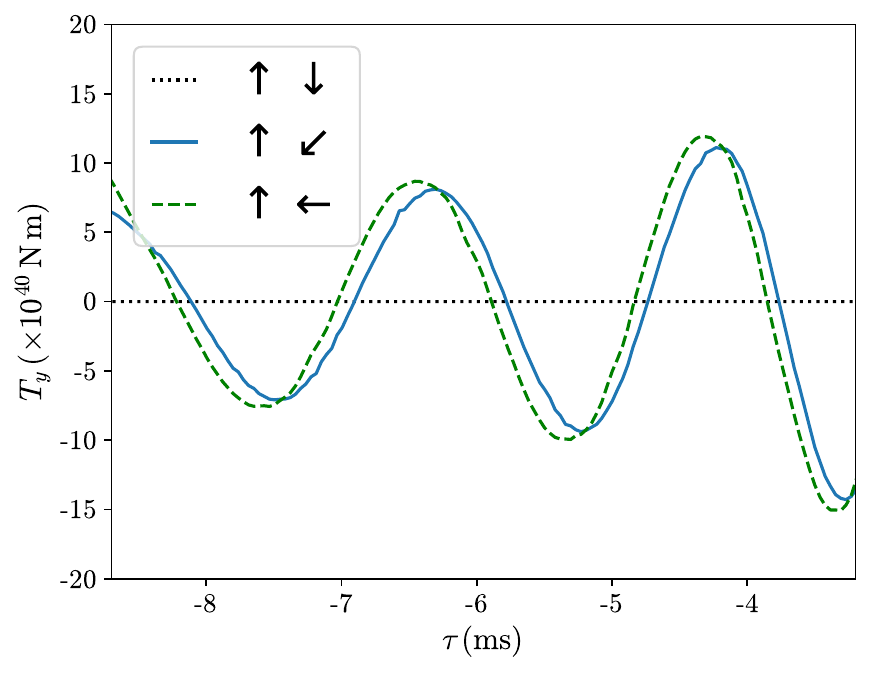}\hfill
    \includegraphics[width=0.5\textwidth]{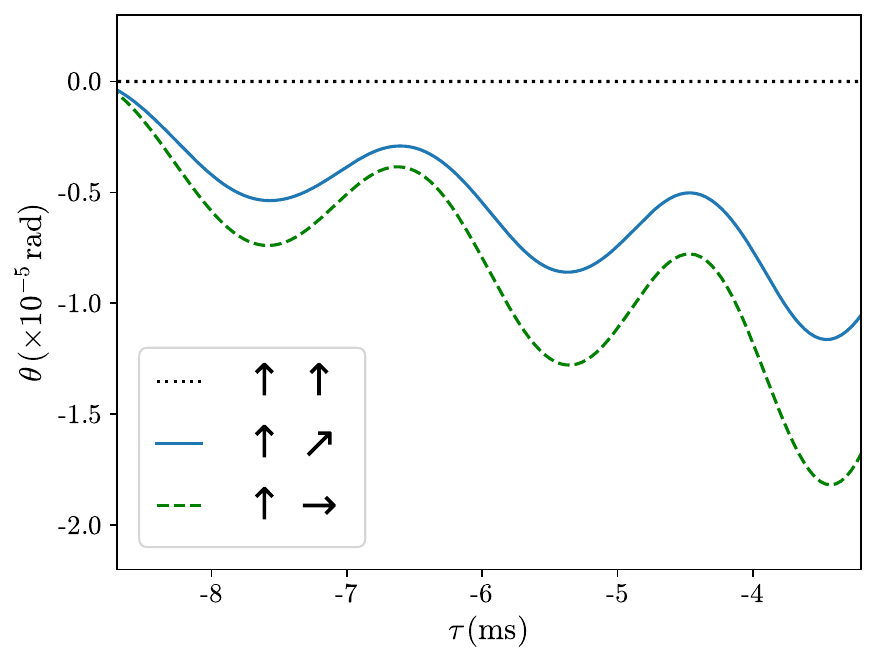}\hfill
    \includegraphics[width=0.5\textwidth]{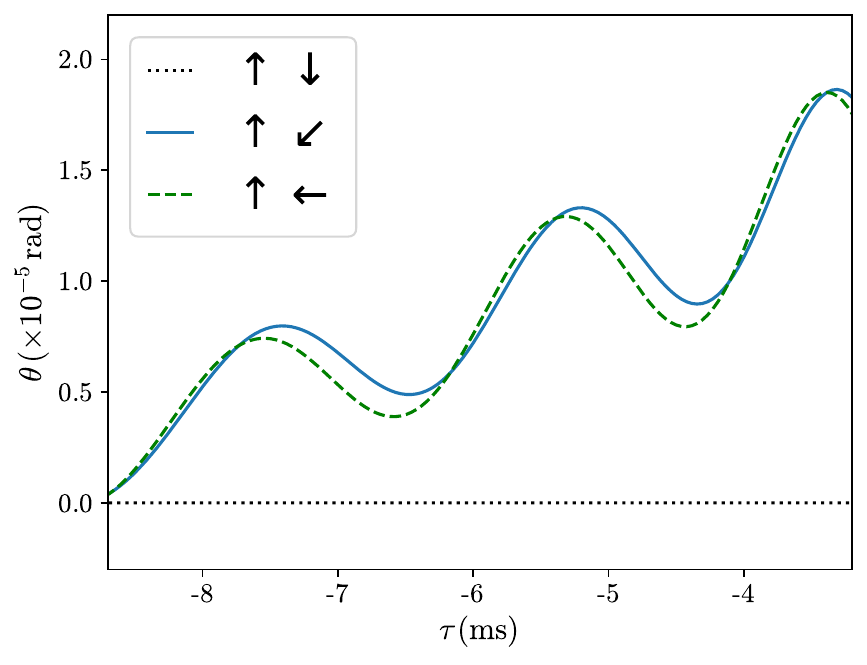}\hfill
    \caption{ \textbf{Top row:} evolution of the torque in the $y$-direction of the corotating frame over a short period before merger for positive (\textbf{left}) and negative (\textbf{right}) $\lambda$, with $\rm B_\star=3.16\times10^{15}G$. The cases correspond to $\alpha_1=0^\circ$ with $\alpha_2=0^\circ$ (dotted line), $\alpha_2=45^\circ$ (solid line), and $\alpha_2=90^\circ$ (dashed line). \textbf{Bottom row}: evolution of the angle $\theta$ of the orbital plane's deviation from its fixed position, resulting from the torque in $y$-direction, for negative (\textbf{left}) and positive (\textbf{right}) $\lambda$, following the same cases as in the top row.}
    \label{Figure19}
\end{figure*}

The evolution of the torque and its effects are shown in Figure \ref{Figure19}. The top row shows the torque along the $y$-axis for different magnetic inclination angles $\alpha_2$ ($0^\circ$, $45^\circ$, and $90^\circ$) and magnetic moment ratios $\lambda$ (both positive and negative values). The bottom row presents the corresponding evolution of the orbital plane deviation angle, $\theta$, induced by this torque. For stars with magnetic fields of $\sim 3 \times 10^{15}$~G, the deviation angle increases up to $\sim 10^{-5}$~rad during the final $7.7$~ms of the system's evolution. While this angle appears small, it demonstrates a clear trend of continuous growth over time. Considering that the total inspiral duration significantly exceeds $7.7$~ms, the cumulative effect of this torque from earlier times could potentially leave an observable imprint on the corresponding GW signal. However, it is crucial to highlight that these calculations were conducted in postprocessing and did not account for the dynamical feedback of the magnetosphere on the orbital motion of the stars. For a more accurate and comprehensive assessment, the evolution should be treated self-consistently by dynamically incorporating this feedback into simulations, updating the stars’ positions at each time step. Investigating the full significance of this EM backreaction physics is reserved for future work.

\subsubsection{Electromagnetic Crustal Deformations and Stresses}
\label{cc}

During the rapid evolution of the field in the magnetosphere, the forces and torques will result in nonuniform stresses on the NS crusts. This, if sufficiently large, could provoke a failure of the crustal Coulomb lattice. In order to determine whether the nonelastic response of the crust exceeds its yield limit, we follow the procedure described in \citet{Lander2015}. 

Assuming that the hydrostatic equilibrium state remains largely unchanged over the evolution time, the strain, $\kappa$, at a surface just outside each star is computed solely from Maxwell’s stress tensor (Eq.~\ref{eq:Maxwellstresstensor}). It is defined as $\mathbf{\kappa}=(\mathbf{\sigma^M_0-\sigma^M})/\mu$, where $\mu \sim 10^{30}$ dyn cm$^{-2}$ is the shear modulus \citep{Ruderman1969,1991ApJ...375..679S}, and $\sigma_0^M$ represents the Maxwell stress tensor at an initial relaxed equilibrium state. 

Considering a maximum breaking strain $\kappa_{\rm \max}\sim 0.1$ \citep{2009PhRvL.102s1102H}, the crustal lattice will yield (``crack") if the von Mises criterion $K=\sqrt{\kappa_{ij}\kappa^{ij}/2}\geq \kappa_{\rm \max}$ is satisfied. 

Using as an example the case with $\alpha_1=0^\circ,\alpha_2=45^{\circ}, \lambda=-1$, and $\Delta\Phi=0^\circ$ (\AngledUpArrow{0}\AngledDownArrow{-45}) at $\tau\approx -30$ ms,  we find that $K$ ranges from $\sim0.2$ to $\sim0.8 \times 10^{-6} B_{\rm 12}^2$ for the zero-inclination star while reaching $\sim1.2 \times 10^{-6} B_{\rm 12}^2$ in the case of the inclined star. These values are nonuniform over the surfaces of the stars and vary with time with the orbital motion. The distribution of $K$ on the star surface could be seen for the case with $\alpha_1=0, \alpha_2=45^\circ, \lambda=1$, and $\Delta\Phi=0$ (\AngledUpArrow{0}\AngledUpArrow{-45}) in Figure~\ref{Figure18}. The stresses on the crust thus vary on a timescale $\sim \pi/\Omega$. At late times in the inspiral, the motion of stresses may approach or exceed the typical shear wave propagation speed in the crust, possibly exciting crustal normal modes \citep{2007MNRAS.374..256S}.  

These values suggest that if one of the NSs has a magnetic field of $\sim 10^{15} \rm G$, the resulting EM stresses could induce significant deformations, potentially leading to catastrophic crustal failure. Such an event would likely produce an X-ray transient similar to a magnetar giant flare, while the excitation of crustal normal modes could give rise to additional observational signatures \citep[e.g.,][]{2024ApJ...973L..37M}.

\subsection{High-energy Emission}
\label{sec:HE_emission}

In this section, we focus on exploring the high-energy emission patterns generated in binary systems and identifying characteristic photon energies at which this emission might occur. By analyzing the emission anisotropy and time dependence of the emission, we aim to uncover key observational signatures and assess the potential detectability of these systems. This exploration builds on the framework established in previous observational and theoretical research on isolated pulsars, which demonstrated that gamma-ray emission primarily originates from particle acceleration near current sheets in the outer magnetosphere via synchro-curvature radiation. Notably, the GeV band represents the most efficient and prolific energy output channel for isolated pulsars \citep{3PC}, where the spectral energy distribution (SED), $\epsilon^2 dN/d\epsilon$, peaks. As a result, a significant fraction of the dissipated energy is expected to be emitted in gamma rays.

In particular, studies of global pulsar magnetospheres, including FF \citep{2010ApJ...715.1282B,2010MNRAS.404..767C}, dissipative \citep{Kalapotharakos2012,Kalapotharakos2014,Li2012,2019ApJ...874..166C}, and kinetic plasma models 
\citep{2014ApJ...795L..22C, 2016MNRAS.457.2401C,2018ApJ...858...81B,Kalapotharakos2018,2018ApJ...855...94P,Kalapotharakos2023,2024arXiv241202307C}, have established that the primary region for particle acceleration is near the equatorial current sheet beyond the light cylinder. In this region, particles are accelerated by accelerating electric field components $\textbf{E}_{\rm acc}$, producing high-energy photons.  

\citet{Kalapotharakos2019,Kalapotharakos2022,Kalapotharakos2023} demonstrated that the {\it Fermi}-LAT gamma-ray pulsars follow a ``fundamental plane" gamma-ray luminosity relation ($L_{\rm\gamma}=10^{14.3\pm1.3}\epsilon_{\rm cut}^{1.39\pm0.17}B_\star^{0.12\pm0.03}\dot{\cal{E}}^{0.39\pm0.05} {\rm erg\,s^{-1}}$, where $\epsilon_{\rm cut}$, the spectral cutoff energy, is measured in MeV, $B_\star$, the surface magnetic field, is measured in G, and $\dot{\cal{E}}$, the spin-down power, is measured in ${\rm erg\,s^{-1}}$). Remarkably, this correlation of observables aligns closely with the theoretical relations predicted for curvature radiation ($L_{\rm\gamma}\propto \epsilon_{\rm cut}^{4/3}B_\star^{1/6}\dot{\mathcal{E}}^{5/12}$). Thus, curvature radiation is adopted as the dominant emission mechanism in this work.

\subsubsection{Maximum Particle Energies}
\label{sec:specutoff}
A key feature of the curvature radiation spectrum of the pulsar gamma-ray emission is the characteristic photon energy, $\epsilon_c$, which marks the turnover point where the SED begins to decline. This energy reads
\begin{equation} \label{ecutoff}
    \epsilon_{\rm cut}=\frac{3}{2}c\hbar\frac{\gamma_L^3}{R_C},
\end{equation}
where $\gamma_L$ is the Lorentz factor of an electron/positron, and $R_C$ is the radius of curvature of the particle trajectory.

The Lorentz factor, $\gamma_L$, is determined by the balance between energy gain due to acceleration by the $E_{\rm acc}$ and energy loss from curvature radiation. This balance (in the ultrarelativistic limit) is governed by the equation
\begin{equation} \label{dgamma}
    \frac{d\gamma_L}{dt}=\frac{q_e E_{\rm acc}}{m_ec}-\frac{2q_e^2\gamma_L^4}{3R_C^2m_ec},
\end{equation}
where $m_e$ and $q_e$ are the electron rest mass and charge, respectively. 

The value of $\gamma_L$ at radiation-reaction limited acceleration is obtained by setting $\dot{\gamma_{\rm L}}=0$ in Eq.~\eqref{dgamma}, 
\begin{equation} 
    \gamma_L=\left(\frac{3R_C^2E_{\rm acc}}{2q_e}\right)^{1/4}.
    \label{eq:gammaL}
\end{equation}

For the double NS system, we now {\it conjecture} that particles have reached the curvature radiation-reaction limit. Furthermore, we consider that the radius of curvature, $R_C$, is the same order of magnitude as the radius of the orbital lengthscale $R_C \sim R_{\rm LC}$, and that $E_\parallel$ is assumed to be comparable to the magnetic field at the light cylinder, i.e., $E_{\rm acc} \sim B_{\rm LC}\approx B_\star(r_\star/R_{\rm LC})^3$. Then, $\gamma_L$ can be derived using Eq.~\eqref{eq:gammaL}. Substituting this value into Eq.~\eqref{ecutoff} yields the characteristic photon energy of the curvature radiation spectrum. 

\begin{figure*}
    \includegraphics[width=0.5\textwidth]{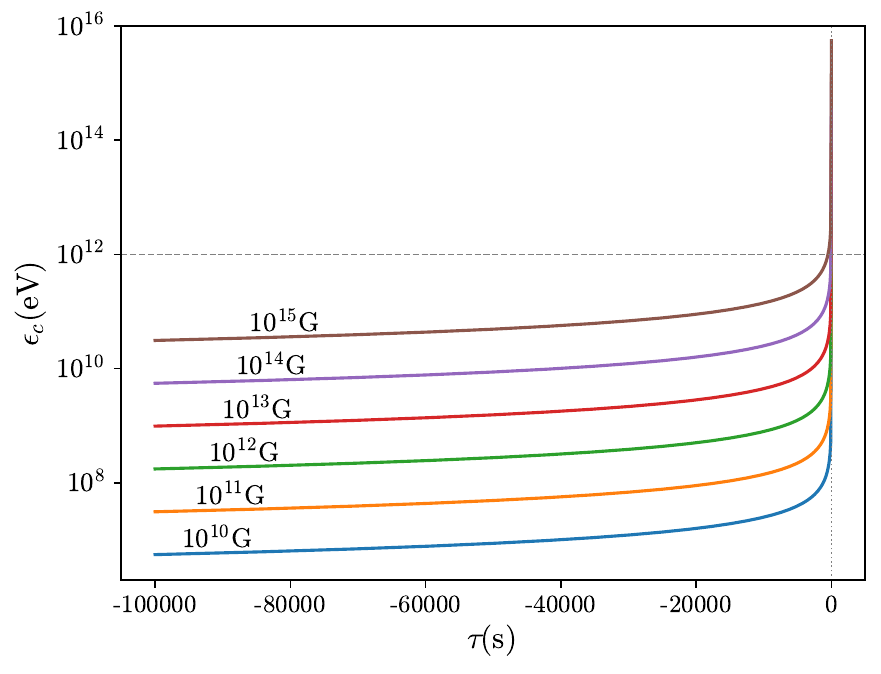}\hfill
    \includegraphics[width=0.5\textwidth]{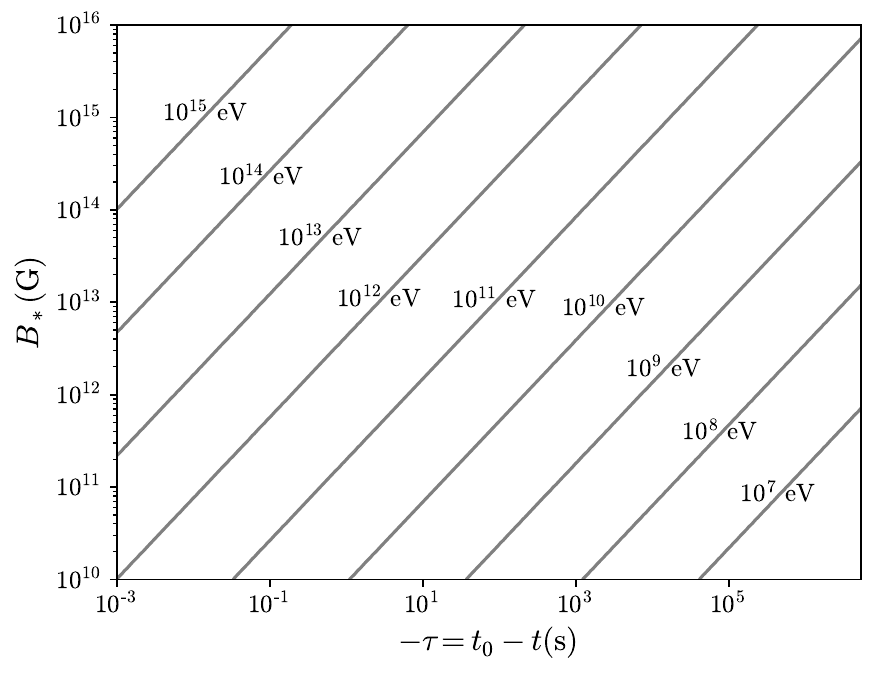}\hfill
    \caption{(\textbf{Left}) Evolution of the curvature radiation cutoff energy prior to BNS merger for different typical magnetic field strengths. (\textbf{Right}) A contour plot of the characteristic cutoff energies in the $B_{\star} - \tau$ log-log plane. Horizontal slices at fixed surface magnetic field values illustrate the evolution of cutoff energies throughout the BNS system's inspiral.}
    \label{Figure17}
\end{figure*}

Unlike single pulsars, where the light cylinder radius remains practically constant (changing on the spin-down timescale),  the binary system undergoes a continuous and fast decrease in orbital separation, leading to an increase in orbital frequency. Consequently, $R_{LC}$ shrinks over time, causing the cutoff energy to rise as $\epsilon_c \propto R_{\rm LC}^{-7/4}B_\star^{3/4}$ or equivalently $\epsilon_c \propto \Omega^{7/4}B_\star^{3/4}$. Considering the dependence of orbital frequency on time, i.e., Eq.~\ref{eq:omega}, the evolution of the cutoff energy is evident in the left panel of Figure~\ref{Figure17} for BNSs with magnetic field strengths of $10^{10}-10^{15}$~G. 

Remarkably, the photon energies exceed $\sim 10$~MeV as early as a day before the merger (albeit at low luminosity). This raises the possibility of coherent soft gamma-ray searches for such signals, which may be unattenuated hours or days before the merger (see discussion).

During the last few milliseconds before coalescence, the putative cutoff energy attains $\sim 10$~TeV. For systems with stronger magnetic fields, the implied cutoff energy can be even higher. As highlighted in the right panel of 
Fig.~\ref{Figure17}, even for stars with $10^{10}$~G, the cutoff energy might attain  $\sim1$~TeV just $1$~ms prior to the merger, while for stars with $10^{15}$~G it may exceed 1~PeV. However, this is not the full story as gamma-ray opacities to electron-positron pair production (single photon or two photons) rise drastically close to coalescence, either inhibiting the escape of photons of these energies or screening $E_{\rm acc}$ (see \S\ref{sec:opacity}).

\subsubsection{Gamma-ray Emission Anisotropy and Light Curves}
\label{sec:gammaraypatterns}

For isolated pulsars, skymaps of high-energy emission are constructed by assuming that the particle trajectories follow the asymptotic Aristotelian velocity flow, given by \citep{Gruzinov2012,Gruzinov2013,Kelner2015}
\begin{equation}\textbf{v}=\frac{\textbf{E}\times\textbf{B}\pm (B_0\textbf{B}+E_0\textbf{E})}{B^2+E_0^2},
\end{equation}
where the $\pm$ branches correspond to the two signs of particle charge. These trajectories determine the particle velocity, which in turn defines the photon emission direction, as well as the local $R_C$. By integrating along these trajectories, the energy gains and losses due to curvature radiation can be accounted for, allowing the emissivity at each point to be calculated. This approach works well in single pulsar systems because the polar cap rim, or ``separatrix", can be clearly identified as field lines connecting to the current sheet acceleration zone, and the solid-body rotation of the magnetosphere ensures temporal symmetry.

With the assumption of synchronized NSs, where the stars have the same orbital and spin frequencies, the magnetosphere structure corotates with the system, and thus, the above approaches from single pulsar studies are still applicable. For such a case, in \cite{Ortiz2022}, the authors considered a spherical conducting surface that encloses the system and identifying there the polar caps and the ``separatrix", and they construct high-energy skymaps and light curves following the prescription in \cite{2010ApJ...715.1282B}. 

In more generic binary systems, however, the situation becomes significantly more complex. The solid-body rotation assumption no longer holds due to the orbital motion and the continuous reconfiguration of the magnetic field structure caused by the motion of the irrotational stars. As a result, the field evolves dynamically on timescales comparable to the orbital period. Additionally, identifying the stellar surface regions where high-energy particles originate at each time step is nontrivial. In the isolated pulsar case, particles in the current sheet can reliably be traced to the polar cap rim, but in binary systems, determining these regions does not guarantee that the particles will reach the dissipative regions of the magnetosphere. A proper treatment would require integrating particle trajectories originating from all points on the stellar surfaces at every time step, which is computationally expensive given the rapid evolution of the magnetic field.

To overcome these challenges, we adopt an alternative approach that exploits the conjecture of the radiation-reaction limit regime. In this approach, the high-energy emission is assumed to occur in regions with strong accelerating electric field components, $E_{\rm acc}$, which are the primary drivers of particle acceleration and high-energy photon production.

\begin{figure*}
\centering
    \includegraphics[width=1.0\textwidth]{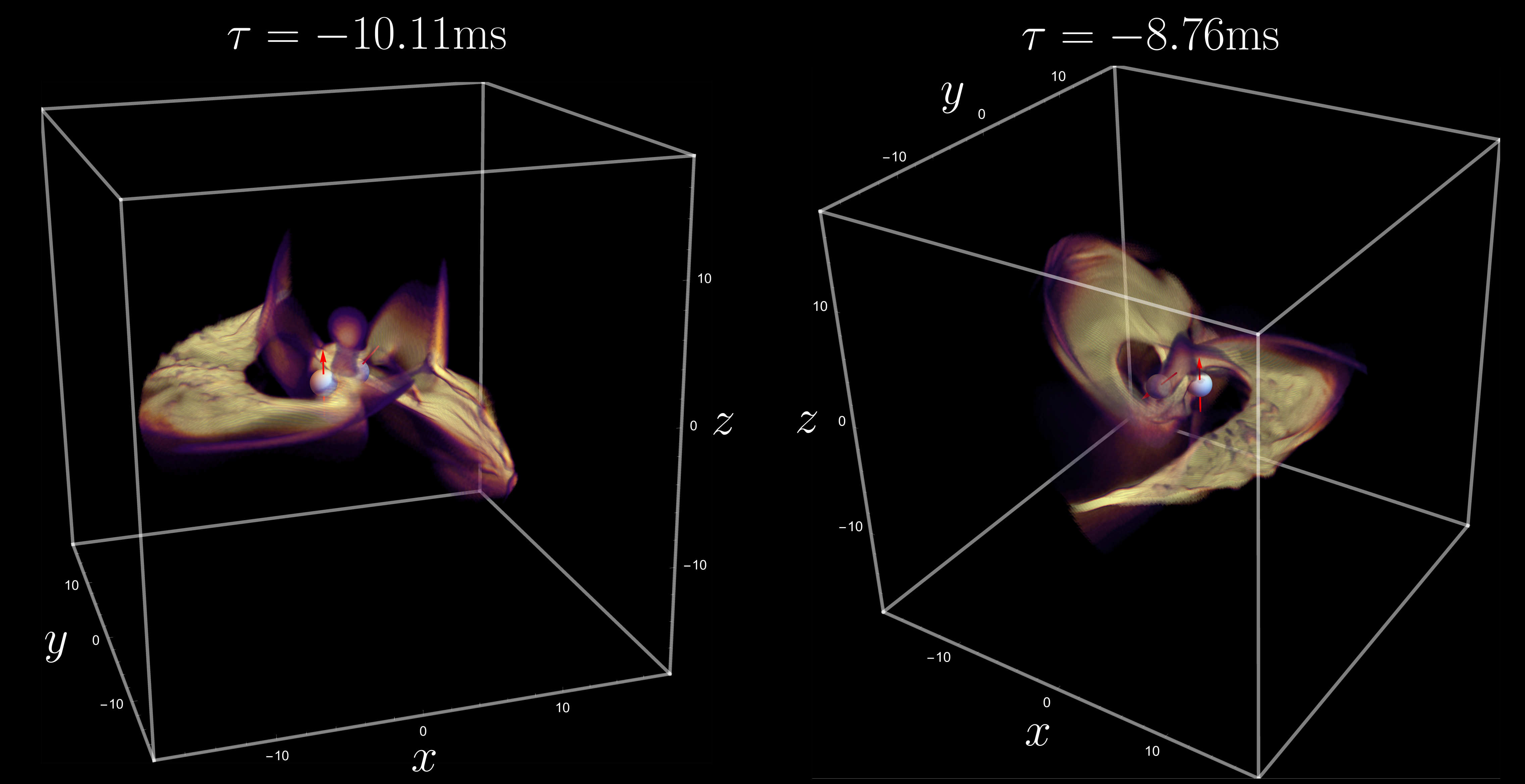}
    \caption{Regions of high-energy emission origin for the case with $\alpha_1=0^\circ,,\alpha_2=45^\circ,,\lambda=-1$, and $\Delta\Phi=0^\circ$ (\AngledUpArrow{0}\AngledDownArrow{-45}), similar to the middle row of Figure~\ref{Figure5}. The two snapshots, taken at $\tau=-10.11$~ms and $\tau=-8.76$~ms, are separated by half an orbital rotation. The color scale represents the strength of the parallel electric field, $E_{||}$.}
    \label{Figure22}
\end{figure*}

More specifically, we calculate $E_{\rm acc}$ in a dense grid of spherical shells ($dr_{\rm shell}=0.1r_\star$) spanning from $0.9 R_{\rm LC}$ to $2 R_{\rm LC}$. Since $R_{\rm LC}$ decreases as the orbital frequency increases, the volume of the shell evolves over time. In FF solutions, $E_{\rm acc}$ is negligible, so we follow the prescription of \cite{Kalapotharakos2014} to derive $E_{\rm acc}$, i.e., the parallel (to the magnetic field) electric field component. In this framework, $E_{\rm acc}$ is derived from
\begin{equation} \label{dEpar}
    \frac{d\textbf{E}_{acc}}{dt}=c\mathbf({\nabla}\times\textbf{B})_{||}-4\pi\textbf{J}_{||}
\end{equation}
and expressed as 
\begin{equation}
    \textbf{E}_{acc}=\frac{c\mathbf({\nabla}\times\textbf{B})_{||}}{4\pi\sigma}.
\end{equation}

where $\sigma$ is the plasma conductivity, assumed to be large. This effectively rescales the accelerating electric field. In our study, we adopt $\sigma = 1 \equiv \Omega_{\text{i}}$, a value chosen to ensure that the derived $E_{\rm acc}$ reflects regions where particle acceleration can realistically occur. In single pulsar studies, this method has been shown to produce results consistent with dissipative magnetosphere models that incorporate finite $\sigma$ values, particularly near the equatorial current sheet.

To avoid inadvertently selecting regions where $(\nabla \times \mathbf{B})_{\parallel}$ appears high simply because they are located in proximity to strong magnetic fields (e.g., near the stellar surfaces), we apply additional constraints to ensure the identified accelerating regions are physically meaningful. Specifically, we identify locations where $E{\rm_{acc}}$ exceeds 10\% (to avoid noise in the simulation) of the local electric field magnitude, $E$, as the regions where high-energy emission originates. Figure~\ref{Figure22} illustrates these regions at two-time snapshots, separated by half an orbital period, for $\alpha_1 = 0^\circ$, $\alpha_2 = 45^\circ$, $\Delta\Phi=0^\circ$, and $\lambda = -1$ (\AngledUpArrow{0}\AngledDownArrow{-45}). These regions are predominantly located near the current sheets, consistent with single pulsar expectations, and they evolve dynamically as the magnetosphere reconfigures during the system’s inspiral.

Once these emission regions are identified, we assume that particle trajectories at these points follow the asymptotic Aristotelian velocity flow, which depends only on the local electric and magnetic fields. The particle velocity determines the direction of photon emission. Additionally, assuming curvature radiation in the radiation-reaction limit, we find that the emissivity of a particle, i.e., $\propto \gamma_L^4/R_C^2$, is proportional to $E_{\rm acc}$ (see Eq.~\ref{dgamma}).

A major challenge in approaches involving test particle trajectories lies in determining how many physical particles are represented by each test particle. This uncertainty, which can vary from one test particle to another, persists even in single pulsar studies and can only be partially resolved with more self-consistent PIC simulations. To mitigate this issue for binary systems, we adopt a simpler prescription where the emitting points lie on spherical surfaces corresponding to a specific Poynting flux value. The relative weights of these emission points are determined by the strength of their local $E_{\rm acc}$ components. To include the high-energy efficiency of each surface, we scale the total emissivity relative to the Poynting flux power crossing that surface. Note here that, in order to preserve causality, we use the flux that crosses the spherical surface containing the point where emission occurs at the time of the emission.

\begin{figure*}
\centering
    \includegraphics[width=0.65\textwidth]{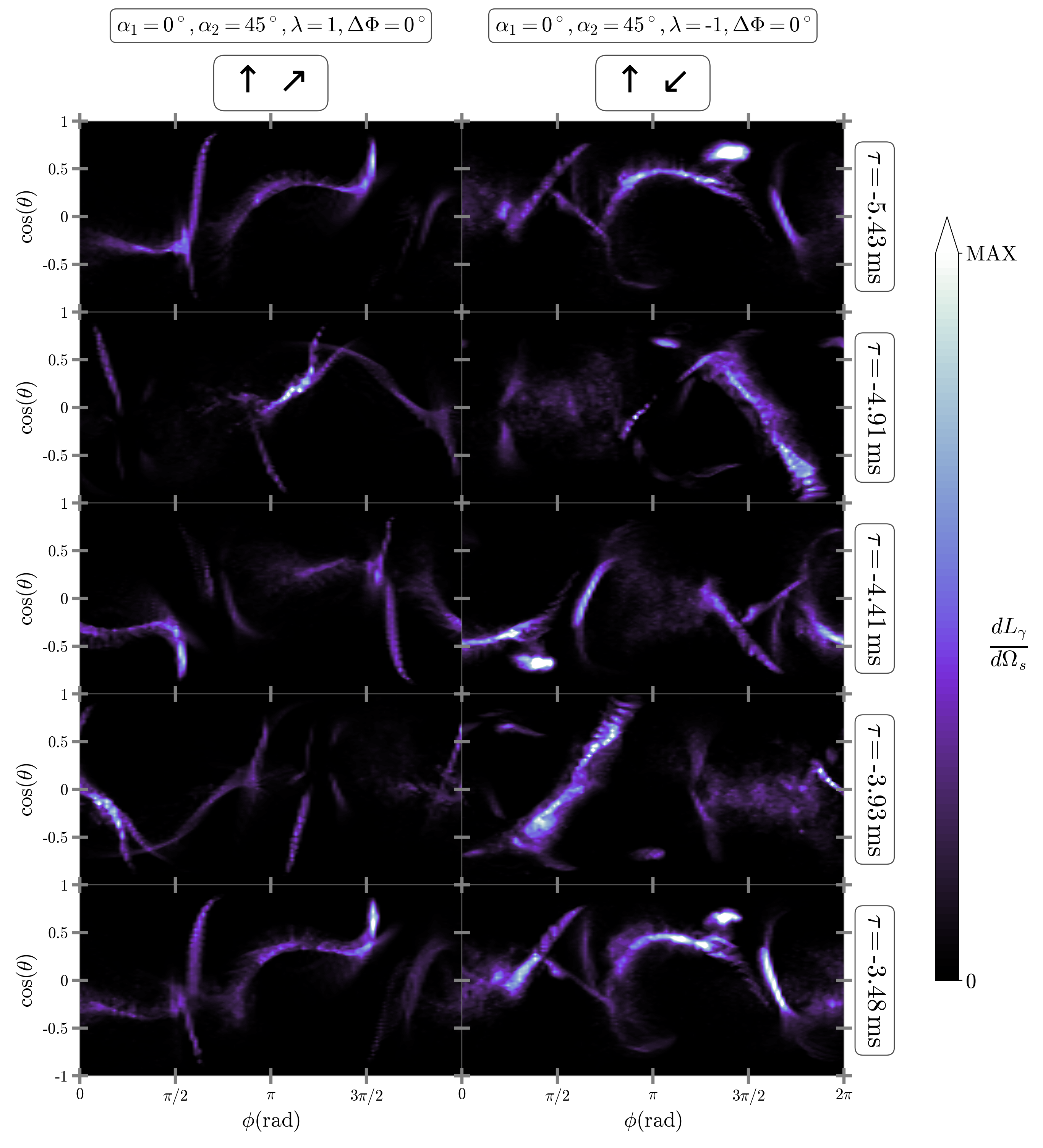}
    \caption{The evolution of high-energy emission skymaps for BNS systems with $\alpha_1=0$, $\alpha_2=45^o$, $\Delta\Phi=0$, and $\lambda=1$(\AngledUpArrow{0}\AngledUpArrow{-45}), \textbf{left}, and $\lambda = -1$ (\AngledUpArrow{0}\AngledDownArrow{-45}), \textbf{right}. Notably, the horizontal axis labeled $\phi$ does not represent phase, as is typically the case in skymaps of isolated pulsars. Instead, $\phi$ corresponds to the azimuthal angle measured from the $x$-axis, meaning each point in these diagrams represents a fixed absolute direction in the sky relative to the BNS system. Additionally, the values of $\tau$ indicate the relative time at which these skymaps will reach the observational sphere, referenced to a photon emitted from the center of the system at the moment of the merger.}
    \label{Figure20}
\end{figure*}

For the purposes of this study, we explore two approaches for determining the emissivity weight. First, we assume that the high-energy photon luminosity is proportional to the local Poynting flux, consistent with previous isolated pulsar studies where $L_{\gamma} \propto \dot{E}$. Alternatively, we adopt the empirical relation $L_{\gamma} \propto \sqrt{\dot{E}}$, derived from observations of gamma-ray pulsars \citep{3PC}. Both approaches yield similar emission pattern results, and for the skymaps presented in this work, we adopt the second prescription.

Algorithmically, at every time step, we divide the region between $0.9 R_{\rm LC}$ and $2 R_{\rm LC}$ into spherical shells centered on the system’s center. For each shell, we calculate the Poynting flux at its midradius surface and identify the emission points where $E_{\rm acc}/E$ exceeds the defined threshold. The emission from each sampled point is then weighted proportionally to $\sqrt{L_{\rm shell}} / \sum_{N_{\rm shell}} E_{\rm acc}$, where $L_{\rm shell}$ is the Poynting flux for the shell, and $N_{\rm shell}$ is the number of sampled emission points within the shell.

Unlike the isolated pulsar case, where the nearly constant spin period allows for phase folding and coherent photon contributions, the rapidly evolving orbital motion in the binary system prevents the existence of a well-defined or coherent phase \textbf{\citep{Ortiz2022}}. Instead, the relative detection time for each photon must be calculated at a distant surface.

\begin{figure*}
\centering
    \includegraphics[width=0.98\textwidth]{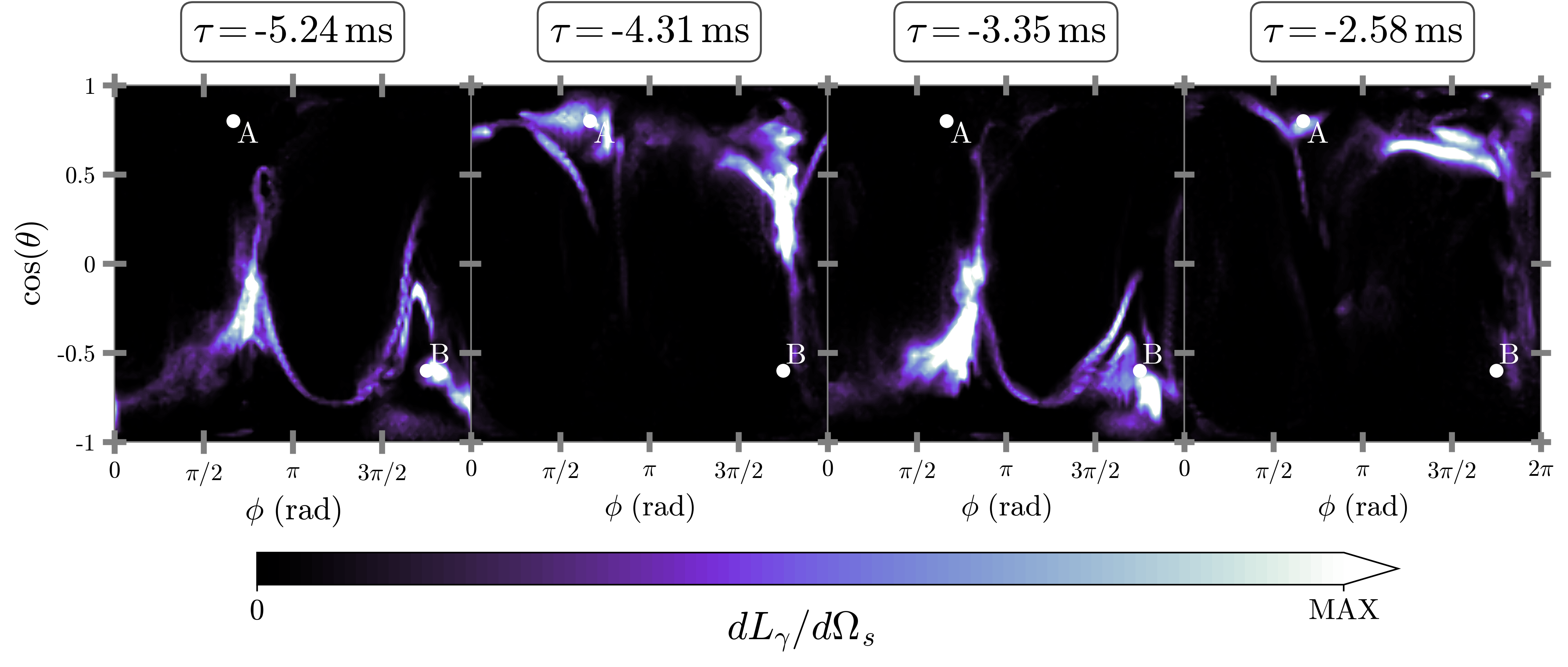}
    \includegraphics[width=0.49\textwidth]{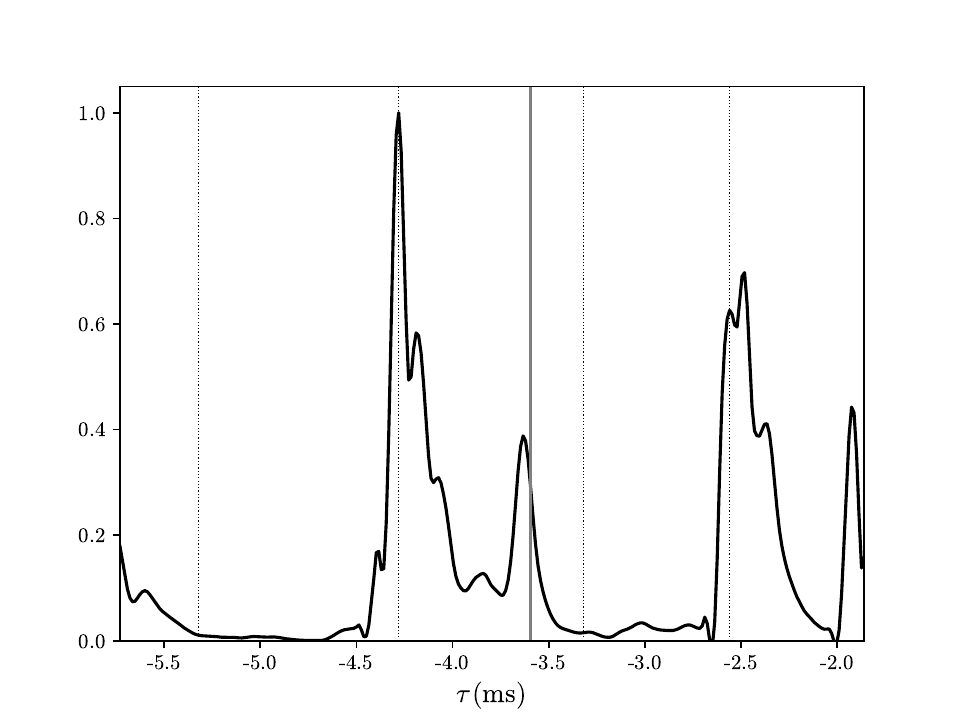}
    \includegraphics[width=0.49\textwidth]{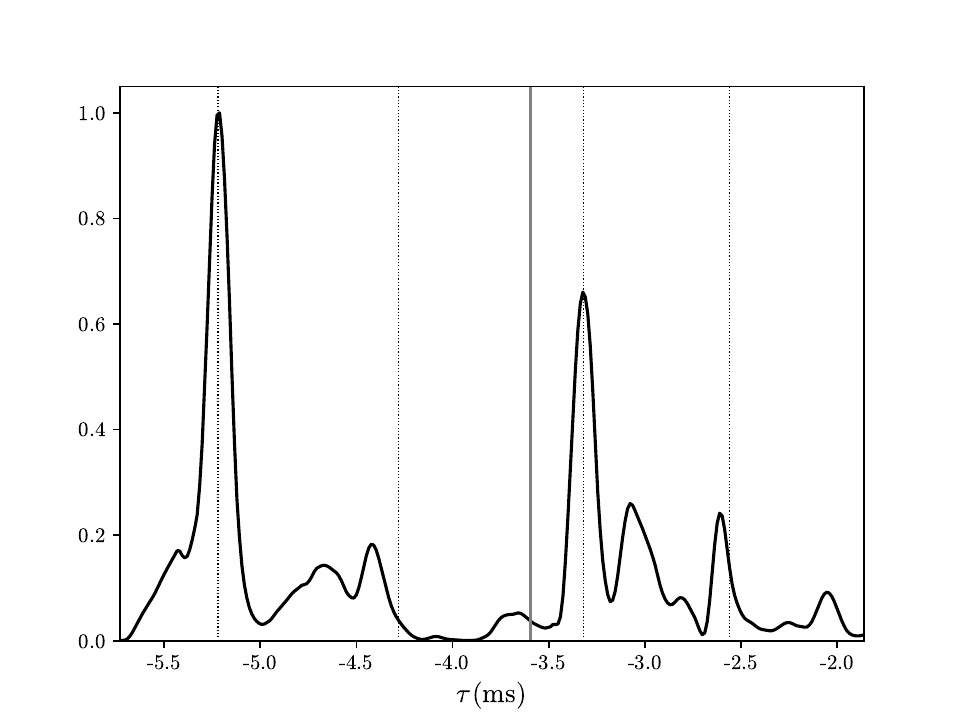}
    \caption{In the first row the high-energy emission skymaps for $\alpha_1=45^o$, $\alpha_2=90^o$, $\Delta\Phi=0$, and $\lambda=1$ (\AngledUpArrow{-45}\AngledUpArrow{-90}). The two white points, labeled A and B, correspond to two chosen directions for the construction of the respective light curves. In the second row, the normalized light curves from the direction A (\textbf{left}) and B (\textbf{right}) for two rotations of the stars. The dotted vertical lines correspond to the instances of the first row, while the gray line serves as a separation of the two rotations.}
    \label{Figure21}
\end{figure*}

To account for the time delays caused by photons traveling different distances to a distant observer, we compute the delay time as
\begin{equation}
    t_{\rm delay}=-\frac{\textbf{r}\cdot\boldsymbol{v}}{c|\boldsymbol{v} |},
\end{equation}
where $r$ is the position vector of the emission point, and $\boldsymbol{v}$ is the particle velocity. Since we assume equal masses, the center of mass coincides with the origin of the coordinate system. The relative detection time, measured with respect to a fiducial photon emitted from the system's center of mass at the beginning of the simulation, at which each photon contributes to the observed emission, is then given by
\begin{equation}
    t=t_{\rm emission}+t_{\rm delay}.
\end{equation}
This method ensures that the relative arrival times of photons are accurately accounted for, capturing the dynamically evolving nature of the binary system.

We validate the generalized approach detailed above by applying it to the isolated pulsar scenario, where it successfully reproduces skymaps consistent with previous studies using global magnetosphere simulations and PIC models \citep{Kalapotharakos2012,Kalapotharakos2014,Kalapotharakos2018,Kalapotharakos2023}.

\begin{figure*}
    \includegraphics[width=0.5\textwidth]{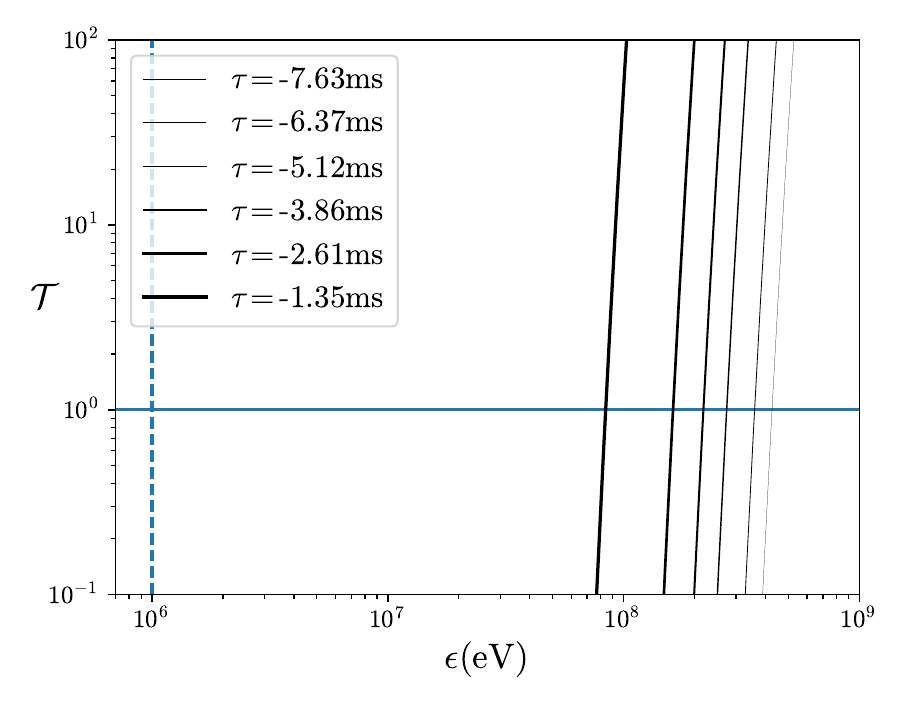}\hfill
    \includegraphics[width=0.5\textwidth]{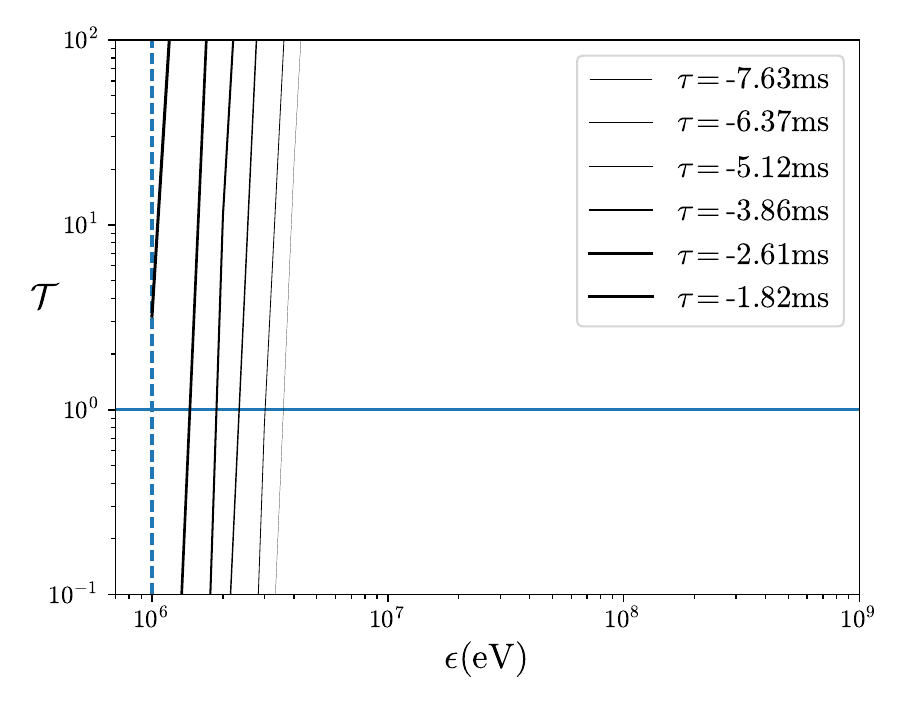}\hfill
    \caption{Optical depth for the case of $\alpha_1=0,\alpha_2=45^\circ, \lambda=-1$, and $\Delta\Phi=0$ (\AngledUpArrow{0}\AngledDownArrow{-45}) with $10^{12}$~G (\textbf{left}) and $10^{14}$~G (\textbf{right}) magnetic field stars. A wider line indicates a later stage in the evolution. The vertical dashed line highlights the minimum pair-production threshold of $1$~MeV while the blue horizontal line is the ${\cal T}=1$ line. Note that, in the right-hand panel, $\mathcal{T}$ may be overestimated near $\epsilon = 1$ MeV (see main text for discussion), potentially permitting the escape of marginally higher-energy photons.}
    \label{Figure23}
\end{figure*}

Figure~\ref{Figure20} presents the evolution of high-energy emission skymaps for BNS systems with $\alpha_1=0$, $\alpha_2=45^\circ$, and $\Delta\Phi=0$, comparing cases with $\lambda=1$ (\AngledUpArrow{0}\AngledUpArrow{-45}, left) and $\lambda=-1$ (\AngledUpArrow{0}\AngledDownArrow{-45}, right). These skymaps illustrate how the directional intensity and beaming patterns of high-energy emission evolve over time, revealing differences in anisotropy and variability between the two configurations. The time values $\tau$ do not directly map to specific stages of the inspiral but rather indicate when the emission reaches the observational sphere relative to a photon emitted from the system's center at the merger. As expected, the emission strength increases as the merger time approaches, following the overall behavior of the Poynting flux. However, unlike single pulsars, where emission follows a steady-state and well-defined structure, the BNS system exhibits far more dynamic and intricate patterns. This complexity arises from the continuous interaction and reconfiguration of current sheets as the two stars orbit each other, leading to highly variable and evolving emission profiles.   

In those figures, the vertical and horizontal axes denote the polar $\theta$ and azimuthal $\phi$ angles on a distant sphere around the system. The color indicates the high-energy flux density that is beamed toward the direction defined by the angles $\theta$ and $\phi$ at the indicated time. 

Finally, Figure~\ref{Figure21} presents the high-energy emission skymaps (top row) for a BNS system with $\alpha_1=45^\circ$, $\alpha_2=90^\circ$, $\Delta\Phi=0$, and $\lambda=1$ (\AngledUpArrow{-45}\AngledUpArrow{-90}). The skymaps illustrate the emission distribution at the indicated times, with two selected observer directions labeled A and B, marked by white points. The corresponding light curves (bottom row) depict the temporal variation of high-energy emission for these two observer directions over two full orbital rotations. The dotted vertical lines indicate the times corresponding to the skymap times, while the gray line separates the two orbital rotations. The rapid evolution of the magnetosphere and the orbital motion of the stars result in sharp and narrow peaks in the light curves, some $\lesssim 0.5$~ms duration. These features are direct signatures of the dynamic magnetospheric structure in the final stages of BNS inspiral.

\subsubsection{Pair Production and Gamma-ray Attenuation}
\label{sec:opacity}

As mentioned above, the emitted high-energy photons are not guaranteed to escape the magnetosphere. In regions of strong magnetic fields, these photons may interact with the field and produce electron-positron pairs, a process that can occur without violating the energy and momentum conservation. 

This magnetic pair production is particularly significant in environments where the magnetic field strength approaches or exceeds the critical value, $B_c = 4.414 \times 10^{13}$~G, at which the cyclotron energy equals the rest mass energy of an electron. 
A photon with the energy $\epsilon$, moving at an angle $\theta_{\rm kB}$ relative to the local magnetic field $\textbf{B}$, can produce a pair when the condition $\epsilon^2\sin^2\theta_{\rm kB}/\bar{B}\gg1$ is satisfied for pairs produced in high Landau states above the threshold $2 m_e c^2/\sin\theta_{\rm kB}$   \citep{1966RvMP...38..626E,1983ApJ...273..761D} where $\bar{B}=B/B_{c}$.

In those cases, the attenuation coefficient, ${\cal R}$, for pair production can be calculated using the expression \citep{1966RvMP...38..626E,1974PhRvD..10..492T,2006RPPh...69.2631H}
\begin{equation}
    {\cal R}=\frac{\alpha}{2\lambdabar}\bar{B}\sin\theta_{\rm kB} A(\chi)
    \label{eq:R}
\end{equation}
where
\begin{equation}
    A(\chi)\approx 4.74\chi^{-1/3}\rm{Ai}^2(\chi^{-2/3}),
    \label{eq:pairT}
\end{equation}
where Ai is the Airy function, $\chi= \epsilon\sin^2\theta_{\rm kB}\bar{B}/2$, $\alpha$ is the fine structure constant, and $\lambdabar$ is the reduced electron Compton wavelength. The expression in Eq.~\ref{eq:pairT} remains accurate for fields $\bar{B} \lesssim 0.1$; however, for higher values, incipient pair states introduce a pronounced sawtooth structure and a strong dependence on photon polarization \citep{1983ApJ...273..761D}. In particular, in the regime $\bar{B} \gtrsim 0.1$, Eq.~\ref{eq:pairT} tends to significantly overestimate attenuation near the pair-production threshold \citep{1988MNRAS.235...51B,1991A&A...249..581B,2007PhRvD..75g3009B}. In these cases, Equation ~\ref{eq:R} will not be valid, and corrections or approximate expressions are required \citep{1983ApJ...273..761D,1991A&A...249..581B}.

As a result, photons with sufficiently high energy propagating in high-strength magnetic fields will not survive. In order to determine the highest energy a photon can retain while escaping the magnetosphere, we compute the optical depth,
\begin{equation}
    {\cal T}(l)=\int_0^l {\cal R} \, ds,
    \label{eq:tauopacity}
\end{equation}
where $l$ is the path length along the photon trajectory.

A photon with the energy $\epsilon$ will escape the magnetosphere if its optical depth satisfies ${\cal T}(\epsilon,l) \ll 1$. A similar calculation as Eq.~\ref{eq:tauopacity} has previously been considered for magnetospheric opacity of gamma-rays for isolated NSs; see, e.g., \cite{1997ApJ...476..246H},\cite{2001ApJ...547..929B},\cite{2014ApJ...790...61S},\cite{2019MNRAS.486.3327H},\cite{2022ApJ...940...91H} and references therein. 

The photon escape energy can be proximity estimated by assuming that the mean path length is of the order of the lengthscale $c/\Omega$. Additionally, we consider average values for the magnetic field and the angle $\theta_{\rm kB}$ at the high-energy emission regions every time step. The attenuation coefficient was calculated from Equations~(\ref{eq:R})--(\ref{eq:pairT}), and then, Equation(\ref{eq:tauopacity}) was calculated via an integration for $l=c/\Omega$ for different $\epsilon$. The cutoff energy is considered the upper limit for $\epsilon$. Note that $\chi$, the attenuation coefficient $\cal{R}$, and consequently the optical depth and the escape energy vary over time as the orbital separation decreases due to the orbital inspiral. This evolution causes the outer magnetosphere to become increasingly opaque. 

This behavior is evident in Figure~\ref{Figure23}, which shows the optical depth as a function of the photon energy for the case of $\alpha_1=0,\alpha_2=45, \lambda=1$, and $\Delta\Phi=0$, assuming stars with magnetic fields of $10^{12}$~G, left, and $10^{14}$~G, right. The line thickness represents the progression of time. The escape energy, defined as the energy where each line crosses the horizontal ${\cal T}=1$ threshold, decreases over time. For strong magnetic fields, the escaping energy decreases below the 1~MeV threshold (indicated by the vertical dashed line) required for pair production.

This analysis confirms that magnetic pair production imposes a stringent upper limit on escaping photon energies, effectively confining detectable high-energy emission to the MeV range in the final stages of inspiral, an important consideration for the observability of precursor signals.

\section{Discussion and Conclusions}  
\label{sec:summary}

In this work, we have employed numerical FF MHD simulations to investigate the evolving structure and interactions of BNS magnetospheres, tracking their temporal evolution, scaling laws, and outgoing anisotropy emission as the orbital separation decreases toward the merger.  Unlike studies that assume a fixed orbital separation, our approach follows a representative inspiral trajectory, capturing the dynamic changes in magnetospheric interactions. 

The FF framework assumes ample plasma supply to screen accelerating electric field components while neglecting particle inertia, making it a well-suited approximation for studying large-scale magnetospheric EM interactions in the premerger phase. Our findings reveal strong Poynting flux emission anisotropies, which in turn induce forces on the two NSs, possibly affecting their orbital dynamics.

We have systematically explored the parameter space of magnetic field configurations of two irrotational stars, varying both the magnitudes and orientations of their magnetic moments, as well as their relative phase offsets. Additionally, we present the first systematic study of high-energy emission in this context, investigating its anisotropy, evolution, and characteristic photon energies. This detailed survey uncovers a rich phenomenology with significant physical consequences, many of which are explored here for the first time.

\subsection{Global Magnetospheric Structure and Poynting Flux Evolution}

Our results demonstrate that the interaction of the two stellar magnetospheres leads to a complex and dynamically evolving magnetic topology governed by the relative orientations and magnitudes of the magnetic moments. The overall structure of the magnetospheres and the distributions of the Poynting flux presented here exhibit similarities with other works presented in the literature \citep{Palenzueala2013b,2014PhRvD..90d4007P,Most&Philippov2022,Mahlmann&Beloborodov}. 

The topology of the magnetosphere varies significantly with the alignment of the magnetic moments. In the aligned case (\AngledUpArrow{0}\AngledUpArrow{0}), field lines primarily bend toroidally, preventing direct connections between the two stars.
In the antialigned case (\AngledUpArrow{0}\AngledDownArrow{0}), field lines link the two stars, forming flux tubes that evolve dynamically as the orbit tightens.
In both cases, the structure evolves with the time as expected from \cite{Palenzueala2013b} where the full Einstein-Maxwell-hydrodynamic equations are solved. Following the decrease of the separation between the two objects, the global structure shrinks as the merger time approaches. 

For oblique configurations, where at least one of the stars has a magnetic moment inclined with respect to the orbital axis, the magnetosphere undergoes rapid reconfigurations. Field lines continuously attach and detach from the companion, affecting the balance between open, shared, and closed flux regions. Shared flux tubes can exhibit helical structures in some cases (e.g., Fig.~\ref{Figure5}). This significantly impacts polar cap structure and outgoing EM flux, introducing periodic modulations in energy release. 

For the cases with $\lambda<0$ and $\alpha_1=0^\circ$, the magnetospheric structure exhibits flux tubes connecting the two stars, with topology determined by the orientation of their magnetic moments, as presented in the simulations of \cite{Most&Philippov2022} where the stars are held at a constant separation. However, in Figures 1 and 3, presented in \cite{Most&Philippov2022}, the magnetic field lines become increasingly twisted, forming detached closed loops leading to a solar flare-like eruption. These features were postulated to potentially result in FRBs. In Figures \ref{Figure4}, \ref{Figure8} and \ref{Figure13}, for our cases whose configurations most closely match, we do not observe this behavior on the $x-z$ plane \citep[similar to][]{Palenzueala2013b}. However, we do see eruptive behavior off the $x-z$ plane, as exhibited by time-dependent Poynting flux patterns in misaligned cases.

The total Poynting flux exhibits strong anisotropies and varies significantly with the orbital phase. Compared to the aligned case, the antialigned configuration systematically produces higher flux, while cases with misaligned moments lead to periodic oscillations, with luminosity modulating at half the orbital period \citep{2014PhRvD..90d4007P,Most&Philippov2022,Mahlmann&Beloborodov}. These variations arise from the continuous restructuring of the magnetosphere and the evolution of polar caps along the inspiral. The luminosity reaches peak values of up to a few $10^{44} \, B_{12}^2 \, r_{12\rm km}^2$~erg s$^{-1}$. Notably, in the presence of a highly magnetized NS with $B \gtrsim 10^{15}$ G, Poynting luminosities can exceed $10^{50}$ erg s$^{-1}$. Given the strong emission anisotropy, certain sight lines will experience isotropic-equivalent luminosities of $L_{\rm iso} \lesssim 10^{52} B_{15.5}^2/f_{\Omega, -1}$ for a beaming factor of $f_{\Omega, -1} \sim 0.1$ (e.g., see Figs.~\ref{Figure7},~\ref{Figure10},~\ref{Figure12},~\ref{Figure33}, and \ref{Figure15}) and have been conjectured as the source of short GRBs \citep{1992Natur.357..472U}. These results suggest that previous pessimistic evaluations regarding the detectability of premerger EM signals may need to be reconsidered.

Although directly comparing the Poynting flux values presented here with those reported in other studies in the literature is difficult, as different NS parameters and assumptions were adopted,  approximate scaling can be used to assess relative values. Furthermore, when comparing with simulations that assume constant separation, i.e., corresponding to earlier inspiral stages, one must account for the specific power-law dependence of the Poynting flux on orbital frequency (see subsection \ref{sec:fit}). Additionally, the simulations in other studies adopt different mass-to-radius ratios, leading to distinct orbital dynamics. As a result, discrepancies in the reported Poynting flux magnitude naturally arise. For example, when comparing the Poynting fluxes at the same time before merger, our results are higher by a factor of 2-3 for the antialigned case and 3.5-10 for the aligned case relative to the flux evolution reported in \cite{Palenzueala2013b}, consistent with our higher initial orbital frequency. Conversely, our fluxes are lower by a factor 1.8-2.5 compared to the aligned cases in \citep{Mahlmann&Beloborodov} as our simulations correspond to lower orbital frequency at their separation. Finally, relative to the antialigned cases presented in Figure~10 of \cite{Most&Philippov2022}, our fluxes are lower by roughly a factor of 10. We note, however, that the fluxes in that figure have been explicitly rescaled for visualization (by factors of 1/3 and 1/6), and it remains unclear whether further normalization or intercase scaling was applied, potentially contributing to the apparent discrepancy.

The degree of NS magnetization at the merger is uncertain, as it depends on evolutionary factors such as field decay, accretion history, and the merger timescale. Accretion in recycled MSP is known to suppress surface magnetic fields \citep[e.g.,][]{1974SvA....18..217B,1990Natur.347..741R,1997MNRAS.284..311K,2004MNRAS.351..569P}, and similar but weaker effects may occur in double NS systems, where limited accretion could reduce the field strength \citep{2017ApJ...846..170T,2023pbse.book.....T}. However, recent studies suggest that some magnetars retain their strong magnetic fields for Myr or longer \citep{2023MNRAS.520.1872B}, indicating that highly magnetized NSs may still be involved in BNS mergers. If such a magnetar participates in a merger, it could drive significant precursor EM radiation and introduce pronounced asymmetry in the magnetic moments of the two stars. The existence of ``fast" but rare merger channels may also allow younger magnetars to merge before significant field decay occurs \citep{2024ApJ...966...17B}. While the overall prevalence of such systems remains uncertain \citep{2022LRR....25....1M}, multimessenger observations may provide constraints on their frequency.

For highly asymmetric magnetic moments ($|\lambda| \gg 1$), the magnetosphere is predominantly shaped by the more strongly magnetized star. The Poynting flux becomes highly concentrated near the equatorial plane, and the interface of the interacting magnetospheres shifts toward the weaker star.

Azimuthal phase offsets ($\Delta\Phi$) introduce time differences in flux evolution. As the orbital separation decreases, these offsets modify the final magnetic field geometry at the merger, potentially influencing postmerger magnetospheric activity and the resulting EM signals.

The scaling of EM luminosity with orbital frequency typically expressed as $\langle L \rangle \propto \Omega^p$ has been the subject of both theoretical and numerical studies, yielding a range of predictions that differ from $p=4$ for an isolated NS with a dipole field. Theoretical estimates for a magnetized star interacting with an unmagnetized companion suggest $p=13/3$ \citep{Lai2012}, assuming an energy dissipation model driven by flux tube reconnections. Models that account for shielding effects from a magnetized companion lead to a slightly shallower scaling of $p=10/3$ \citep{Palenzueala2013b}, while models considering the sum of the two luminosities lead to a scaling of $p=14/3$ \citep{Medvedev&Loeb}. Numerical simulations have reported a variety of power-law indices. Simulations tracking full inspirals found $p\approx 3/2$ \citep{Palenzueala2013b} for equal dipole moments, deviating from their theoretically expected $p=10/3$. In a highly asymmetric case ($\lambda = 100$), the power law steepens to $p=14/3$ at earlier inspiral stages and reaches $p \approx 12$ just before the merger \citep{Palenzueala2013b}. Other simulations \citep{Most&Philippov2020,Most&Philippov2022} using fixed orbital separations obtained $p \approx 7/3$ for peak luminosity scaling.

Our simulations reveal that the power-law index is not universal but depends sensitively on the magnetospheric configuration. Across our parameter survey, we find a wide range of exponents spanning $p \sim 1 - 6$ (Figure~\ref{Figure16}). However, regardless of inclination, we find a coarse scaling law of the average (over the different magnetic orientations) index $\overline{p} \sim (6\pm 0.2) - (3.6 \pm 0.1)|\lambda|^{-0.21 \pm 0.02}$, i.e., larger p is realized for larger $|\lambda|$ (Eq.~\ref{eq:Lfit}).

For highly magnetized primary stars ($|\lambda| > 100$) with moments aligned to the orbital axis, we find the index reaches $p \approx 4.3-4.6$, closely matching the $p =13/3$ \citet{Lai2012} prediction.
For cases where the dominant magnetic moment is inclined, the exponent deviates significantly, demonstrating that inclination effects alter the energy dissipation mechanism. In the equal aligned dipole case, we obtain $p \approx 3.6$, consistent with the expected $p =10/3$ for symmetric magnetized binaries \citep{Palenzueala2013b}.  

These results suggest that the premerger magnetospheric state plays a crucial role in shaping the overall evolution of EM luminosity. As discussed earlier, the variability in $p$ arises from the intrinsically complex nature of magnetospheric interactions, which themselves exhibit significant variation across the parameter space, leading to the continuous restructuring of open flux tubes throughout the inspiral. Our simulations confirm that these magnetospheric dynamics directly govern the evolution of open flux tubes, aligning with the observed luminosity variations.

\subsection{Electromagnetically Induced Torques, Crustal Stresses, and Potential Gravitational-wave Imprints}

In cases where at least one of the stars has a misaligned magnetic moment, the regions of peak outward EM flux periodically shift between the northern and the southern hemispheres during each orbital cycle. However, due to the ongoing inspiral, the total flux in each hemisphere progressively increases with each successive peak. This asymmetric flip-flopping leads to the {\it cumulative} buildup of a torque in a direction perpendicular to the plane defined by the orbital axis and the line connecting the centers of the two stars. Consequently, this effect can alter the orbital motion of the BNS system, causing the stars to oscillate up and down, with an average angle increasing over time. 

If magnetars are involved in such a BNS coalescence, this EM-induced deviation from a purely equatorial inspiral could imprint growing discrepancies in the gravitational waveform relative to standard unmagnetized NS merger models. Such systematic effects may be detectable with third-generation GW observatories, such as the Cosmic Explorer \citep{2017CQGra..34d4001A,2021arXiv210909882E,2023arXiv230613745E,2024FrASS..1186748C} or the Einstein Telescope \citep{2010CQGra..27s4002P,2011CQGra..28i4013H,2025arXiv250312263A}. These instruments are expected to achieve exceptionally high signal-to-noise ratios, exceeding $10^3$ on a nonnegligible subset of BNS mergers \citep{2024PhRvD.110h3040B}, potentially revealing these previously unexplored magnetospheric effects on the GW signal.

However, while our postprocessed analysis suggests the presence of this accumulating angular momentum transfer, a fully self-consistent treatment is required to confirm its impact. Future simulations should incorporate the modified orbital dynamics caused by EM forces and evolve the system dynamically, allowing for feedback between the magnetosphere, the NS spins, and the binary’s motion and eccentricity.  Furthermore, the model of irrotational NSs inspiraling along quasi-circular orbits becomes increasingly approximate in the presence of strong magnetic interactions. As shown in \cite{2000ApJ...537..327I}, where the equations of motion are derived from the Euler-Lagrange equations, including the interaction term of static magnetic dipoles in a vacuum, circular orbit solutions may not exist when the magnetic moments are misaligned with the orbital axis. Although our simulations assume an FF magnetosphere rather than vacuum fields, these findings suggest that a more rigorous treatment of the orbital dynamics is necessary to accurately capture the backreaction effects of EM forces. Additionally, the approximation of spherical stars neglects magnetically induced deformations, i.e., ellipticities, which may be nonnegligible in highly magnetized systems \citep[see][and references therein]{Chandrasekhar1981,2000ApJ...537..327I}. Such an approach will be crucial in assessing whether these effects leave indeed measurable imprints on GW signals (potentially hours before the merger) and determining their potential detectability.

Beyond influencing the binary’s orbital dynamics, the strong EM stresses induced by the evolving magnetosphere also act directly on the NS crusts, generating time-dependent, localized strain patterns. Our analysis reveals that these stresses vary over an orbital period, and for highly magnetized stars ($B \sim 10^{15}$ G), these stresses may become sufficiently strong to deform or even fracture the crust, potentially triggering magnetar-like bursting activity. If the crust fails, a transient X-ray burst similar to a magnetar giant flare could be produced, and the excitation of crustal normal modes could introduce additional observable signatures. These effects suggest that EM interactions in merging BNSs not only may modify the inspiral trajectory but also could lead to precursor high-energy transients detectable prior to coalescence.

\subsection{High-Energy Signals}

We also investigated the production of high-energy photons via curvature radiation from particles accelerated near the equatorial current sheets. We applied methodologies successfully used in pulsar studies, demonstrating how skymaps and light curves of high-energy emissions from BNSs can be constructed. These emission luminosities scale with the square root of the Poynting flux, following the heuristic dependence of $L_\gamma$ on the spin-down power at high $\dot{E}$ \citep{3PC}. Compared to the isolated pulsar case, we find a significantly more intricate emission structure, with light curves featuring extremely narrow peaks on the order of $<0.5$ ms. Detecting such rapid variability requires high temporal resolution and sensitivity, making Imaging Atmospheric Cherenkov Telescopes (IACTs) such as Cherenkov Telescope Array (CTA) strong candidates for observing these signals \citep{2019scta.book.....C, cherenkov_telescope_array_observatory_2021_5499840}. IACTs offer 6 to 8 orders of magnitude better gamma-ray sensitivity (at higher energies) on short timescales compared to Fermi-LAT. However, advanced triplet-based photon analyses in Fermi-LAT might enable a statistically significant association \citep{2021NatAs...5..385F,2023A&A...675A..99P} at lower photon energies.  The periodic nature of these signals also allows for photon-based coherent search techniques \citep[e.g.,][]{1989A&A...221..180D} hours or days before the merger, which can significantly enhance detectability. This requires precise premerger warnings and accurate localizations are available, capabilities that may only be feasible with third-generation GW networks or decihertz detectors.

Assuming that particles reach the radiation-reaction limit, we estimated the cutoff energy of the curvature radiation spectrum, finding a broad range of values: from TeV-scale energies for $B=10^{10}$ G up to PeV-scale energies for $B=10^{15}$ G. However, this assumption likely overestimates the maximum escaping photon energies because it does not account for the very efficient pair production in high-field regimes, which can suppress particle acceleration before reaching these extreme energies.

Our analysis indicates that B-$\gamma$ pair production is highly efficient in the strong magnetic fields present during the final inspiral stages. This leads to significant attenuation of high-energy photons, effectively limiting the maximum escaping energy to the MeV range in the last moments before the merger. As a result, while TeV to PeV emission is theoretically possible under radiation-reaction limit assumptions, most escaping photons in the late inspiral phases are likely to be limited to $\lesssim 100$ MeV energies where competition between pair production and particle acceleration might approach a dynamic equilibrium. This suggests that the MeV band, where numerous instruments are currently selected or proposed \citep[e.g.,][]{2019BAAS...51g.245M,2022JATIS...8d4003C,2022JCAP...07..036O,2024NuScT..35..149P,2024icrc.confE.745T,2025arXiv250214841S},  is the most promising range for detecting these emissions. Furthermore, as these pairs must be created in high Landau states, these pairs will subsequently radiate below 1 MeV, likely in the X-ray regime.

Although magnetic pair production dominates attenuation, photon-photon ($\gamma\gamma$) pair production could also contribute\footnote{For instance, \cite{2016MNRAS.461.4435M} crudely estimated the compactness parameter and found it to be large.}, particularly in the earlier inspiral stages when the magnetosphere is more transparent. This process, owing to the threshold kinematics and cross section, however, depends on the presence of a sufficient density of low-energy photons, which could arise from sources such as resonant crust shattering \citep{2012PhRvL.108a1102T,2022A&A...664A.177S}, tidal heating or crust melt \citep{2012ApJ...749L..36P,2019MNRAS.486.1424A,2020PhRvL.125t1102P,2021MNRAS.504.1273P}, or pair synchrotron radiation. However, due to the strong suppression of high-energy photons by B-$\gamma$ attenuation, $\gamma\gamma$ pair production is expected to be secondary except possibly for $\sim 0.1-1$~GeV photons, which should interact efficiently with possibly extant local X-rays. 
More critically, an open question remains regarding how much of the energy originally carried by high-energy photons is ultimately redistributed at lower frequencies by the pairs produced in both B-$\gamma$ and $\gamma\gamma$ interactions. Understanding the reprocessing efficiency of these pairs is crucial for assessing potential X-ray and optical emission, which may provide additional observational signatures beyond the MeV range.

Unraveling the full complexity of these effects requires simulations that self-consistently incorporate the interplay between particle acceleration, pair production, and photon transport processes. The FF approximations used in this study provide valuable insights, but more detailed kinetic plasma simulations (e.g., PIC approaches) that incorporate detailed microphysics will be essential for fully resolving these open questions.

A potential but uncertain source of precursor emission is coherent radio emission from these pair discharges, similar to polar cap radio generation in isolated pulsars \citep[e.g.,][]{2010MNRAS.408.2092T,2013MNRAS.429...20T,2020PhRvL.124x5101P,2024A&A...691A.137B}. If such emission occurs, it could be detectable far beyond the GW horizon \citep[e.g.,][]{Cooper2023}. However, rate estimates suggest that BNS mergers contribute only a small fraction of apparently nonrepeating FRBs \citep{2019NatAs...3..928R,2023RvMP...95c5005Z,2024ApJ...961...10M,2024ARNPS..74...89Z}, making this an unlikely but intriguing possibility.

\subsection{Postmerger Signals and Afterglow Echos} 

Because the EM energy escaping a BNS system is strongly anisotropic, most observers will not be situated where the bulk of the outflow is directed. However, even if an observer’s line of sight does not coincide with the initial direction of the outflow, interactions with the surrounding medium may lead to secondary or echo signals. In particular, magnetically dominated precursor outflows could generate a pulsar-wind-like shock \citep[e.g.,][]{2021MNRAS.501.3184S} before the actual merger, modifying the surrounding circumburst medium well in advance of the final coalescence. This interaction may later contribute to delayed EM signals that are distinct from standard postmerger afterglow components \citep[e.g.,][]{1998ApJ...494L..49S,1999ApJ...525..737R,2002ApJ...570L..61G}.

Unlike standard GRB afterglow echoes, where a delayed signal arises due to the expansion and deceleration of a jet-aligned outflow, premerger EM outflows may be emitted in directions distinct from both the observer's line of sight and the GRB jet. If these precursor-driven outflows later interact with a dense region of circumburst material, they can decelerate, expand, and reprocess their energy, leading to secondary radiation that eventually reaches the observer with a delay relative to the GW signal. The total observed delay of such an echo consists of two components: a geometric delay due to the additional distance traveled before encountering the medium and a dynamical delay associated with the deceleration and expansion of the outflow.

The geometric delay, $t_{\rm echo}^{\rm geo}$, arises from the precursor outflow propagating a distance $\ell$ before interacting with the medium and is given by $t_{\rm echo}^{\rm geo} \sim \frac{\ell}{c} \left(1-\cos\theta_{\rm 1}\right)$ where $\theta_{\rm 1}$ is the angle between the precursor outflow direction and the GRB jet axis. Additionally, the deceleration and expansion of the shocked outflow introduce a second-time delay, $t_{\rm peak}^{\rm off-axis}$, which depends on the dynamics of the outflow-medium interaction and the offset angle $\theta_{\rm 2}$ between the interaction site and the observer. This is approximately given by $t_{\rm peak}^{\rm off-axis}~\sim 100 \left(\frac{E_{\rm pre}}{10^{47} \, {\rm erg}}\right)^{1/3} \left(\frac{n_0}{10^{-3} \, {\rm cm^{-3}}}\right)^{-1/3} \theta_{\rm 2}^{8/3}~\text{days}$ where $n_0$ is the circumburst medium density, and $E_{\rm pre}$ is the total energy carried by the precursor Poynting-flux-dominated outflow (see Fig.~\ref{Figure14}). The total observed time delay of the echo is then given by $t_{\rm echo}^{\rm tot}/(1+z) \sim t_{\rm peak}^{\rm off-axis} + t_{\rm echo}^{\rm geo}$. If the precursor-driven outflow interacts with a dense clump of circumburst material ($n_0 \gg 10^{-3} \, {\rm cm^{-3}}$) far from the merger symmetry axis, the echo could be significantly earlier, appearing weeks or even sooner after the direct GW and merger-associated EM signals.

Such precursor-driven echoes may contribute to late-time rebrightening in short GRBs. For instance, GW170817 exhibited prolonged X-ray afterglow rebrightening up to 1000 day postmerger \citep{Ruan2018, Troja2020}, and some short GRBs have shown late-time radio activity of unknown origin \citep[e.g.,][]{2024ApJ...970..139S}. While these events are often attributed to prolonged central engine activity or structured jets, pre-merger EM outflows could present an alternative or complementary explanation. Notably, precursor-driven echoes could also appear as orphan afterglows associated with GW events at unusually large inclination angles.

Yet, several uncertainties persist. The presence of a sufficiently strong premerger outflow requires at least one highly magnetized NS. The environment at off-axis angles must allow efficient shock formation, and the resulting emission must be strong enough to be detectable. Hence, while premerger echoes offer an intriguing explanation for some late-time GRB rebrightenings, their actual contribution and detectability remain to be tested by future deeper multiwavelength observations, particularly in the radio and X-ray bands.

\section{Summary and Future Directions}\label{sec:Summary and Future Directions}

In this study, we performed FF simulations of BNS magnetospheres, exploring their evolution, EM outflows, and potential observational signatures. We analyzed the structure of the Poynting flux, high-energy emission distributions, and the development of EM torques, which may have subtle but cumulative effects on the orbital motion. In particular, by simulating a wide range of magnetic configurations, we confirmed several robust features of BNS magnetospheric interactions while also uncovering novel trends. We systematically quantified the strong anisotropy of the Poynting flux distributions and established, for the first time, a configuration-dependent power-law relationship between the Poynting flux luminosity and the orbital frequency. The power-law index was found to vary significantly with the relative alignment and strength of the stellar magnetic moments. Furthermore, we introduced a new method to construct high-energy emission skymaps and light curves for BNS systems by generalizing techniques originally developed for isolated pulsars. Finally, our analysis of the photon spectra revealed that, although extremely high-energy photons (up to TeV-PeV) can be produced in the radiation reaction limit, magnetic pair production severely attenuates them. As a result, the escaping emission is constrained to the MeV range, identifying this as a promising observational window for precursor high-energy signals. While our results provide insight into the fundamental mechanisms shaping magnetospheric interactions, several aspects remain to be explored.

In particular, our analysis neglects tidal distortions and magnetosphere-induced forces on the NS surfaces, which could contribute to additional dynamics, such as crustal deformations or resonant crust-shattering events. Although these effects are unlikely to alter the qualitative structures derived here, they may influence the fine details of precursor EM signals. Future studies should incorporate these effects self-consistently, alongside kinetic plasma simulations and more advanced MHD treatments, to refine the modeling of pair production, particle acceleration, and radiation transport. Additionally, incorporating full general relativistic effects will be necessary to capture the intricate coupling between magnetospheric evolution and the strong-gravity environment near the merger. These improvements will be essential for more accurate predictions of precursor signals and their detectability in upcoming multimessenger observations.

\begin{acknowledgments}
We would like to thank the anonymous referee for useful comments that helped to improve the manuscript. We also acknowledge insightful discussions with Ioannis Contopoulos, Paz Beniamini, Lauren Rhodes, and Vikram Ravi. The material is based upon work supported by NASA under award numbers 80GSFC21M0002 and 80GSFC24M0006, 80NSSC21K1999, 22-ADAP22-0142, 22-TCAN22-0027, 21-ATP21-0116, and 22-FERMI22-0035. Resources supporting this work were provided by the NASA High-End Computing (HEC) Program through the NASA Advanced Supercomputing (NAS) Division at Ames Research Center. This work has made use of the NASA Astrophysics Data System. 

\end{acknowledgments}

\software{}
For the analysis and visualization Mathematica \citep{Mathematica}, Matplotlib \citep{Hunter:2007}, Plotly \citep{plotly}, PyVista \citep{sullivan2019pyvista}, CMasher\citep{2020JOSS....5.2004V}, Cartopy \citep{Cartopy}, NumPy \citep{harris2020array}, Numba \citep{10.1145/2833157.2833162} and Scipy \citep{2020SciPy-NMeth} are used.

\appendix
\section{Resolution Tests and Robustness of Results}
\label{AppendixA}

\begin{figure}
\centering
    \includegraphics[width=.5\textwidth]{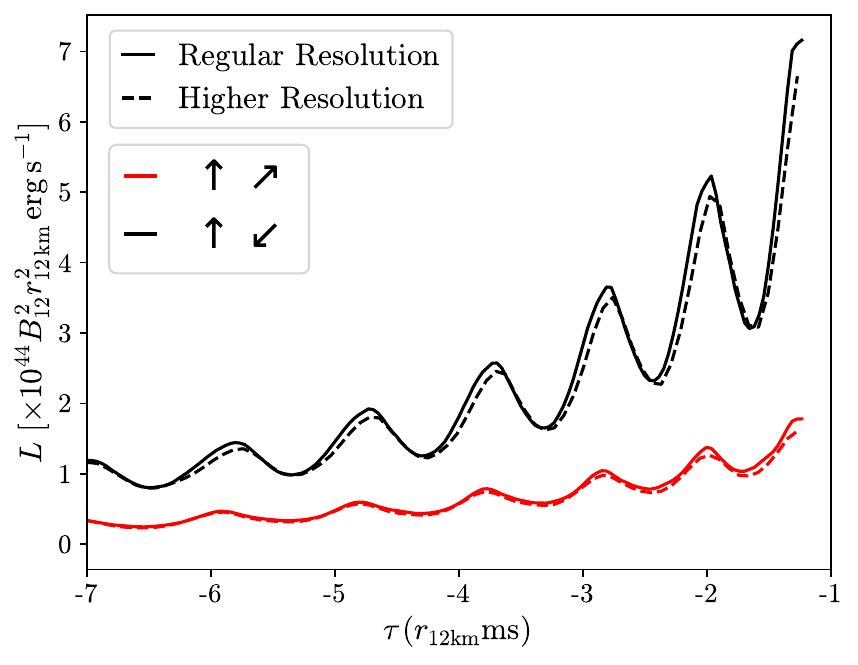}\hfill
    \caption{The Poynting flux evolution for the cases with $\alpha_1=0^\circ, \alpha_2=45^\circ, \Delta\Phi=0^\circ$, and $\lambda=1$ (\AngledUpArrow{0}\AngledUpArrow{-45}) in red, and $\lambda=-1$ (\AngledUpArrow{0}\AngledDownArrow{-45}) is in black with the regular (solid line) and double (dashed line) resolution. }
    \label{FigureA1}
\end{figure}

\begin{figure}
\centering
    \includegraphics[width=1.\textwidth]{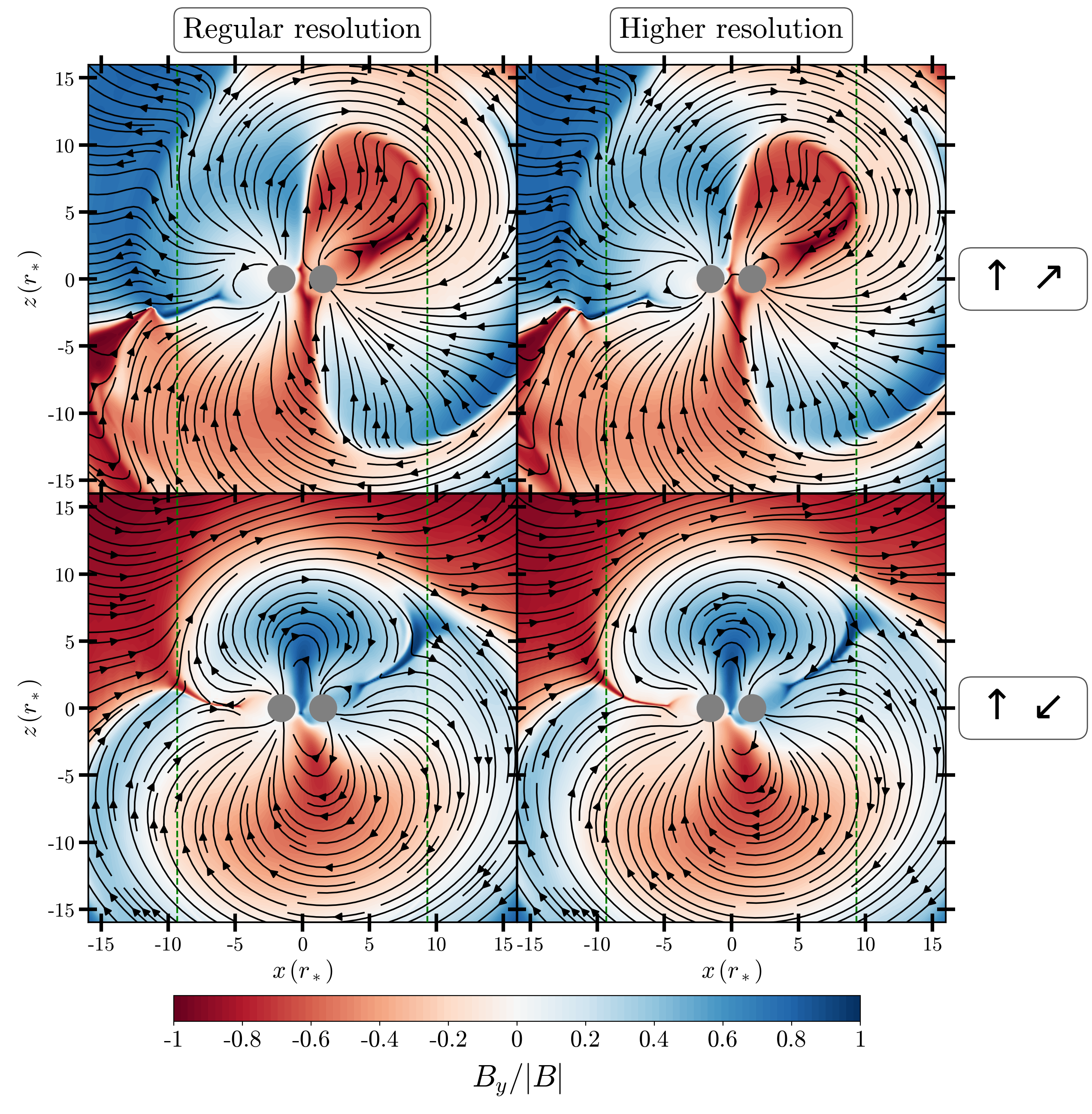}\hfill
    \caption{Similar to Figure \ref{Figure8} the structure of the magnetosphere, at $\tau\simeq-7ms,$ for the cases with $\alpha_1=0^\circ, \alpha_2=45^\circ, \Delta\Phi=0^\circ$, and $\lambda=1$ (\AngledUpArrow{0}\AngledUpArrow{-45}) in \textbf{top} row while $\lambda=-1$ (\AngledUpArrow{0}\AngledDownArrow{-45}) in \textbf{bottom} row. \textbf{Left} are the simulations with the resolution used in the main text while \textbf{right} are the higher-resolution simulations.}
    \label{FigureA2}
\end{figure}

\begin{figure}
\centering
    \includegraphics[width=0.96\textwidth]{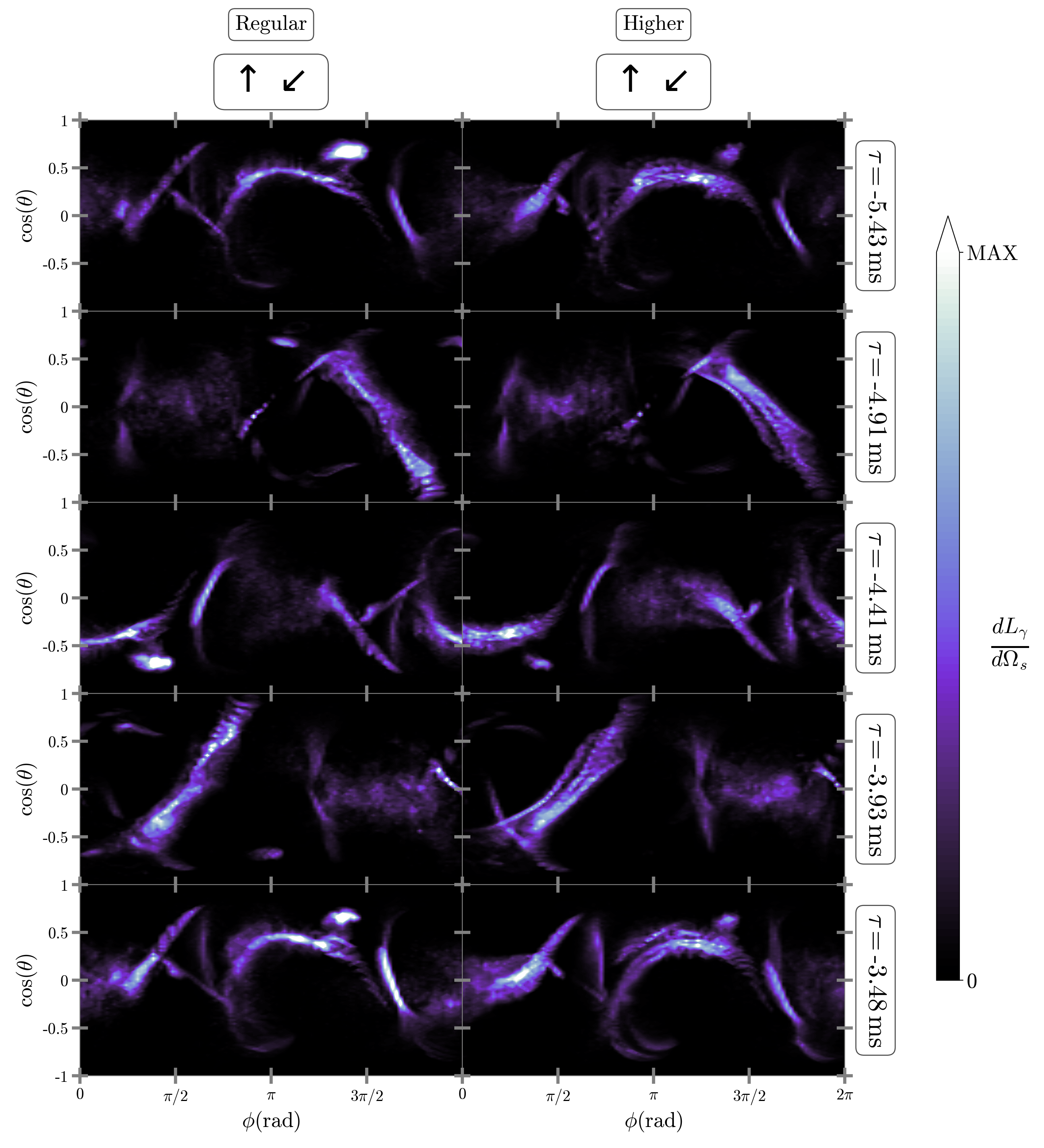}\hfill
    \caption{Similar to Figure \ref{Figure20}. Snapshots of the evolution of high-energy emission skymaps for the case with $\alpha_1=0^\circ, \alpha_2=45^\circ, \Delta\Phi=0^\circ$, and $\lambda=-1$ (\AngledUpArrow{0}\AngledDownArrow{-45}) in higher resolution. }
    \label{FigureA4}
\end{figure}

To assess the robustness of our results, higher resolution were conducted for selected cases. For these simulations, a double spatial and time resolution was employed relative to the baseline simulations. Specifically, the grid size was reduced to $0.05r_\star$ - half compared to the one used for the main results - and the time step to $dt=0.02r_\star/c$ in order to maintain the same Courant number. 

In Figure \ref{FigureA1}, the Poynting flux evolution is shown for the cases with $\alpha_1=0^\circ, \alpha_2=45^\circ, \Delta\Phi=0^\circ$, and $\lambda=1$ (\AngledUpArrow{0}\AngledUpArrow{-45}) in red while $\lambda=-1$ (\AngledUpArrow{0}\AngledDownArrow{-45}) is in black. The solid lines correspond to the simulations with the baseline resolution used in the rest of this study, while the dashed lines correspond to simulations with double the resolution.

In both cases, the Poynting flux evolution obtained from the higher-resolution simulations is in great agreement with that of the baseline simulations presented in the main text, indicating that our main results are not sensitive to the resolution. As expected from such flux curves, the values of the power-law index of the higher-resolution runs are indistinguishable in both resolutions, $3.3$ and $2.9$ for the \AngledUpArrow{0}\AngledUpArrow{-45} and \AngledUpArrow{0}\AngledDownArrow{-45} case, respectively.

\begin{figure}{H}
\centering
    \includegraphics[width=0.5\textwidth]{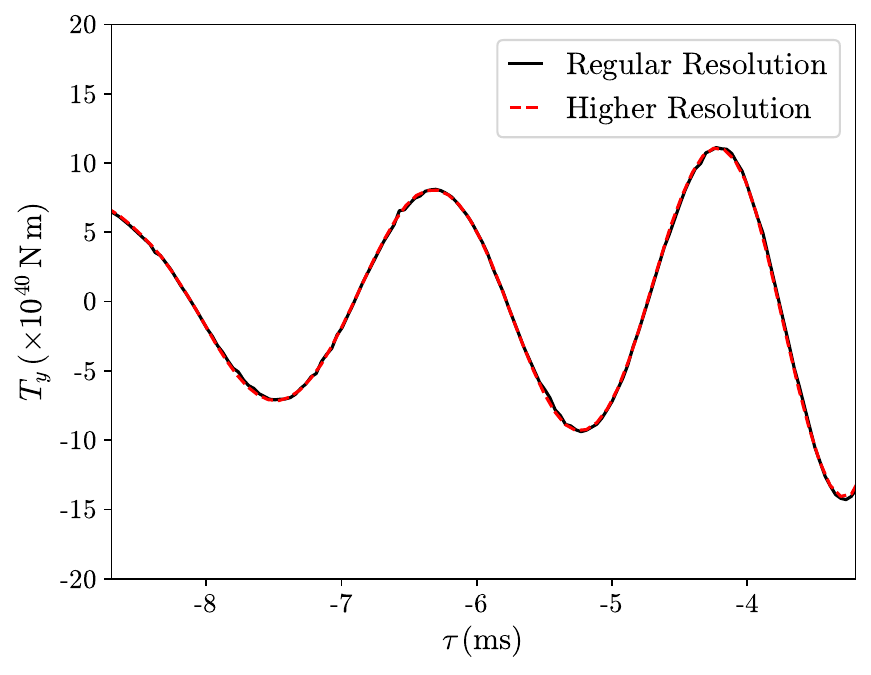}\hfill
    \caption{The evolution of the torque for the case with $\alpha_1=0^\circ, \alpha_2=45^\circ, \Delta\Phi=0^\circ$, and $\lambda=-1$ (\AngledUpArrow{0}\AngledDownArrow{-45}) in the regular (solid line) and higher resolution (dashed line). }
    \label{FigureA3}
\end{figure}

To further verify the fidelity of our adopted resolution in the main text, we present in Figure \ref{FigureA2}, the magnetospheric structure at $\tau\simeq-7ms$ for both configurations across both resolutions. In addition, Figure~\ref{FigureA4} compares the evolution of the high-energy emission skymaps from both resolutions of the (\AngledUpArrow{0}\AngledUpArrow{-45}) case. 
Lastly, Figure~\ref{FigureA3} displays the evolution of the EM torque for both resolutions.

Across all diagnostics, the global magnetospheric structures exhibit only minor differences. The torque evolution is nearly identical, further supporting the convergence of our results. The high-energy emission skymaps,  while exhibiting some small discrepancies, particularly in finer details, remain consistent in their overall morphology and dominant features. These differences are understandable, as the skymap construction depends on localized regions of strong accelerating electric fields, which are more sensitive to resolution-dependent magnetic topology variations.

Overall, the higher-resolution simulations validate the accuracy and robustness of our primary results. The modest differences observed suggest that the resolution used in the main analysis is sufficient to capture the essential features without incurring significantly higher computational cost.

\clearpage
\bibliography{bibfile}{}
\bibliographystyle{aasjournal}

\end{document}